\documentclass[openany,doublespaced,twoside,a4paper]{amsbook}
\numberwithin{equation}{chapter}
\numberwithin{figure}{chapter}
\numberwithin{table}{chapter}
\numberwithin{section}{chapter}

\renewcommand{\baselinestretch}{1.5}
\usepackage{fixltx2e}
\usepackage{graphicx}
\usepackage{pdfpages}
\begin{document}
\frontmatter
\title[Machine Learning for Astrophysics]{Machine Learning Techniques for Astrophysical Modelling and Photometric Redshift Estimation of Quasars in Optical Sky Surveys}
\author[N.\,D. Kumar]{N. Daniel Kumar}
\address{St Anne's College, Oxford, OX2 6HS, United Kingdom}
\email{daniel.kumar@st-annes.ox.ac.uk}
\date{September 2008}

\begin{abstract}
Machine learning techniques are utilised in several areas of astrophysical research today. This dissertation addresses the application of ML techniques to two classes of problems in astrophysics, namely, the analysis of individual astronomical phenomena over time and the automated, simultaneous analysis of thousands of objects in large optical sky surveys. Specifically investigated are (1) techniques to approximate the precise orbits of the satellites of Jupiter and Saturn given Earth-based observations as well as (2) techniques to quickly estimate the distances of quasars observed in the Sloan Digital Sky Survey. Learning methods considered include genetic algorithms, particle swarm optimisation, artificial neural networks, and radial basis function networks.

The first part of this dissertation demonstrates that GAs and PSO can both be efficiently used to model functions that are highly non-linear in several dimensions. It is subsequently demonstrated in the second part that ANNs and RBFNs can be used as effective predictors of spectroscopic redshift given accurate photometry, especially in combination with other learning-based approaches described in the literature. Careful application of these and other ML techniques to problems in astronomy and astrophysics will contribute to a better understanding of stellar evolution, binary star systems, cosmology, and the large-scale structure of the universe.
\end{abstract}

\includepdf{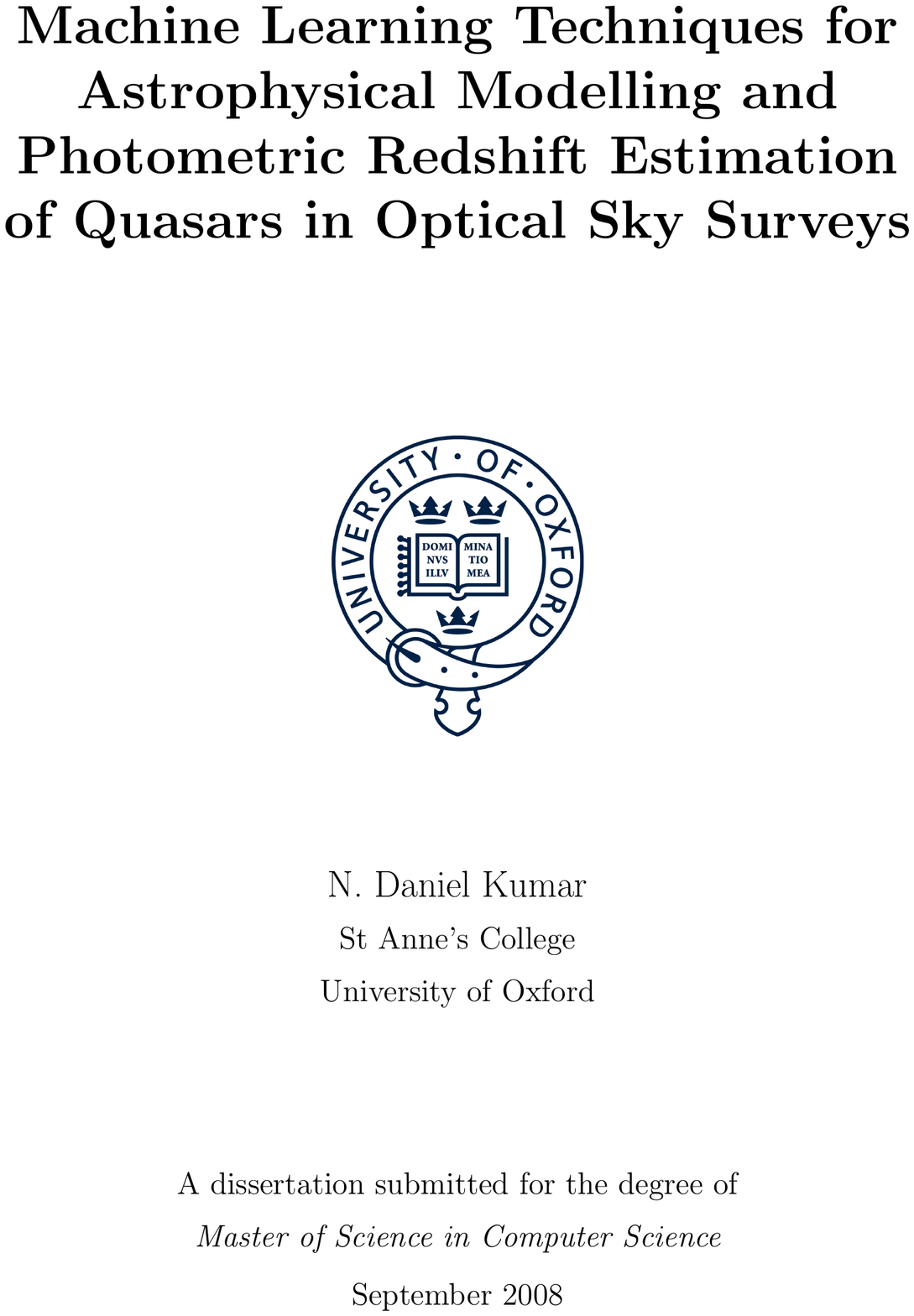}
\includepdf[pages=2]{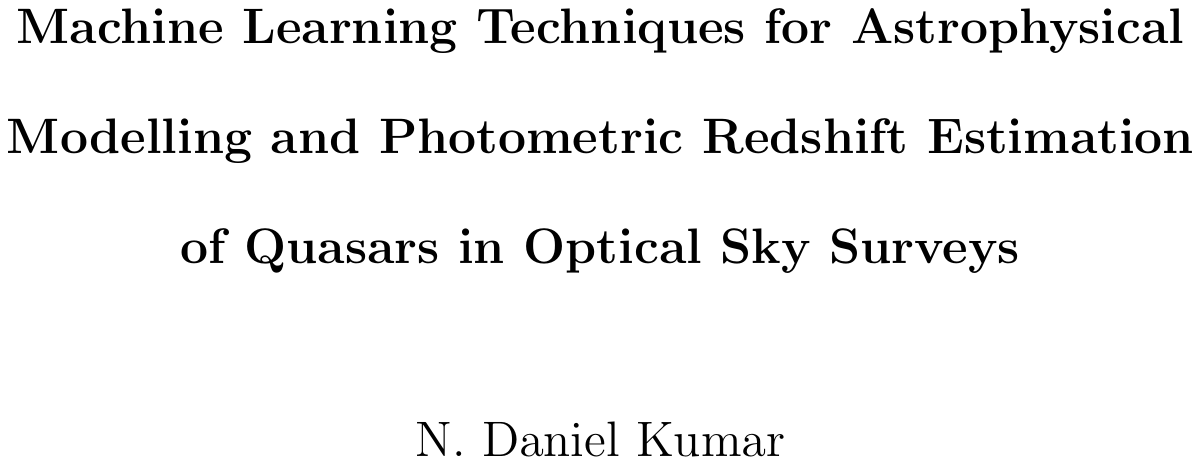}

\tableofcontents
\listoffigures
\listoftables

\chapter*{Acknowledgements}
I gratefully acknowledge the assistance of \textbf{Dr Vasile Palade}, Oxford University Computing Laboratory, for his introducing me to the field of machine learning as well as his supervision of this project; \textbf{Dr Somak Raychaudhury}, School of Physics and Astronomy, University of Birmingham, for his suggestions about photometric redshift estimation; \textbf{Mr Andrew Foster}, UNC-Chapel Hill, for his useful contributions regarding the implementation of evolutionary algorithms and modelling techniques; \textbf{Dr Daniel Reichart}, Department of Physics and Astronomy, UNC-Chapel Hill, for his introducing me to astrophysical and statistical modelling and his providing me with initial ideas about the moons of Jupiter; \textbf{Dr Daniel Vanden Berk}, Department of Astronomy and Astrophysics, Penn State University, for his assistance in interpreting measurement errors in the SDSS quasar dataset; and \textbf{Dr Ani Calinescu}, Oxford University Computing Laboratory, for her attentive guidance throughout the year.

I do not, of course, neglect to express my gratitude and indebtedness to my friends, my family, and especially my parents, without whose continuous support and encouragement the following dissertation could not have been written.

\begin{flushright}
\ \\
NDK \\
Oxford \\
September 2008
\end{flushright}

    \chapter{Introduction: Astrophysical and Astronomical Applications}
        \section{Motivation}
            \subsection{Modelling Astrophysical Phenomena}
            One basic problem that has always confronted astrophysicists is that of coming to understand the physical dynamics behind a set of astronomical observations over time, i.e., trying to learn what causes the behaviour we see from our telescopes on Earth. On a superficial level, however, we often observe behaviour that is at best indirectly related to the actual dynamics of a system, or even seemingly counterintuitive. For example, we might observe a single pulsating beam of light in the sky, progressing from bright to very bright to dim to very bright, and so on: what we are actually seeing is two stars of unequal magnitude rotating around each other in our line of sight, eclipsing one another in turn as they progress.

            Astrophysical modelling allows us to take theories of expected astrophysical behaviour, and, combining them with order-of-magnitude estimates, apply them to a particular phenomenon to determine whether our observations are consistent with our expectations. When we understand precisely the physical behaviour in an observed system (e.g., the simple orbit of a moon around a planet), we are able to determine specific parameters of the system (say, the elements that completely specify the moon's orbit) by modelling on our observations.

            \subsection{Data Mining in Large Astronomical Sky Surveys}
            A newer problem beginning to confront astronomers is the need to analyse and make sense of data obtained in large sky surveys such as the Sloan Digital Sky Survey (SDSS) \cite{dY00}, the 2dF QSO Redshift Survey (2QZ), and the forthcoming Panoramic Survey Telescope and Rapid Response System (Pan-STARRS) and Large Synoptic Survey Telescope (LSST). Once in operation, these newer surveys will collect terabytes \cite{iW05} of data each night, orders of magnitude more than could be analysed by any team using traditional techniques.

            In order to conduct useful science on these tera- and petascale survey databases, more efficient and accurate data mining techniques are needed---especially those that require little or no human intervention. As survey databases continue to grow by orders of magnitude, our analysis techniques will have to keep pace, and machine learning techniques are particularly well suited for the task.

        \section{Traditional vs. Learning-Based Approaches}
        Learning algorithms may be preferred for their accuracy, efficiency, and/or scalability. For example, linear or quadratic polynomial fitting or nonlinear regression may be straightforwardly used to model a system or estimate a relation,\footnote{Cf. \cite{mW06} and especially \cite{bH05}.} but can often lead to ``large systematic deviations'' \cite{dW08}. A nonlinear solution estimated with a learning technique may be harder for humans to make sense of (as with `black box' techniques), but can be arbitrarily more complex---and thus sometimes arbitrarily more precise.

        The traditional method of photometric redshift estimation before learning techniques were introduced had been spectral energy distribution (SED) template-fitting: a theoretical composite SED is developed by averaging the spectra of hundreds or thousands of objects that are thought to be parametrically similar (e.g., high-$z$ quasars, Seyfert galaxies, or Type Ia supernovae). Absorption and emission lines are identified and shifted into place, and the resulting SED is then redshifted to several values to form a set of templates \cite{eH00}. Photometric measurements are then compared to these templates, and a best-fit redshift is determined by minimisation over some chi-square distribution. Unfortunately, this method requires a set of representative templates, and cannot be reasonably utilised for astronomical objects that cannot be safely classified\footnote{Cf. \cite{sK04}} (e.g., quasars, which were until recently not well understood, and which are difficult enough to distinguish from stars in colour-colour space \cite{gR01a}).

        Particularly useful also is the ability to adjust a learning algorithm to suit a particular problem-specific need (say, for efficiency over accuracy, or for distributed computation over an unknown number of systems), which is often not possible with traditional statistical approaches.

        \section{Project Aims and Dissertation Structure}
        The aims of this project are two-fold: we aim firstly to demonstrate the applicability of learning techniques to two subsets of current astrophysical problems, and secondly to compare the advantages and disadvantages of each learning method considered. It is hoped that interdisciplinary investigations such as this will help further cooperation between those with domain-specific knowledge of astrophysical problems and those with a grasp of the theoretical foundations of machine learning techniques---a cooperation that is becoming increasingly important for the future of astrophysical research.

        This dissertation is divided into two parts: Part I considers astrophysical modelling of orbiting satellites, and Part II considers the problem of photometric redshift estimation of quasars. Within Part I, Chapter 2 deals with the fundamentals of orbital mechanics and the calculation of satellite position according to Kepler's laws. Chapter 3 discusses the genetic algorithm approach to this modelling problem, and Chapter 4 compares this to an approach based on particle swarm optimisation. In Part II, Chapter 5 presents the topic of redshift estimation of quasars. Chapters 6 and 7 discuss artificial neural networks and radial basis function networks as techniques for redshift estimation and present analyses of the results, and Chapter 8 is discussion. Finally, by way of conclusion, Chapter 9 elaborates on the implications of this research and on possible future directions.

\mainmatter

\part{Astrophysical Modelling}
    \chapter{Background}
        \section{The Orbits of the Satellites of Jupiter}
\begin{figure}
        \centering\includegraphics[width=126mm]{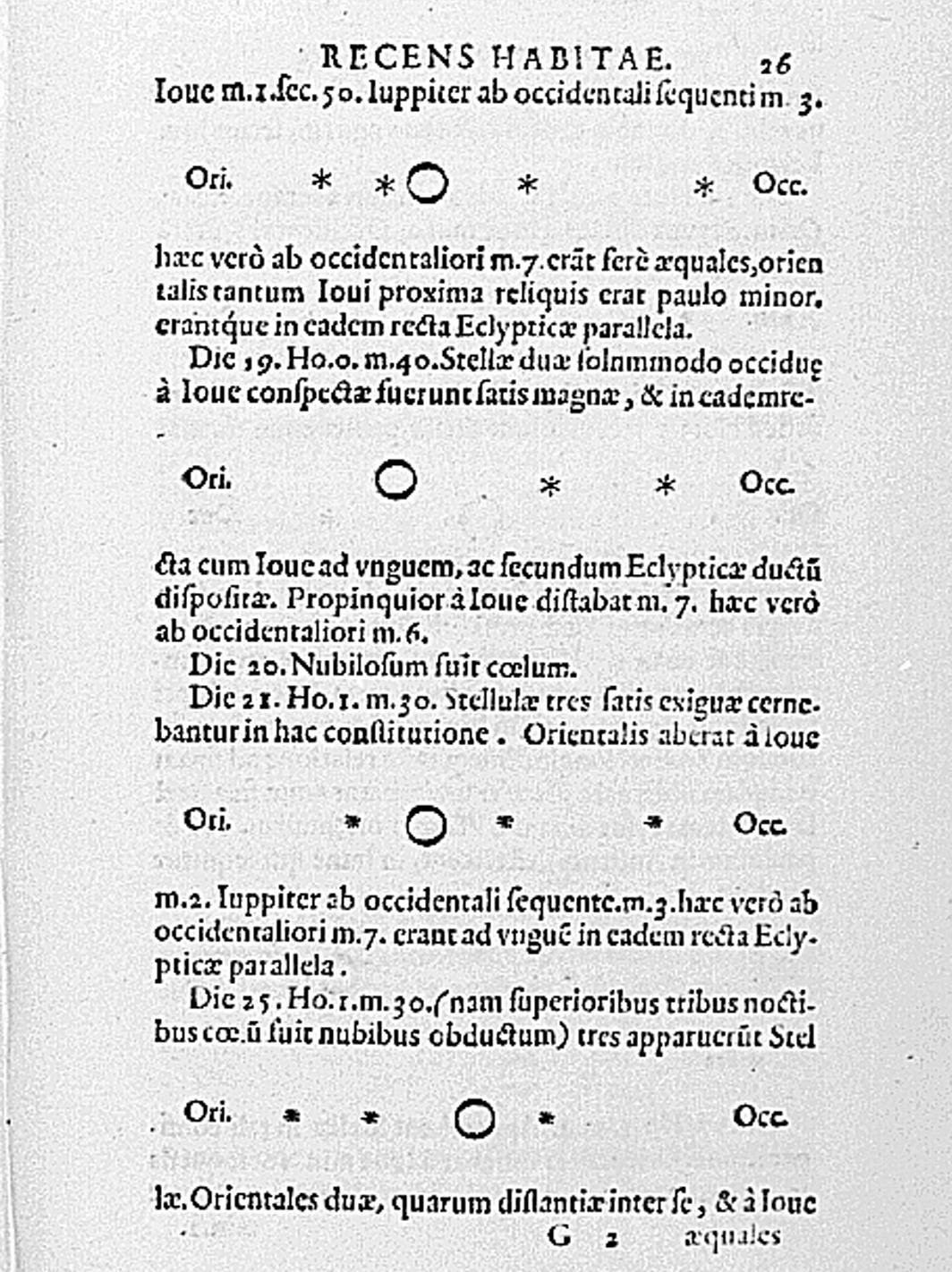}
        \caption[Galileo's observations of Jupiter in \emph{Sidereus Nuncius}. Image courtesy of the Master and Fellows of Trinity College, Cambridge.]{Galileo's observations of Jupiter in \emph{Sidereus Nuncius}. From \texttt{http://www.hps.cam.ac.uk/starry/galileo.html}; image courtesy of the Master and Fellows of Trinity College, Cambridge.}\label{F:sidnunc}
        \end{figure}

        The Galilean Satellites of Jupiter, so-called because of their discovery by Galileo in 1610, were some of the first subjects of astrophysical modelling. Galileo could not directly observe the four moons' orbits around the planet, but he observed their change in position around Jupiter over the course of a month (Figure \ref{F:sidnunc}). From these observations he was able to induce an elementary theory about what was causing the motions of the little `stars': they were actually orbiting around the major planet. This theory had implications that fit the observed evidence, and it was deemed to be the most likely explanation for the phenomenon.

        The problem undertaken here is, in principle, to determine all of the six orbital elements \cite[p. 58]{rB71} that precisely describe the orbit of a given satellite around Jupiter, using only observations available from telescopes on Earth. Given the well defined Keplerian theory of celestial mechanics, this problem is essentially one of function approximation in six dimensions. As such, it is similar in form to any other classical modelling problem in astrophysics, and demonstrates the applicability of learning techniques to the approximation of highly parametric models.

        More immediately, however, this solution of approximating celestial orbits can be used to study binary star systems, comets with highly eccentric orbits, and planetary star systems, as well as astrodynamical problems such as optical or lossy radar-based tracking of missiles and satellites.

        \section{Data}
        Accurate ephemerides (calculated positions of astronomical objects) are available from the HORIZONS System\footnote{\texttt{http://ssd.jpl.nasa.gov/?horizons}} at NASA's Jet Propulsion Laboratory. This system was used to simulate observations of Jupiter and one of its moons from a ground-based optical telescope at Oxford, England, over a period of 31 days, at intervals of 2 hours. For simplicity, observations were simulated during daylight as well as at night. The data provided were in the form of right ascension ($\alpha$) and declination ($\delta$) coordinates---references to an astronomical coordinate system independent of the rotation of the Earth \cite[p. 56]{rB71}.

        These two-dimensional coordinates essentially show the location of Jupiter and one of its moons on an XY-plane, and, together with an arbitrary\footnote{Using only point coordinates in two-space, it is impossible to calculate certain orbital elements and target distance simultaneously. The semi-major axis (see \S\ref{S:orbitelem}) of the moon's orbit would vary linearly with distance.} distance, are sufficient to describe a satellite's orbit precisely over time. However, in order to save time and minimise computational rounding errors, we opted to use data describing the relative location of a moon to Jupiter in three dimensions. While this is not just a simple geometric transformation of two-dimensional data into three-space (it does contain more information per se), the techniques should be similarly effective on the 2-D data: they would simply take longer to converge in light of the additional ambiguity.

        We first used a test dataset with simple orbital elements to optimise the GA and PSO parameters, and then we applied the techniques with these parameters to actual satellite ephemerides.
        \section{Classical Orbital Elements}\label{S:orbitelem}
\begin{figure}
        \centering\includegraphics[width=126mm]{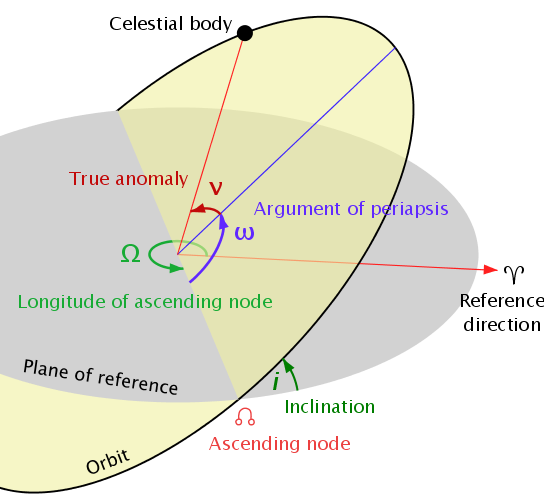}
        \caption[Classical angular orbital elements.]{Classical angular orbital elements. From \texttt{http://en.wikipedia.org/wiki/Orbital\_elements}.}\label{F:orbitelem}
        \end{figure}

        Six quantities \cite[p. 58]{rB71} are enough, using classical mechanics, to describe the orbit of a satellite around a major body:
        \begin{enumerate}
            \item $a$, \emph{semi-major axis}, half the length of the longest line that can be drawn across the orbital ellipse,
            \item $e$, \emph{eccentricity}, varying from 0 (circular orbit) to nearly 1 (almost parabolic),
            \item $i$, \emph{inclination}, the angle at which the orbital plane is removed from the equatorial (fundamental) plane,
            \item $\Omega$, \emph{longitude of the ascending node}, the angle at which the satellite `ascends' (in a northerly direction) across the equatorial plane measured anticlockwise from the reference direction ($\gamma$, where $\alpha = 0^\circ$) when viewed from the north side of the major body,
            \item $\omega$, \emph{argument of periapsis}, the angle in the orbital plane between the ascending node and periapsis---the point at which the satellite is closest to the major body, and
            \item \label{I:M0}$M_{epoch}$, \emph{mean anomaly at epoch},\footnote{For comparison to mean orbital elements at \texttt{http://ssd.jpl.nasa.gov/?sat\_elem\#jupiter}, this approach deviates from \cite{rB71}: Bate et al. (1971) refer instead to $T$, the \emph{time of periapsis passage}. $M_{epoch}$ can be derived from $T$ given the epoch, current time $t$, current mean anomaly $M$, and $\omega$.} roughly the angle the satellite has travelled about the centre of the major body at a predefined time \emph{epoch},\footnote{The term epoch is used because orbital relationships slowly deviate (precess) over time, and an approximate time frame must therefore be specified.} measured anticlockwise from periapsis when viewed from the north.
        \end{enumerate}
        See Figure \ref{F:orbitelem} for an illustration of the angular elements. In the figure, true anomaly $\nu$ can be thought of as representing mean anomaly at a given time $M$, although it is calculated according to \eqref{E:nu}.

        \section{The Kepler Problem}\label{S:keplerian}
        The calculation of position in an orbit (in terms of the orbital elements described in \S\ref{S:orbitelem}) as a function of time is known as the Kepler problem,\footnote{The term `Keplerian problem' is occasionally used, where the Kepler problem is taken to mean the two-body problem in classical mechanics; cf. \texttt{http://en.wikipedia.org/wiki/Keplerian\_problem}.} as it requires the solution of Kepler's equation \cite[pp. 185, 193]{rB71}:
        \begin{equation}\label{E:kepler's}
        M = M_{epoch} + n \cdot \Delta t = E - e \cdot sin(E)
        \end{equation}
        where $n$ is the \emph{mean motion},\footnote{The mean motion is the average rate of progression of a satellite around its major body, $\sqrt{\mu/a^{3}}$, where the standard gravitational parameter $\mu = GM$, the gravitational constant times the mass of the major body.} $\Delta t$\footnote{We deviate from the presentation of Bate et al. (1971) again, for consistency with \eqref{I:M0} above.} is the time elapsed since the current epoch, $E$ is the \emph{eccentric anomaly} \cite[p. 183]{rB71}, another measure of progression in orbit, and $M_{epoch}$ and $e$ are as above.

        Beginning with hypothetical values of $M_{epoch}$, $n$, and $\Delta t$ that we want to test, we approximate $E$ (as described below), and then solve for the \emph{true anomaly} $\nu$ and \emph{distance} $\rho$, simple polar coordinates representing the position of the satellite about the orbital plane:
        \begin{subequations}\label{E:nu-rho}
        \begin{equation}\label{E:nu}
        \nu = 2 \cdot \arctan{(\sqrt{\frac{1 + e}{1 - e}} \cdot \tan{\frac{E}{2}})}
        \end{equation}
        \begin{equation}\label{E:rho}
        \rho = a \cdot \frac{1 - e^2}{1 + e \cdot cos(\nu)}.
        \end{equation}
        \end{subequations}
        From $\nu$ and $\rho$, a simple transformation into Cartesian coordinates in 3-space---where the x-axis is aligned with the reference direction in the equatorial plane ($\alpha = 0^\circ$), the y-axis is positive at $\alpha = 90^\circ$, and the z-axis is positive northwards---allows us to compare our calculation with our `observational' data.

        As we cannot isolate the value of $E$ in Kepler's equation (it is `transcendental' in $E$), we must use a numerical solution to approximate it. This we have done with nine iterations (beginning with $E_{0} = M$) of $E_{i+1} = M + e \cdot sin(E_{i})$.\footnote{\texttt{http://en.wikipedia.org/wiki/Eccentric\_anomaly}}

    \chapter{Genetic Algorithms for Astrophysical Modelling}\label{C:GAs}
        \section{Fundamentals of GAs}
        The inspiration for \emph{genetic algorithms} (GAs) comes from the biological processes of evolution and natural selection \cite[p. 222]{mN02}. In constructing a GA, one first initialises a population of randomised \emph{organisms}, each of which represents the set of parameters one wants to approximate. The population is then evolved through many generations wherein organisms are allowed to \emph{crossover} with each other, thereby creating offspring organisms that take parameters from both parent organisms.

        In each generation, given some definition of \emph{fitness}, the healthier organisms are (probabilistically) more likely to survive, allowing the overall population to approach the desired characteristics. A fraction of organisms are \emph{mutated} in each generation as well, to reduce the likelihood of convergence on a local (rather than global) minimum. A description of a standard GA is given in Table \ref{T:GAalg}.

        \begin{table}[th]
        \fbox{\parbox{126mm}{\small GA($Fitness$, $p$, $s$, $m$) \\
        $Fitness$: Function that calculates accuracy. \\
        $p$: Size of the population. \\
        $s$: Fraction of the population `selected' between generations; others are replaced by crossover. \\
        $m$: Mutation rate.
        \noindent
        \begin{description}
        \item[Initialise] $P$ $\leftarrow$ Randomly generate $p$ organisms.
        \item[Evaluate] For each $o$ in $P$, find $Fitness(o)$.
        \item[Create a new generation, $P_S$] \
            \begin{enumerate}
            \item{Select:} Probabilistically select $s \cdot p$ members of $P$ to add to $P_S$. The probability Pr($o$) of selecting organism $o_i$ from $P$ is
                \begin{equation*}
                \text{Pr}(o_i) = \frac{Fitness(o_i)}{\sum_{j=1}^p Fitness(o_j)}.
                \end{equation*}
            \item{Crossover:} Probabilistically select $\frac{(1-s) \cdot p}{2}$ pairs of organisms from $P$ according to Pr($o_i$) above. Crossover each pair, adding offspring to $P_S$.
            \item{Mutate:} Choose $m$ percent of the members of $P$ with uniform probability, randomly modifying their parameters slightly.
            \item{Update:} $P \leftarrow P_S$.
            \item{Evaluate:} For each $o$ in $P$, find $Fitness(o)$.
            \item{Repeat:} Repeat if minimum fitness or generation threshold has not been reached, or until manually halted.
            \end{enumerate}
        \item[Return] Output the organism $o$ in $P$ with the highest fitness.
        \end{description} }}
        \\[5Pt]\caption{A canonical genetic algorithm. Adapted from Mitchell (1997) \cite[p. 251]{tM97}.}\label{T:GAalg}
        \end{table}

            \subsection{Advantages of GAs}
            As GAs are flexible and easily implemented, they provide a straight-forward solution to the modelling problem presented here. Simple modification of the fitness function would allow us to encourage the GA to converge on some orbital elements more strictly than others, if desired; additionally, incorporation of error margins into the fitness function would enable us to account for measurement errors and systematic deviations in our observational training data.\footnote{This idea was suggested by Dr Daniel Reichart, Dept of Physics and Astronomy, UNC-CH.}

            Stochastic elements in the GA induce a ``randomised, parallel beam search,'' \cite[pp. 252, 259]{tM97} which is important to keep in mind when approximating highly nonlinear functions. In particular, although the GA is not immune to the problem of local minima in the error function (equivalent to local maxima in the fitness function), it avoids some types of local minima that afflict particle swarm optimisation (see Chapter \ref{C:PSO}), as its crossover behaviour tends to create hypotheses that are drastically different from those in the existing population. We also discuss some simple methods of parallelising GAs (\S\ref{SS:parallelising}), for example, for use on high-performance computing clusters for terascale and larger astronomical applications.
        \section{Implementation of GAs}
        The GA was encoded\footnote{Both methods presented here were coded in C++ in a Windows\textsuperscript{\texttrademark} environment, under Microsoft\textsuperscript{\textregistered} Visual Studio 2008. See Appendix \ref{A:src} for the relevant source.} and parameterised as follows.
            \subsection{Organisms and Population}
            An organism is simply a data structure representation of the six orbital elements described in \S\ref{S:orbitelem}. Angular values are represented in radians, distance is represented in metres (as a \verb+long long+) and \verb+long doubles+\footnote{The data type \texttt{long double} is equivalent to \texttt{double} in our C++ implementation.} are used whenever possible to minimise floating-point errors. Initial values are selected uniformly at random from appropriate ranges, and population sizes from 100 to 100,000 are tested.
            \subsection{Crossover and Mutation}
            Crossover is handled more stochastically than in the canonical algorithmic implementation and is similar to a uniform crossover \cite[pp. 254-255]{tM97}. First, from the pool of organisms that have been selected from the previous generation, two are chosen uniformly at random with replacement. A crossover probability $c$ in the interval $[0,1]$ is defined (in our case as 0.25), and two children organisms are created such that they are identical copies of their parents---except with the independent probability $c$ that any of their orbital elements have been swapped. That is, with probability $c$, the two children have swapped their parents' semi-major axis values, and with the same probability, they have swapped their parents' inclination values, and so on.

            An overall mutation rate is defined (for us, $m_o = 0.15$), and this percentage of organisms may be mutated after crossover. With another independent probability (again $m_i = 0.15$), an individual orbital parameter is mutated: half of these are multiplied by a \verb+double+ selected uniformly from $[0.75,1.25]$, and half of these are completely randomised, as at initialisation. This technique allows some values to be slightly adjusted after the population has begun to converge. Also, note that the independent mutation rate for an individual parameter is $(0.15)^2 = 2.25\%$, and the probability that one of the six parameters is changed is $1-(1-0.0225)^6 = 12.76\%$, a very high rate,\footnote{Cf. Mitchell (1997) \cite[p. 256]{tM97}, who suggests 0.1\%, and Negnevitsky (2002) \cite[p. 226]{mN02}, who suggests 0.1\% to 1.0\%.} which we have chosen in light of this problem's rapid convergence to local error minima and severe nonlinearities apparent in several dimensions.
            \subsection{Fitness}\label{SS:fitness}
            Fitness of an organism is determined by calculating the estimated position of the satellite (given its hypothetical orbital elements) relative to Jupiter at every time step $t_i$ for which there is an observational datum. The Euclidean distance $\Delta p(i)$ between the satellite's estimated position and its observed position is calculated:
            \begin{equation}
            \Delta p(i) = \sqrt{(x_e(i) - x_o(i))^2 + (y_e(i) - y_o(i))^2 + (z_e(i) - z_o(i))^2}
            \end{equation}
            and the fitness of the organism is set to the inverse of this distance, $\frac{1}{\Delta p(i)}$, averaged over all time steps $t_i$. Thus, organisms that predict satellite positions closer to the relevant observations have higher fitness values, and all fitness values are in the set
            $\mathbb{R}^+$.
            \subsection{Selection}\label{SS:selection}
            Before crossover, a predefined percentage (called the \emph{replacement} or \emph{selection rate}) of organisms (we use $s = 0.7$) are probabilistically selected according to their fitness values; specifically, an organism $o$ has probability
            \begin{equation}
            probselect_o = \frac{fitness_o}{\sum_{i \in unselected} fitness_i}
            \end{equation}
            of being selected to persist in the next generation \cite[p. 251]{tM97}, and this selection is repeated from the original population (without replacement) until $s \cdot popSize$ elements have been chosen. Note that, owing to the inverse linear relationship established for fitness in \S\ref{SS:fitness}, organisms with lower $\Delta p$ values are significantly more likely to be selected between generations, creating a greedier algorithm, as seen in Table \ref{T:fitness}.
            \begin{table}
            \begin{tabular}{ | r | c | }
            \hline
            $\Delta p$          & Relative Fitness                  \\ \hline
            $p_0$               & ${1}/{p_0} \equiv fitness_0$  \\ \hline
            $0.75 \cdot p_0$    & $1.\overline{3} \cdot fitness_0$  \\ \hline
            $0.5 \cdot p_0$     & $2 \cdot fitness_0$               \\ \hline
            $0.25 \cdot p_0$    & $4 \cdot fitness_0$               \\ \hline
            $0.1 \cdot p_0$     & $10 \cdot fitness_0$              \\ \hline
            \end{tabular}
            \\[5pt]\caption{Fitnesses associated with declining $\Delta p$ values.}\label{T:fitness}
            \end{table}
            \subsection{Termination Conditions}\label{SS:termination}
            One of the difficulties of using GAs and PSO is that convergence to global minima in the error function is not guaranteed; therefore predefined termination conditions are often difficult to set or even inappropriate, if the likelihood of convergence is not well understood. As such, no automated termination conditions were imposed; the algorithm was halted at will once useful data had been obtained (typically when $4 \cdot 10^7 < popSize \cdot generations < 8 \cdot 10^7$).
        \section{Results and Analysis}
        \subsection{Artificial Dataset}\label{SS:gaArtData}
        We first tested our algorithm and observed convergence rates on an artificial dataset with known orbital elements $[a = 10^8, e = 0, i = \pi, \Omega-(\omega+M_{epoch}) \equiv 0 \text{ (mod } 2\pi)]$ to decide which GA parameters to apply to our real data. To analyse the performance of our algorithm, we measured best fitness, worst fitness, mean fitness $\mu_{fitness}$, and standard deviation of fitness $\sigma_{fitness}$ over $G$ generations as described in \S\ref{SS:termination}. We did not measure the mean and variance of each of the six orbital elements individually, as we were mainly concerned with the relationship between fitness, $\mu$, and $\sigma$ in light of the termination problem, not with the algorithm's movement over our particular six-dimensional search space.

        We tested GAs with population sizes of 100, 1,000, 10,000, and 100,000, although populations of 100,000 did not evolve sufficiently quickly on our hardware\footnote{Most simulations were run on a 2 GHz Intel\textsuperscript{\textregistered} Core\textsuperscript{\texttrademark} 2 Duo processor with 2 GB RAM.} to be useful. A brief overview of best results achieved with these populations is presented in Tables \ref{T:gaADresults} and \ref{T:gaADanswers}.
        \begin{table}
        \begin{tabular}{ | r | r | r | r | r | }
            \hline
            Size          & Best Fitness & ...at Gen. & Total Gen's & Runtime per $10^5$ Gen's \\ \hline
            100           & 4.78825e-05 & 40669 & 300000 & 396.23 s  \\ \hline
            1000    & 3.1155e-06 & 11715 & 40000 & 6002.8 s  \\ \hline
            10000     & 1.74912e-06 & 345 & 2000 & 150530 s               \\ \hline
        \end{tabular}
        \\[5pt]\caption[GA performance on the artificial dataset.]{GA performance on the artificial dataset. We use best fitness at the final generation, as in the canonical GA presented in Table \ref{T:GAalg}, but we include the generation at which this fitness or better was first achieved. Note that runtime is related to dataset size.}\label{T:gaADresults}
        \end{table}

        \begin{table}
        \begin{tabular}{ | r | r | r | r | r | }
            \hline
            Size          & $a$ & $e$ & $i$ & $\Omega-(\omega+M_{epoch})$ \\ \hline
            100           & 99996000 & 0 & 3.14154 & 0.290308  \\ \hline
            1000          & 99983731 & 1.62e-06 & 3.14159 & 1.18074  \\ \hline
            10000         & 100021587 & 1.02e-04 & 3.1414 & 4.7178   \\ \hline
        \end{tabular}
        \\[5pt]\caption{GA results on the artificial dataset.}\label{T:gaADanswers}
        \end{table}

\begin{figure}
        \centering\includegraphics[width=126mm]{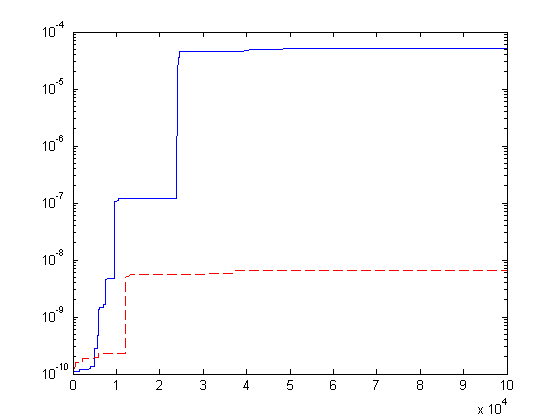}
        \caption[Overall best fitness of two GAs of size 100 over 100,000 generations.]{Overall best fitness of two GAs of size 100 over 100,000 generations. The dashed-line GA has fallen into a local error minimum.}\label{F:GA100fail-success}
        \end{figure}

\begin{figure}
        \centering\includegraphics[width=126mm]{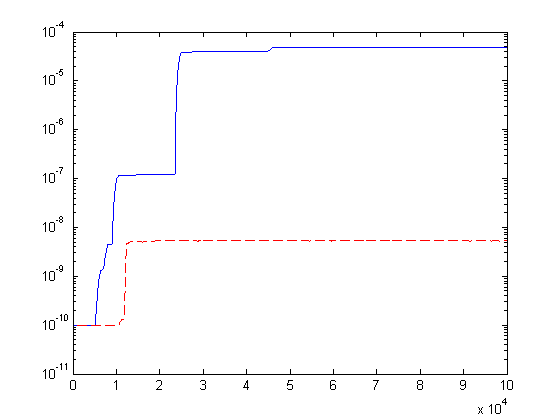}
        \caption[$\mu_{fitness}$ of the two GAs in Figure \ref{F:GA100fail-success}.]{$\mu_{fitness}$ of the two GAs in Figure \ref{F:GA100fail-success}. Mean fitness, smoothed over 500 generations, tends to approximate the fitness of the current (surviving) best organism.}\label{F:GA100meanfit}
        \end{figure}

\begin{figure}
        \centering\includegraphics[width=126mm]{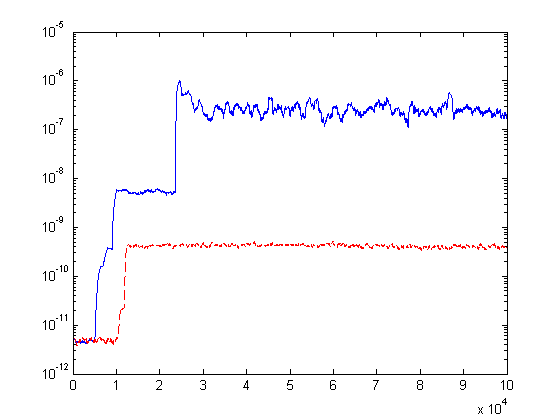}
        \caption[$\sigma_{fitness}$ of the two GAs in Figure \ref{F:GA100fail-success}.]{$\sigma_{fitness}$ of the two GAs in Figure \ref{F:GA100fail-success}, smoothed over 500 generations.}\label{F:GA100stdevfit}
        \end{figure}

\begin{figure}
        \centering\includegraphics[width=126mm]{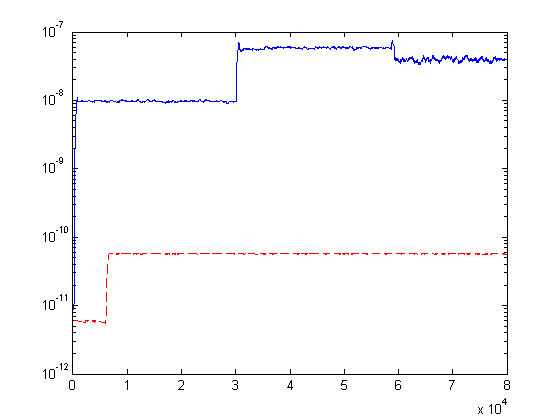}
        \caption[$\sigma_{fitness}$ of two GAs of size 1,000 over 80,000 generations.]{$\sigma_{fitness}$ of two GAs of size 1,000 over 80,000 generations. The dashed line describes a GA that has fallen into a local minimum; note its low variance in early generations.}\label{F:GA1000stdev-localmin}
        \end{figure}

        Although the GA with population size 100 achieved the best result, it was also most susceptible to falling into local minima in the search space. Because the local minima for this dataset are well known, GA runs that converged on local solutions could be manually excluded. However, on a real dataset for which the search space is poorly understood, these local minimum errors cannot be automatically corrected; a GA approaching a local minimum is theoretically identical in behaviour to one approaching the global minimum, as seen in Figures \ref{F:GA100fail-success}, \ref{F:GA100meanfit}, and \ref{F:GA100stdevfit}. However, the lower standard deviation in early generations (early convergence) on the poorer GA in Figure \ref{F:GA100stdevfit} is possibly an indication of its having fallen into a local minimum, judging qualitatively from similar plots such as Figure \ref{F:GA1000stdev-localmin}.


        GAs with smaller populations have the advantage that their organisms evolve more quickly and can therefore arrive at more `finely tuned' solutions near the global minimum without expensive computation time; however, they often fail to generate the early variation necessary to approach the global minimum. Accordingly, we proceed primarily with GAs of size 1,000, as these were much less susceptible to problems of premature convergence, yet still evolved quickly enough to yield acceptably precise solutions.
            \subsection{Jupiter's Moons: Io and Himalia}\label{SS:gaJup}
            We first looked to Io, as one of the four Galilean satellites with a stable, near-circular orbit. However, because most moons with stable circular orbits have an inclination near 0, the dimensionality of our search space for Io was reduced---or, rather, our algorithm had to explore a very flat search space over the parameters $\Omega$, $\omega$, and $M_{epoch}$ in the region of $i=0$. We were still able to converge on some of Io's orbital parameters, as illustrated in Table \ref{T:gaIo}, but for the others, the search area around the global minimum was too flat for our algorithm to efficiently approach the correct values.

            \begin{table}
            \begin{tabular}{ | c | c | c | c | }
            \hline
            Element       & GA Approximation  & Actual Value & Error\\ \hline
            $a$           & 421075000 & 421800000 & 0.00172 \\ \hline
            $e$          & 0.00555 & 0.0041 & 0.00145 \\ \hline
            $i$         & 9.68e-05 & (0.036) & (0.0057)  \\ \hline
            \end{tabular}
            \\[5pt]\caption[Results on Io for a GA with population 1,000, after 15,000 generations.]{Results on Io for a GA with population 1,000, after 15,000 generations. The actual inclination $i$ is referred to the Laplace plane, but for a large moon like Io this is roughly the same as Jupiter's equatorial plane.}\label{T:gaIo}
            \end{table}

            We then chose Himalia, a smaller moon with a more eccentric orbit and a significant inclination. However, although our GA should have been able to converge on more of the orbital elements of Himalia, it became apparent that we would not be able to compare our estimated $i$, $\Omega$, $\omega$, or $M_{epoch}$ values to real data: the values of these parameters presented in the literature \cite{rJ00} for all of Jupiter's moons are referred to the Laplace plane,\footnote{The orbits of small satellites tend to precess more rapidly than those of larger satellites, meaning that the pole of the satellite's orbital plane moves gyroscopically about another pole---the pole of its Laplace plane. The Laplace plane thus represents something of an `average' orbital plane if one integrates over the full precession period of a satellite's orbit.} while our code was designed to calculate orbital elements referred to a planet's equatorial plane. Moreover, it is impossible to translate between Laplace plane values and equatorial ones without more information about the precise orientation of a satellite's orbit in a given epoch. Still, we present our best converged values of $a$ and $e$ for Himalia in Table \ref{T:gaHimalia}.

            \begin{table}
            \begin{tabular}{ | c | c | c | c | }
            \hline
            Element       & GA Approximation  & Actual Value & Error\\ \hline
            $a$           & 11315960111 & 11461000000 & 0.0127 \\ \hline
            $e$          & 0.100526 & 0.1623 & 0.0618 \\ \hline
            \end{tabular}
            \\[5pt]\caption[GA results on Himalia with population 1,000, after 40,000 generations.]{GA results on Himalia with population 1,000, after 40,000 generations. Slightly better (2.8\% gain in fitness) results were achieved with a GA of size 100 after 200,000 generations, although it suffered from the local minimum problem described in \S\ref{SS:gaArtData}.}\label{T:gaHimalia}
            \end{table}

            \subsection{Saturn's Moon: Atlas}\label{SS:gaSat}
            The moons most similar to Jupiter's with data that can be referred to a planetary equatorial plane are those of Saturn. Accordingly, we chose one of Saturn's major moons, Atlas, for a complete application of our learning algorithms. Atlas, like all of Saturn's inner satellites, has a low inclination and so has a very flat search space in the region of the global minimum, but we can at least compare our GA results to all six of its orbital elements. The orbits of Saturn's smaller, outer satellites, being more inclined and eccentric, are not measured in reference to the planet's equatorial plane.\footnote{See \texttt{http://ssd.jpl.nasa.gov/?sat\_elem\#saturn}.}

            The best solution for Atlas's orbital elements found by our GA is presented in Table \ref{T:gaAtlas}. We see that the GA converged on very accurate values of $a$, $e$, and $i$, but, as expected, was unable to effectively traverse the flat search area around the global minimum with respect to the other three parameters.

            \begin{table}
            \begin{tabular}{ | c | c | c | c | }
            \hline
            Element       & GA Approximation  & Actual Value & Error\\ \hline
            $a$           & 137450238 & 137670000 & 0.0127 \\ \hline
            $e$          & 4.275e-08 & 0.0012 & 0.0012 \\ \hline
            $i$          & 9.470e-08 & 0.003 & 0.0005 \\ \hline
            $\Omega$      & 4.890 & 0.0087 & 0.777 \\ \hline
            $\omega$      & 3.439 & 5.786 & 0.374 \\ \hline
            $M_{epoch}$    & 5.441 & 2.753 & 0.428 \\ \hline
            \end{tabular}
            \\[5pt]\caption{GA results on Atlas with population 1,000, after 80,000 generations.}\label{T:gaAtlas}
            \end{table}

\begin{figure}
            \centering\includegraphics[width=126mm]{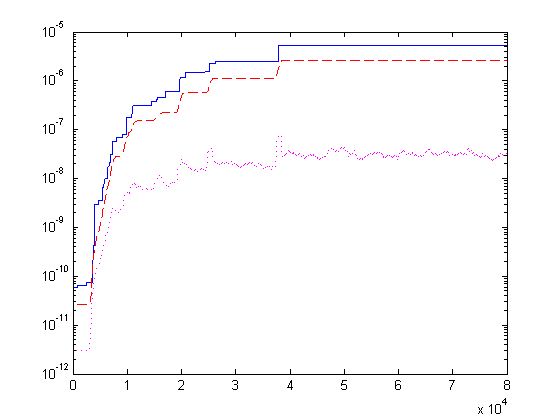}
            \caption[Log plot of overall best fitness, $\mu_{fitness}$, and $\sigma_{fitness}$ of a size 1,000 GA running on Atlas for 80,000 generations.]{Log plot of overall best fitness, $\mu_{fitness}$ (dashed curve), and $\sigma_{fitness}$ (dotted curve) of a size 1,000 GA running on Atlas for 80,000 generations. For clarity, $\mu_{fitness}$ is transposed downwards by a factor of 2.}\label{F:AtlasGA1000osm}
            \end{figure}

\begin{figure}
            \centering\includegraphics[width=126mm]{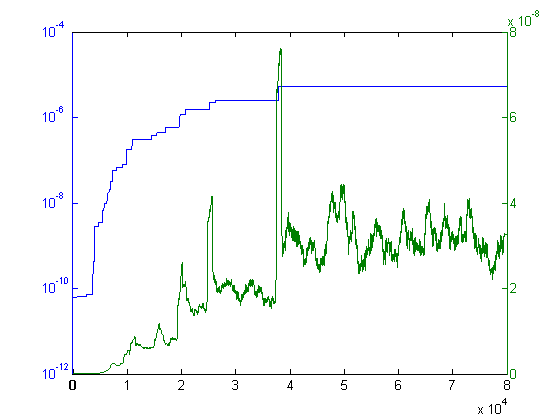}
            \caption[Linear plot of the GA in Figure \ref{F:AtlasGA1000osm}, overall best fitness with $\sigma_{fitness}$.]{Linear plot of the GA in Figure \ref{F:AtlasGA1000osm}, overall best fitness with $\sigma_{fitness}$. $\sigma_{fitness}$ is smoothed over 1,000 generations, yet still indicates great increases in fitness variance over the population of 1,000 when the overall best fitness increases.}\label{F:AtlasGA1000os}
            \end{figure}

            Figure \ref{F:AtlasGA1000osm} shows the overall best fitness over time along with the mean fitness in the population and the standard deviation of fitness (after selection and before and after crossover, $\mu_{fitness}$ and $\sigma_{fitness}$ remain almost unchanged), where the latter two values are smoothed over 1,000 generations. In the GA implementation, the mean fitness of the population is nearly identical (when averaged over some hundreds of generations) to the best fitness observed; the population does not periodically decrease in fitness. Note that this does not mean that our algorithm behaves greedily; it merely indicates that a single organism attaining a higher fitness tends to bring a majority of the population up to its fitness level. This does not in itself indicate convergence: the entire population may have the same fitness but may still be dispersed about the search space. That increases in best fitness quickly bring other organisms to higher fitness levels is also illustrated in Figure \ref{F:AtlasGA1000os}, where the sharp rises and falls of $\sigma_{fitness}$ are consistent with these shifts in fitness.

        \section{Discussion}
            \subsection{Difficulties of Convergence}\label{SS:diffConvergence}
            As discussed earlier, large moons tend to orbit their major planets at very small inclination, such that $i \sim 0 \text{ (mod } \pi)$. This not only creates a relatively flat search space in the region of $i=0$, but also induces a dependent relationship between the three other angular orbital parameters. That is, $\Omega$, $\omega$, and $M_{epoch}$ will have higher fitness when they vary according to the relation $\Omega + \omega + M_{epoch} = t$ for some value $t \in [0,2\pi)$. This dependence effectively disallows any significant movement in a single dimension; in order to approach optimal values, organisms must either mutate by very small amounts in single dimensions or must mutate in multiple dimensions simultaneously while roughly obeying the given relation.

            As it is, this dependence tends to bring about convergence of $\Omega$, $\omega$, and $M_{epoch}$ in our implementation on parameter values that are insufficiently optimal. Once convergence over these three parameters has occurred, it is very unlikely that the GA will approach more optimal values for them.
            \subsection{Cross-Fertilisation of Populations}
            We have mentioned that smaller populations can approach more optimal solutions in fixed computational time, although larger populations are better suited to avoiding local minima in the search space---particularly in early generations. One approach that could take advantage of both of these strengths is that of cross-fertilisation:\footnote{Cf. Mitchell (1997) \cite[p. 268]{tM97}, who discusses this in the context of parallelisation.}

            Multiple GAs could be run with completely distinct populations of differing sizes, such that from time to time some organisms are migrated between populations. These organisms could be randomly chosen or could be selected by fitness (for a greedier implementation), and they could move between all populations or only from larger populations to smaller ones, to avoid local minima traps.

\begin{figure}
            \centering\includegraphics[width=126mm]{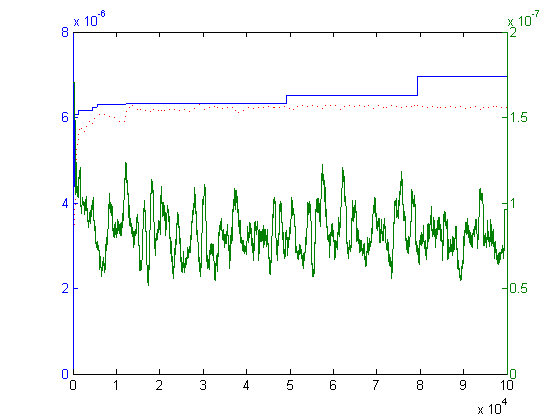}
            \caption[Overall best fitness, $\mu_{fitness}$, and $\sigma_{fitness}$ of the size-100 GA on Atlas after seeding.]{Overall best fitness, $\mu_{fitness}$ (dotted curve), and $\sigma_{fitness}$ of the size-100 GA after seeding, smoothed over 1,000 generations. The absence of order-of-magnitude increases in fitness and the relatively stable $\sigma_{fitness}$ could indicate a numerical impossibility of convergence any closer to the global minimum.}\label{F:Atlas100}
            \end{figure}

            An attempt at mimicking this cross-fertilisation was made by seeding a GA of size 100 with $a$, $e$, and $i$ values found by the size-1,000 GA (as though cross-fertilising the smaller population with the most fit organism from the larger population), as listed in Table \ref{T:Atlas100}. This placed the 100 new organisms in the region of the global minimum, and allowed them to explore the search space over 100,000 additional generations. Fitness improved by 18.8\% (not an order of magnitude increase), and although the smaller GA still converged relatively quickly on $\Omega$, $\omega$, and $M_{epoch}$ as described in \S\ref{SS:diffConvergence}, it still achieved more accurate values of these three parameters as shown in Table \ref{T:Atlas100}. The GA's behaviour is further illustrated in Figure \ref{F:Atlas100}; the lack of subsequent convergence after seeding and after 100,000 generations could also point to numerical difficulties in calculating precise satellite positions along the ephemerides.

            \begin{table}
            \begin{tabular}{ | c | c | c | c | }
            \hline
            Element       & GA Approximation  & Actual Value & Error\\ \hline
            $a$           & 137450238 & 137670000 & 0.0127 \\ \hline
            $e$          & 0.0008922 & 0.0012 & 0.0003 \\ \hline
            $i$          & 9.353e-09 & 0.003 & 0.0005 \\ \hline
            $\Omega$      & 1.157 & 0.0087 & 0.183 \\ \hline
            $\omega$      & 6.112 & 5.786 & 0.0519 \\ \hline
            $M_{epoch}$    & 0.217 & 2.753 & 0.404 \\ \hline
            \end{tabular}
            \\[5pt]\caption{GA results on Atlas with a seeded population of 100, after 100,000 generations.}\label{T:Atlas100}
            \end{table}

            \subsection{Parallelising GAs}\label{SS:parallelising}
            An idea closely related to that of cross-fertilisation is that of parallelisation, which aims primarily at distributing the GA for efficient computation on decentralised systems. Besides the method of running distinct populations on different machines that are then cross-fertilised, one can also create a non-traditional GA that operates not in synchronised generations, but utilises one centrally managed population:\footnote{This idea is due to Mr Andrew Foster, UNC-CH, who implemented it for an astrophysical application.}

            The population could be kept on one central machine, which would be made globally accessible. `Worker' threads on ancillary machines would request two randomly selected members of the central population, breed a new organism, and suggest it to a manager thread for inclusion if its fitness is higher than the lowest fitness in the major population. Probabilistic methods of selection could also be used in the inclusion routine, to avoid greediness: when replacing an old organism in the population with a newer one, the organism that `dies' could be selected at random with weighting according to fitness, instead of having the manager thread deterministically select the least fit organism to be replaced.

            This parallelisation and the cross-fertilisation parallelisation of Mitchell (1997) \cite[p. 268]{tM97} could be run asynchronously on cluster setups with any number of machines of any speed. Such an implementation could be especially useful for solving astrophysical problems without expensive computing resources, as discussed in Chapter \ref{C:Implications}.
    \chapter{Particle Swarm Optimisation for Astrophysical Modelling}\label{C:PSO}
        \section{Fundamentals of PSO}
        \emph{Particle swarm optimisation} (PSO) is another learning technique inspired by a physical process---the behaviour of flocks of animals or swarms of insects. An implementation of PSO consists of an initially randomised \emph{swarm of particles}, each representing a hypothesis in the search space, wherein each particle has its own \emph{velocity vector} over the set of search parameters. Thus, each particle is `moving' independently about the search space.

        At each step of the algorithm, the velocity of each particle is adjusted to approach stochastically the position of the current `best' particle or the overall `best' particle, for some given fitness function. The algorithm must thus remember the overall `best' particle, but this is minimally memory-intensive. The basic update equations are as follows.
        \begin{subequations}
        \begin{align}\label{E:pso1}
        \mathbf{v}[i] = \mathbf{v}[i] &+ 2 \times \text{rand()} \times (\mathbf{p}_{currentbest} - \mathbf{p}[i]) \\ &+ 2 \times \text{rand()} \times (\mathbf{p}_{overallbest} - \mathbf{p}[i])\notag
        \end{align}
        \begin{equation}\label{E:pso2}
        \mathbf{p}[i] = \mathbf{p}[i] + \mathbf{v}[i]
        \end{equation}
        \end{subequations}
        In these equations \cite{jK95}, $\mathbf{p}[i]$ and $\mathbf{v}[i]$ represent the position and velocity vectors of a particle $i$ in the six-dimensional parameter space; $\mathbf{p}_{currentbest}$ and $\mathbf{p}_{overallbest}$ are the `best' particles in the swarm, as described above; and rand() is a random real number in the interval $[0,1]$.
            \subsection{Advantages of PSO}
            Without the crossover and mutation operators of a GA, PSO has fewer parameters to tweak for any particular application. The velocity update feature allows the swarm to accelerate (up to some optionally specified maximum velocity $v_{max}$) towards minima in the search space, while its velocity components themselves, combined with the stochastic element in equation \eqref{E:pso1}, allow the algorithm to avoid converging on local minima. The minimal need for centralised information exchange in a PSO algorithm also make the technique well suited for distributed processing.
        \section{Implementation of PSO}
            \subsection{Particles and Swarm}
            As in the GA implementation, a particle in PSO primarily represents the six orbital parameters in a data structure. Additionally, velocities of the parameters over the search space are represented, with a maximum velocity bound (see \S\ref{SSS:vmax}) imposed on five of the parameters. Lower maximum velocities encourage faster convergence and less erratic particle behaviour, but they also incline the algorithm more towards early convergence to local minima in the parameter space. Swarm sizes are similar to population sizes for the GA implementation, i.e., 100 to 100,000.
            \subsection{Problem-Specific Methods}
                \subsubsection{Circular Search Dimensions}\label{SSS:circular}
                Since the particles in the PSO implementation are `moving' about the search space with certain velocities, a problem arose with the angular value representations of four of the orbital parameters. Since these parameters are only valid in the range $[0,2 \pi)$ radians (or rather since a broader search space where values are congruent modulo $2 \pi$ tends to prevent PSO convergence), one obvious solution would be to impose `physical' limits on the particles to prevent them from exploring outside of this interval. However, this limiting induces the particles to converge prematurely at the edges of the search space: high velocities send a few particles near the boundaries, where they halt; the more fit particles in these clusters attract other particles to the boundaries, and the convergence problem worsens.

                The first solution to this problem is to create a circular search space in the four dimensions with angular values. Such a search space has two important properties:
                \begin{enumerate}
                \item Particles are allowed to move beyond the interval $[0,2 \pi)$, but their radian values are immediately adjusted (mod $2 \pi$) to replace them into the interval before further calculations; and\label{I:mod2pi}
                \item When calculating velocity updates, particles do not simply move towards the linear values of the best-fit particles, but travel the shortest distance (arc length) around the 0-to-$2 \pi$ radian circle.\label{I:arclength}
                \end{enumerate}
                For example, if a particle with position 0 were moving towards a particle with position $3 \pi / 2$, it would not update its velocity with the positive value $(3 \pi / 2 - 0)$, but, by \eqref{I:arclength}, with the negative value $-(2 \pi - (3 \pi / 2 - 0)) = -\pi / 2$, inclining it towards the shorter path around the radian circle. By \eqref{I:mod2pi}, then, its radian value would be adjusted, if necessary, to keep it in the interval $[0,2 \pi)$.
                \subsubsection{Maximum Velocities}\label{SSS:vmax}
                The methods described in \S\ref{SSS:circular} are, however, not enough to guarantee a simpler PSO convergence for this problem. The following example best illustrates the remaining difficulty, and why we impose maximum velocities on the particles in five out of six dimensions.

                Suppose the current best and overall best particles are resting at a position of 0 radians. A particle with high positive velocity around the circle is attempting to approach the 0-radian position, and is currently at $\pi / 2$ radians. Its velocity is updated, therefore, by adding a small negative value to its current large positive velocity. By the next time step, its positive velocity has lessened slightly, but it has continued to move past the $\pi$ radians position and is still attempting to approach the 0 position. At this point, its velocity is updated with a further positive value, and it passes the 0 position by the next time step, thus restarting the process and preventing convergence.

                One solution to this further convergence problem is to impose maximum absolute velocities on the particles, meaning that $|v[i][j]| < v_{max}$ for all particles $i$ in each dimension $j$ at all times. Intuitively, a maximum velocity for one of the four angular dimensions should be less than $\pi / 2$, so we have experimented with $v_{max}$ values between 0.1 and 1.5. We have also imposed a maximum velocity of 0.1 on the eccentricity ($e$) parameter in order to avoid the clustering problem described in \S\ref{SSS:circular}. Simulations run without maximum velocities yielded both the clustering and convergence problems, as expected. No maximum velocity was necessary for the semi-major axis ($a$) parameter, being unbounded (or rather being limited to $\mathbb{Z}^+$) and not susceptible to either of the convergence problems described above.\footnote{Some clustering was observed at $a = 1$, but not to a significant degree.}
            \subsection{Learning Factors}
            Learning factors $c_1$ and $c_2$, which control the velocity updates with respect to the current best and overall best particles, were both set to $c_1 = c_2 = 2$, as recorded in equation \eqref{E:pso1}. This makes it as likely that a particle will move towards the current best particle as towards the overall best particle, but reduces the greediness of the algorithm so as to better avoid local minima in the error function.

            \subsection{Dimensionally Independent Velocity Updates}
            A more stochastic movement for particles was tested by assigning independent random coefficients to each of the orbital elements in the particles' position vectors:
            \begin{align}\label{E:pso-indepmov}
            \mathbf{v}[i] = \mathbf{v}[i] &+ 2 \times \mathbf{r} \otimes (\mathbf{p}_{currentbest} - \mathbf{p}[i]) \\ &+ 2 \times \mathbf{r} \otimes (\mathbf{p}_{overallbest} - \mathbf{p}[i])\notag
            \end{align}
            where $\mathbf{r}$ is a six-dimensional vector of independently randomised reals in $[0,1]$ and the operator $\otimes$ represents element-wise multiplication. However, this modification made the algorithm too erratic; it would converge only minimally, although particles would occasionally stumble upon high-fitness solutions. Note that, even though the statistically expected changes in velocity are the same in each dimension:
            \begin{alignat}{2}
            \text{E}(\Delta\mathbf{v}[i]) &= &\ &2 \times \text{E}(\mathbf{r}) \otimes (\mathbf{p}_{currentbest} - \mathbf{p}[i]) \\
            &&\ {}+{} &2 \times \text{E}(\mathbf{r}) \otimes (\mathbf{p}_{overallbest} - \mathbf{p}[i])\notag \\
            &= &\ &2 \times 0.5 \times (\mathbf{p}_{currentbest} - \mathbf{p}[i]) + 2 \times 0.5 \times (\mathbf{p}_{overallbest} - \mathbf{p}[i])\notag \\
            &= &\ &(\mathbf{p}_{currentbest} - \mathbf{p}[i]) + (\mathbf{p}_{overallbest} - \mathbf{p}[i])\notag \\
            &= &\ &2 \times 0.5 \times (\mathbf{p}_{currentbest} - \mathbf{p}[i]) + 2 \times 0.5 \times (\mathbf{p}_{overallbest} - \mathbf{p}[i])\notag \\
            &= &\ &2 \times \text{E(rand())} \times (\mathbf{p}_{currentbest} - \mathbf{p}[i])\notag \\
            &&\ {}+{} &2 \times \text{E(rand())} \times (\mathbf{p}_{overallbest} - \mathbf{p}[i])\notag
            \end{alignat}
            the net effect is different because the probability of uniform movement towards one `best' particle is significantly lower:
            \begin{alignat}{2}
            P(\text{rand}_1 \gg \text{rand}_2) > P(\mathbf{r}_1[1] &\gg \mathbf{r}_2[1]\quad&\land \\ &\vdots &\land\notag \\ \mathbf{r}_1[6] &\gg \mathbf{r}_2[6]) \text{,}\notag
            \end{alignat}
            where $\text{rand}_i \in [0,1]$ and $\mathbf{r}_i \in [0,1]^6$.

            \subsection{Fitness and Termination Conditions}
            Fitness and termination conditions are as described in \S\S\ref{SS:fitness} and \ref{SS:termination}. Algorithm runs that were observed to converge prematurely were halted immediately.
        \section{Results and Analysis}\label{S:psoResults}
        \subsection{Artificial Dataset}\label{SS:psoArtData}
        Just as with our GA implementation, we tested the PSO technique on the artificial dataset, to begin. We used swarm sizes of 100 and 1,000, and maximum velocities of 0.1, 0.25, 0.5, and 0.75. As shown in Tables \ref{T:psoADresults} and \ref{T:psoADanswers}, there was no plain correspondence between parameter settings and best fitness achieved, but parameters did affect the convergence behaviour of the swarm as regards precision around the global minimum and the local minima problem of \S\ref{SS:gaArtData}.

        \begin{table}
        \begin{tabular}{ | r | r | r | r | r | r | }
            \hline
            Size & $v_{max}$   & Best Fitness & ...at Gen. & Total Gen's & Runtime/$10^5$ G's \\ \hline
            100   & 0.1     & 5.78903e-07 & 6657 & 80000 & 599.88 s  \\ \hline
            100   & 0.25    & 3.14102e-06 & 42467 & 80000 & 606.80 s \\ \hline
            100   & 0.5     & 1.07012e-07 & 76614 & 80000 & 611.58 s \\ \hline
            100   & 0.75    & 2.88145e-07 & 44341 & 80000 & 594.45 s \\ \hline
            1000  & 0.1     & 3.70166e-07 & 10886 & 20000 & 2779.1 s \\ \hline
            1000  & 0.25    & 4.11467e-07 & 8830 & 20000 & 2858.8 s \\ \hline
            1000  & 0.5     & 1.11125e-06 & 9898 & 20000 & 2886.8 s \\ \hline
            1000  & 0.75    & 3.02420e-07 & 6137 & 20000 & 2812.3 s \\ \hline
        \end{tabular}
        \\[5pt]\caption[PSO performance on the artificial dataset.]{PSO performance on the artificial dataset. Best fitness is the overall best fitness observed.}\label{T:psoADresults}
        \end{table}

        Smaller swarms could run through more generations than larger swarms, as with GAs, but many of the PSO runs achieved their best overall fitness at or before running through 50\% of their total generations. This attests to the very erratic behaviour of the particles in the search space; running through more generations often did not improve the algorithm's result, although more generations ought theoretically to yield improvement. It was more important that our algorithms did not converge prematurely at the beginning of their runs (for which higher $v_{max}$ values and larger swarms were better), and that their particles eventually converged once in the region of the search space near the global minimum (for which lower $v_{max}$ values were better). One possible remedy for this $v_{max}$ dilemma is presented in \S\ref{S:psoDiscussion}.

        \begin{table}
        \begin{tabular}{ | r | r | r | r | r | r | }
            \hline
            Size & $v_{max}$  & $a$ & $e$ & $i$ & $\Omega-(\omega+M_{epoch})$ \\ \hline
            100  &  0.1     & 100033396 & 0 & 3.14054 & 3.85625  \\ \hline
            100   & 0.25    & 100009747 & 0 & 3.1414 & 5.57166  \\ \hline
            100   & 0.5     & 100023939 & 0 & 3.14632 & 4.52465  \\ \hline
            100   & 0.75    & 100030855 & 0 & 3.1429 & 4.03221  \\ \hline
            1000  & 0.1     & 99964288 & 0 & 3.14294 & 2.59303  \\ \hline
            1000  & 0.25    & 100032127 & 0 & 3.14362 & 3.95884  \\ \hline
            1000  & 0.5     & 99992860 & 0 & 3.14103 & 0.52695  \\ \hline
            1000  & 0.75    & 99963548 & 0 & 3.14104 & 2.63425  \\ \hline
        \end{tabular}
        \\[5pt]\caption[PSO performance on the artificial dataset.]{PSO results on the artificial dataset. The $e$ values are all nought because no maximum velocity had yet been set on this parameter; the clustering problem of \S\ref{SSS:circular} yielded these values.}\label{T:psoADanswers}
        \end{table}

        \subsection{Himalia}\label{SS:psoJup}
        PSO was not tested on Jupiter's moon Io, but the algorithm achieved results with a 58\% improvement in fitness over our GA's results after 60,000 generations with a swarm of 1,000. However, these improvements do not show through, as most of the accuracy increase was apparently in the four orbital parameters whose correct values we cannot transform out of the Laplace plane. Even still, the error of the $a$ estimate seen in Table \ref{T:psoHimalia} is noticeably high.

        \begin{table}
        \begin{tabular}{ | c | c | c | c | }
        \hline
        Element       & GA Approximation  & Actual Value & Error\\ \hline
        $a$           & 14763515741 & 11461000000 & 0.2882 \\ \hline
        $e$          & 0.2423 & 0.1623 & 0.0800 \\ \hline
        \end{tabular}
        \\[5pt]\caption[PSO results on Himalia with population 1,000 and $v_{max}=0.5$, after 60,000 generations.]{PSO results on Himalia with population 1,000 and $v_{max}=0.5$, after 60,000 generations. Other parameter settings produced more accurate $a$ and $e$ estimates, but the run presented here had the highest overall fitness.}\label{T:psoHimalia}
        \end{table}

        \subsection{Atlas}\label{SS:psoSat}
        For Saturn's moon Atlas, we first present results from PSO runs of size 1,000 and $v_{max}$ between 0.5 and 1.0; smaller swarms and lower $v_{max}$ values either fell too quickly into local minima or converged nearly immediately in a flat region of the search space very far from the global minimum. We then present an analysis of our PSO runs, including a couple runs that fell into local minima, and discuss how the successes and failures are distinguishable by their convergence behaviour. Then, a remedy for the local minimum convergence problem is presented in \S\ref{S:psoDiscussion}.

        Our best PSO results were achieved with a $v_{max}$ setting of 0.5, when the algorithm did not fall into a local minimum. A fitness slightly less that of our GA was achieved, although it was of the same order of magnitude. The approximated orbital elements are presented in Table \ref{T:psoAtlas}. We see from these elements that the PSO, like the GA in \S\ref{SS:gaSat}, had difficulty optimising fitness in the flat search space in the vicinity of the global minimum (where $a$, $e$, and $i$ are close to the actual values).

        \begin{table}
        \begin{tabular}{ | c | c | c | c | }
        \hline
        Element       & GA Approximation  & Actual Value & Error\\ \hline
        $a$           & 137446613 & 137670000 & 0.0016 \\ \hline
        $e$          & 0 & 0.0012 & 0.0012 \\ \hline
        $i$          & 8.579e-05 & 0.003 & 0.0029 \\ \hline
        $\Omega$      & 3.112 & 0.0087 & 0.494 \\ \hline
        $\omega$      & 1.682 & 5.786 & 0.653 \\ \hline
        $M_{epoch}$    & 2.682 & 2.753 & 0.0113 \\ \hline
        \end{tabular}
        \\[5pt]\caption{PSO results on Atlas with population 1,000 and $v_{max}=0.5$, after 80,000 generations.}\label{T:psoAtlas}
        \end{table}

        However, many of our PSO runs did not approach the global minimum, but instead converged prematurely at the beginning of the run (usually with swarms smaller than 1,000 and/or $v_{max} < 0.5$) or fell into a local minimum, such as $a = 40635000, i = \pi$. The premature convergence is easily detected (as variance quickly drops to nought), but the local minima problem involves more subtle swarm behaviour in our particular search space. We present some observations of this behaviour that may be useful in other search spaces with similar local minima features.

        Particle swarms that have fallen into a local minimum exhibit a couple well defined characteristics: (1) when smoothed, $\sigma_{fitness}$ does not show any order-of-magnitude increases over the course of the run (often quickly decreasing from its initial value), and (2) when smoothed, $\mu_{fitness}$ and $\sigma_{fitness}$ have a negative correlation, such that $\mu_{fitness} + \sigma_{fitness}$ remains roughly constant. On the other hand, swarms that approach the global minimum exhibit (3) a sharp drop in $\mu_{fitness}$, (4) a sharp rise in $\sigma_{fitness}$ and (5) a positive correlation between the current best value and $\sigma_{fitness}$, once all values have been smoothed.

\begin{figure}
        \centering\includegraphics[width=126mm]{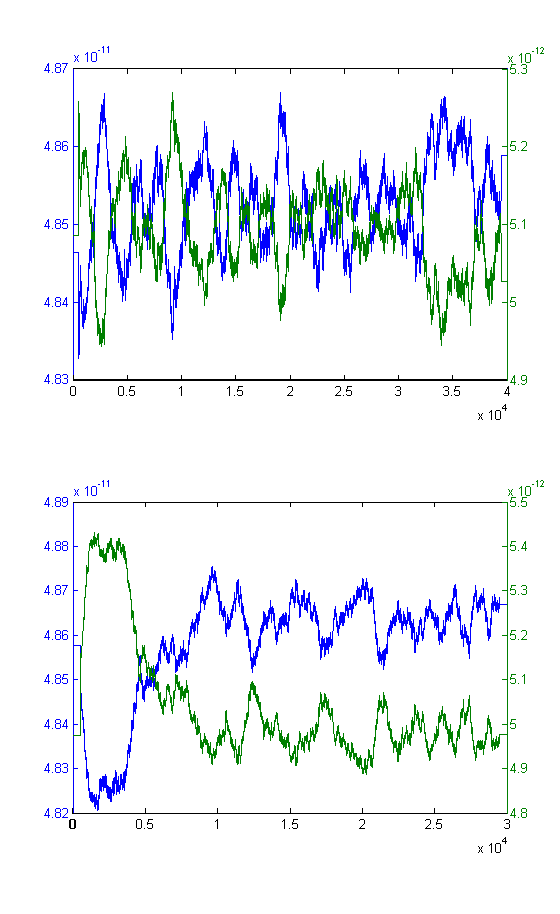}
        \caption[$\mu _{fitness}$ and $\sigma _{fitness}$ of particle swarms with $v_{max} = \text{[}0.5, 1.0\text{]}$ that have fallen into local minima.]{$\mu_{fitness}$ (blue) and $\sigma_{fitness}$ (green) of particle swarms with $v_{max} = [0.5, 1.0]$ that have fallen into local minima. $\sigma_{fitness}$ remains near its initial value, which represents the variance of randomly initialised orbital elements.}\label{F:pso2ms}
        \end{figure}

\begin{figure}
        \centering\includegraphics[width=126mm]{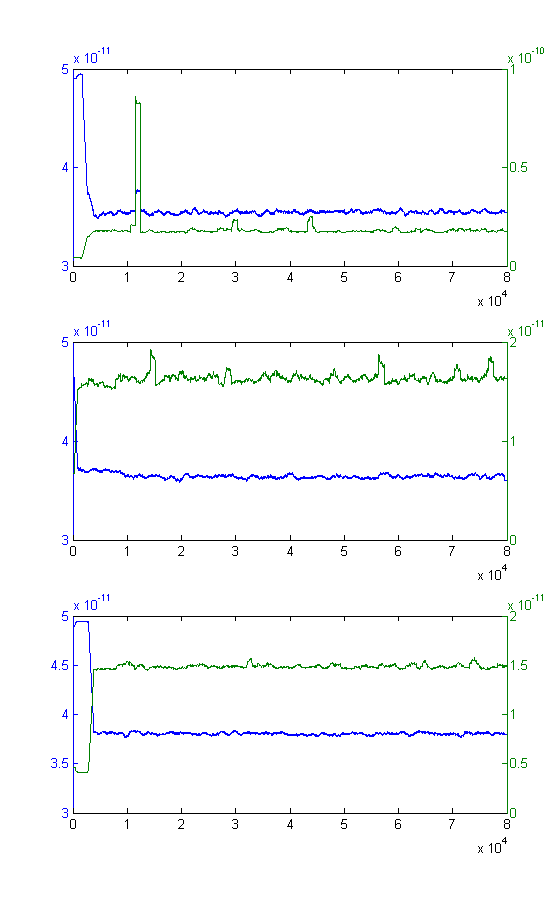}
        \caption[$\mu _{fitness}$ and $\sigma _{fitness}$ of successful particle swarms with $v_{max} = \text{[}0.5, 0.75, 1.0\text{]}$.]{$\mu_{fitness}$ (blue) and $\sigma_{fitness}$ (green) of successful particle swarms with $v_{max} = [0.5, 0.75, 1.0]$. Note the early drop in $\mu_{fitness}$ and the early increase in $\sigma_{fitness}$, as well as their lack of clear correlation.}\label{F:pso3ms}
        \end{figure}

        Characteristics (1) and (2) are illustrated in Figure \ref{F:pso2ms}; the low $\sigma_{fitness}$ value probably attests to early convergence in the search space, or at least convergence towards a value (e.g., $a=1$) that produces similar fitnesses. Eventually, a low $\sigma_{fitness}$ value is desirable, but not at the outset of the run where diversity of parameter values in the swarm is important for a fuller exploration of the search space. The negative correlation between $\mu_{fitness}$ and $\sigma_{fitness}$ (2) in failing swarms could also attest to the small-scale introduction of diversity in the swarm when $\sigma_{fitness}$ rises. This diversity already exists in successful swarms that are approaching the global minimum, so there is no strong negative correlation, seen in Figure \ref{F:pso3ms}.

\begin{figure}
        \centering\includegraphics[width=126mm]{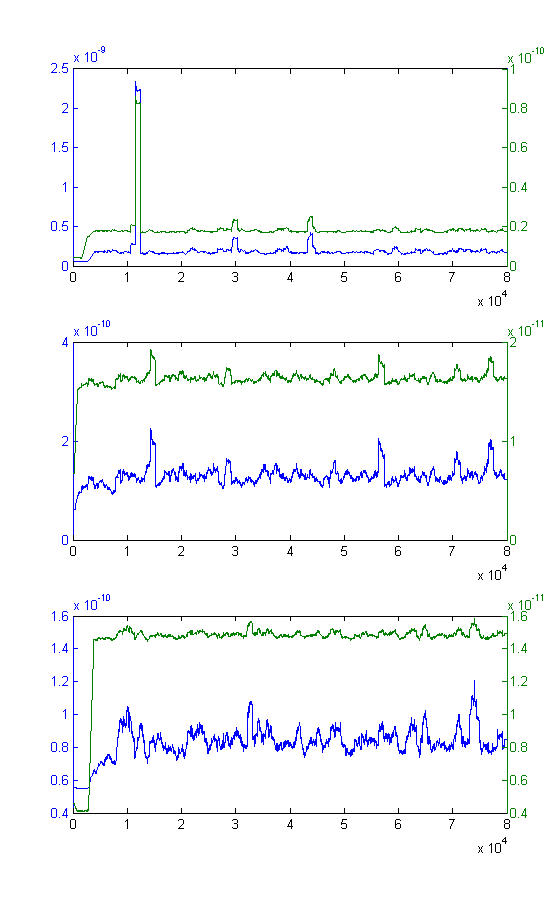}
        \caption[Current best fitness and $\sigma _{fitness}$ of particle swarms sized 1,000 with $v_{max} = \text{[}0.5, 0.75, 1.0\text{]}$.]{Current best fitness (blue) and $\sigma_{fitness}$ (green) of particle swarms sized 1,000 with $v_{max} = [0.5, 0.75, 1.0]$. The major increases in $\sigma_{fitness}$ correspond to enormous leaps in current best fitness values, creating the distinct positive correlation.}\label{F:pso3cs}
        \end{figure}

\begin{figure}
        \centering\includegraphics[width=126mm]{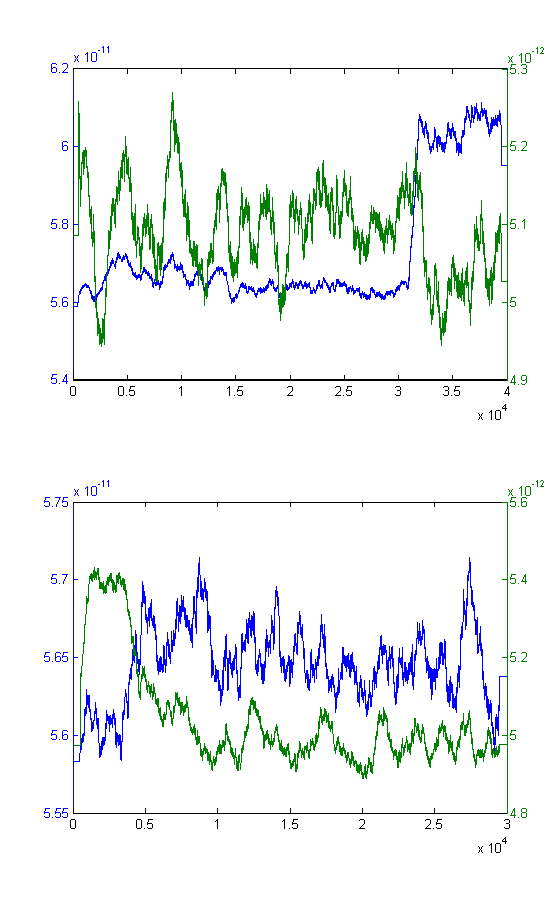}
        \caption[Current best fitness and $\sigma _{fitness}$ of failing particle swarms of size 1,000 with $v_{max} = \text{[}0.5, 1.0\text{]}$.]{Current best fitness (blue) and $\sigma_{fitness}$ (green) of failing particle swarms of size 1,000 with $v_{max} = [0.5, 1.0]$. The lack of positive correlation is due to the lack of order-of-magnitude leaps in current best fitness that would strongly affect $\sigma_{fitness}$.}\label{F:pso2cs}
        \end{figure}

\begin{figure}
        \centering\includegraphics[width=126mm]{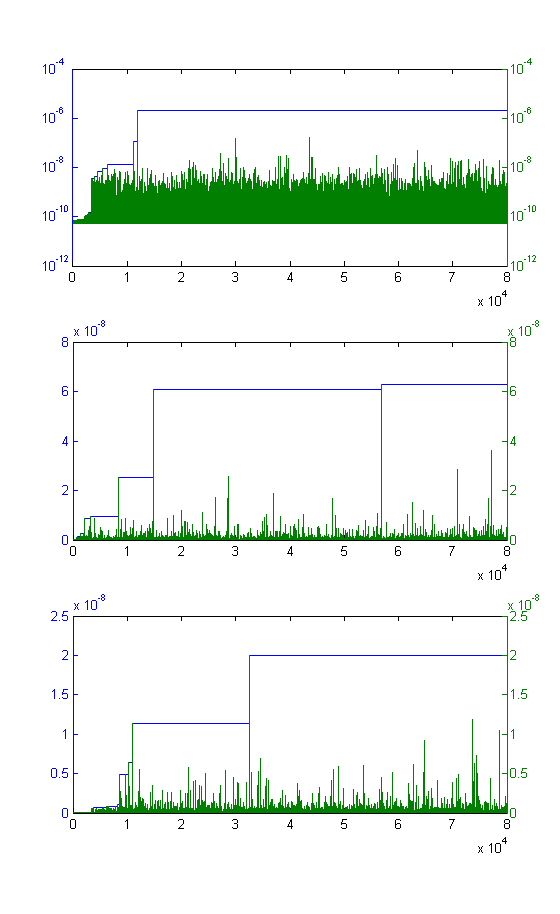}
        \caption[Overall best fitness and current best fitness of particle swarms sized 1,000 with $v_{max} = \text{[}0.5, 0.75, 1.0\text{]}$.]{Overall best fitness and current best fitness, before smoothing, of particle swarms sized 1,000 with $v_{max} = [0.5, 0.75, 1.0]$. The first plot is a log plot.}\label{F:pso3od}
        \end{figure}

        Characteristics (3) and (4), also visible in Figure \ref{F:pso3ms}, are also due to the early introduction of diversity into the successful particle swarms. Surprisingly, $\mu_{fitness}$ drops below its initial value (which represented the mean fitness of randomly initialised particles); this could indicate a search space feature similar to a steep hyperdimensional Mexican hat function (where high-fitness values are surrounded by below-average-fitness values), called the Laplacian of Gaussian function.\footnote{\texttt{http://en.wikipedia.org/wiki/Mexican\_hat\_wavelet}} Finally, the positive correlation between the current best fitness and $\sigma_{fitness}$ (5), seen in Figure \ref{F:pso3cs}, highlights the multiple-order-of-magnitude leaps in the current best fitness relation (see also Figure \ref{F:pso3od}) for successful particle swarms; the lack of this relation in failing swarms is seen in Figure \ref{F:pso2cs}.

        As seen in Figure \ref{F:pso3od}, the smaller $v_{max}$ value of 0.5 brings about many more high-fitness peaks above the $10^{-8}$ fitness level than do larger $v_{max}$ values, by moving around more `slowly' close to the global minimum. Unfortunately, smaller $v_{max}$ values often prematurely converge or fall into local minima, but a solution for this dilemma is presented in \S\ref{S:psoDiscussion}.

        \section{Discussion}\label{S:psoDiscussion}
        As mentioned in \S\ref{S:psoResults}, smaller swarm sizes and smaller values of $v_{max}$ often converge prematurely, but smaller values of $v_{max}$ yield more precise solutions if the swarm approaches the global minimum, and smaller swarm sizes can sometimes be more computationally efficient. In order to leverage the advantages of smaller swarms and smaller $v_{max}$ values while avoiding premature convergence, we implemented a simple, tapered $v_{max}$ setting: this tapered $v_{max}$ is initially very high to avoid early convergence and allow the particles to explore the search space, but it is gradually reduced over the course of the run so that a very small $v_{max}$ is enforced once the swarm has located the region of the global minimum.

        We present three runs: firstly, a swarm of 1,000 with a short taper to 80,000 generations, secondly, a swarm of 1,000 with a longer taper to 100,000 generations, and thirdly, a swarm of 100 with the longer taper to 200,000 generations. Table \ref{T:psoTapers} presents the tapered $v_{max}$ values and their corresponding generation marks.

        \begin{table}
        \begin{tabular}{ | c | c | c | }
        \hline
        $v_{max}$     & From Gen. (short)  & From Gen. (long) \\ \hline
        1.5           & 0     & 0 \\ \hline
        1.0           & 5000  & 10000 \\ \hline
        0.75          & 10000 & 20000 \\ \hline
        0.5           & 15000 & 30000 \\ \hline
        0.25          & 20000 & 40000 \\ \hline
        0.1           & 25000 & 50000 \\ \hline
        0.05          & 30000 & 60000 \\ \hline
        \end{tabular}
        \\[5pt]\caption{Incremental $v_{max}$ values for the tapered PSO algorithm and the generations at which they are enforced.}\label{T:psoTapers}
        \end{table}

\begin{figure}
        \centering\includegraphics[width=126mm]{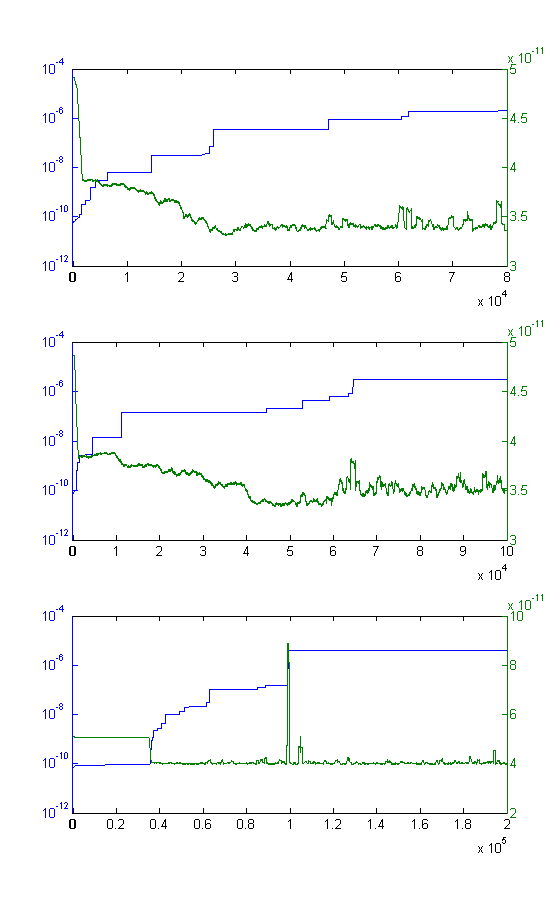}
        \caption[Overall best fitness and $\mu_{fitness}$ of the tapered PSO runs.]{Overall best fitness (blue) and $\mu_{fitness}$ (green) of the tapered PSO runs. $\mu_{fitness}$ drops from its initial value, as expected.}\label{F:pso3tom}
        \end{figure}

\begin{figure}
        \centering\includegraphics[width=126mm]{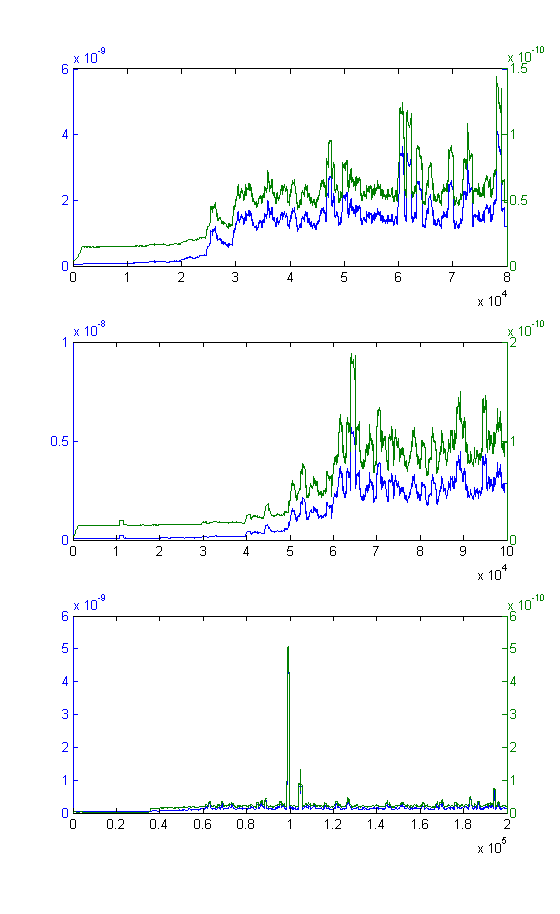}
        \caption[Current best fitness and $\sigma_{fitness}$ of the tapered PSO runs.]{Current best fitness (blue) and $\sigma_{fitness}$ (green) of the tapered PSO runs. Note the same clear positive correlation between current best fitness and $\sigma_{fitness}$, as well as the increasing $\sigma_{fitness}$ values over time. Also, no clear negative correlation between $\mu_{fitness}$ (Figure \ref{F:pso3tom}) and $\sigma_{fitness}$ is apparent.}\label{F:pso3tcs}
        \end{figure}

        All runs of the tapered PSO algorithm (including several not presented here) successfully avoided premature convergence, and all approached the global minimum. Additionally, all achieved significant improvements in fitness over our best performing basic PSO run (although still not reaching the fitness of our GA run), indicating that the smaller $v_{max}$ values in later generations were effective at increasing precision in orbital element estimations. Figures \ref{F:pso3tom} and \ref{F:pso3tcs} display the familiar characteristics of successful PSO runs.

\part{Photometric Redshift Estimation}
    \chapter{Quasar Redshifts in Optical Sky Surveys}
        \section{Quasars and Redshifts}
        Quasars, or quasi-stellar objects (QSOs), are thought to be regions of gas surrounding supermassive black holes at the centres of very distant (\textgreater~800 million light-years) galaxies. Because quasars are at such great distances, they exhibit very high \emph{redshifts} ($z$), translations of observations (on the electromagnetic spectrum) towards longer wavelengths ($\lambda$) as a result of the expansion of the universe:
        \begin{equation}
        z = \frac{\lambda_{observed} - \lambda_{expected}}{\lambda_{expected}}
        \end{equation}
        such that $1+z$ represents the ratio of expansion of the universe between the moment of the object's light emission and our current time:
        \begin{equation}
        1+z = \frac{\lambda_{observed}}{\lambda_{expected}}
        \end{equation}
        That is, if an object is observed at $z=1$, then the universe has expanded by a factor of $1+z=2$ since the light reaching us was emitted.

        Accordingly, since distance on cosmological scales is directly correlated with redshift, and since quasars are the most distant directly observable objects in the universe, it is useful to be able to determine their redshifts accurately and efficiently. Further study into the properties of quasars at different redshifts will contribute to a better understanding of cosmological evolution and the large-scale structure of the universe.

        Unfortunately, the most accurate method of redshift determination (using spectroscopy, see \S\ref{S:zspec}) is very time-consuming and does not scale to keep up with the number of quasars being identified using modern techniques: Ball et al. (to appear) state, ``the number of spectra available typically lags the number of [photometric] images by more than an order of magnitude'' \cite{nB07b}. Thus, we present here methods of photometric redshift estimation that are hundreds of times more efficient than spectroscopic determination and are increasingly consistent with the most accurate spectroscopic results.

        \section{The Sloan Digital Sky Survey}
        The Sloan Digital Sky Survey\footnote{Funding for the SDSS has been provided by the Alfred P. Sloan Foundation, the Participating Institutions, the National Science Foundation, the U.S. Department of Energy, the National Aeronautics and Space Administration, the Japanese Monbukagakusho, the Max Planck Society, and the Higher Education Funding Council for England.} (SDSS) \cite{dY00} is perhaps the most comprehensive sky survey taken to date; it has catalogued nearly 25\% of the sky\footnote{\texttt{http://www.sdss.org/background/}} using five broadband \emph{photometric filters}, which allow its telescope to simultaneously measure the luminosity of objects at different wavelengths, or in different `colours'. The SDSS uses the $u$, $g$, $r$, $i$, and $z$ filters,\footnote{We use the designations $ugriz$ and $u'g'r'i'z'$ (as found in much of the literature) interchangeably; details of their differences can be found in \cite{kA03}, \cite{jG01}, \cite{gR01a}, and especially \cite{cS02}. Note that the $ugriz$ photometric system supersedes the preliminary $u^\ast g^\ast r^\ast i^\ast z^\ast$ system in \cite{gR01a}.} which measure light between 3590 \AA\ and 9060 \AA\footnote{1 \aa ngstr\"{o}m = 0.1 nanometres.} \cite{dY00} as listed in Table \ref{T:ugriz}.\footnote{Cf. also \cite{dS05} and \cite{mF96} for `effective wavelength' figures.}
        \begin{table}
            \begin{tabular}{ | l | l | l |}
            \hline
            Band    & Colour            & Wavelength \\ \hline
            $u$     & near-ultraviolet  & 3590 \AA  \\ \hline
            $g$     & green (visible)   & 4810 \AA  \\ \hline
            $r$     & red (visible)     & 6230 \AA  \\ \hline
            $i$     & far-red           & 7640 \AA  \\ \hline
            $z$     & further-red       & 9060 \AA  \\ \hline
            \end{tabular}
            \\[5pt]\caption{$ugriz$ filters used in the SDSS.}\label{T:ugriz}
        \end{table}
        \section{Spectroscopic Redshifts in the SDSS}\label{S:zspec}
        Photometric data aside, the SDSS has also captured the electromagnetic \emph{spectra}---energy measurements over all `optical' wavelengths from 3800 \AA\ to 9200 \AA\ at a resolution of approximately $\lambda / \Delta\lambda = 1800$ \cite{dY00}\footnote{Cf. \cite{kA05} and \cite {pH04}.}---of some 100,000 quasars. These spectra yield a great deal of information about the quasars inspected, including their chemical contents, temperatures, any intervening gases blocking our line of sight in the interstellar medium, and, most importantly for our purposes, their spectroscopic redshifts ($z_{spec}$). These values are the most accurate estimates of redshift available, as they make use of data points taken across such a large region of the electromagnetic spectrum.
        \section{Magnitudes and Photometric Colours}
        We have noted that the SDSS uses the $ugriz$ photometric system; each of these five measurements is a logarithmic measure of flux ($F$)---a measurement of the amount of light emitted by an object per unit of time---around a particular wavelength. Since the measurements are logarithmic (e.g., $u \propto \log F_u$), and since $\log a - \log b = \log (a/b)$, their differences represent flux ratios; for example, $u-g \propto \log (F_u/F_g)$. It follows from this that the \emph{photometric colours} $u-g$, $g-r$, $r-i$, etc.\ give us information about the colour proportions of a quasar as observed from Earth. We use the notation $C_{xy}$ to denote the colour $x-y$ as in Richards et al. (2001b) \cite{gR01b}.

        \section{Photometric Redshift Estimation}\label{S:PRE}
        There are two major contributing factors in photometric redshift ($z_{phot}$) estimation: firstly, the location of redshifted spectral features in wavelength space relative to the ranges of the broadband filters; and, relatedly, the structure of the \emph{colour-redshift relation} (CZR).

        Independent of redshift, quasar spectra are known to exhibit certain prominent emission and absorption lines, notably Mg II, C III, C IV, and Lyman-$\alpha$ \cite{cS02}. At different redshifts, however, these spectral lines are observable at different wavelengths, and so move in and out of the ranges of the $ugrizJHK$ bands as redshift increases \cite{gR01a}. For redshifts ($z \sim 0.3$) at which the Mg II emission line is observable in the $u$ band, for example, the colour $C_{ug}$ is much bluer than usual. As Mg II moves into the $g$ band at $z \sim 0.6$, $C_{ug}$ becomes redder while $C_{gr}$ becomes bluer (Figure \ref{F:gR01b}).

        However, when these spectral features are not observed in the bandpasses in use, there can be significant degeneracy in the CZR in colour-colour space (see Figure \ref{F:mW04}), whereby the same set of colours corresponds to more than one redshift. Richards et al. (2001b) point out, then, that ``z is not strictly a \emph{function} of color'' \cite{gR01b}. This degeneracy can lead to systematic errors in $z_{phot}$ and regions of `catastrophic failure' \cite{nB07a} in the $z_{phot}$-$z_{spec}$ relation.



        \section{SDSS Quasar Dataset}
        The data we use are from the third SDSS Quasar Catalog \cite{dS05} via the Center for Astrostatistics at Penn State University.\footnote{\texttt{http://astrostatistics.psu.edu/datasets/SDSS\_quasar.html}} This dataset primarily comprises 46,420 quasars measured in $ugriz$ filters with corresponding $z_{spec}$ values, but the quasars are cross-referenced with other sky surveys (FIRST \cite{rB95}, RASS \cite{bA81}, 2MASS\footnote{This publication makes use of data products from the Two Micron All Sky Survey, which is a joint project of the University of Massachusetts and the Infrared Processing and Analysis Center/California Institute of Technology, funded by the National Aeronautics and Space Administration and the National Science Foundation.} \cite{mS06}) when possible. These three additional surveys add matching broadband radio, X-ray, and $JHK$ (shortward of $z$, see Table \ref{T:JHK}) magnitudes for 3,757, 2,672, and 6,192 quasars, respectively.

        \begin{table}
            \begin{tabular}{ | l | l | l | }
            \hline
            Band    & Colour            & Wavelength \\ \hline
            $J$     & near-infrared     & 12350 \AA  \\ \hline
            $H$     & near-infrared     & 16620 \AA  \\ \hline
            $K$     & near-infrared     & 21590 \AA  \\ \hline
            \end{tabular}
            \\[5pt]\caption{$JHK$ filters used in 2MASS \cite{pH04}.}\label{T:JHK}
        \end{table}

        The third edition of the quasar catalog, sourced from SDSS Data Release 3 (DR3), contains $ugriz$ magnitudes that have not been corrected for galactic extinction.\footnote{Also called reddening, extinction is the scattering of emitted light in the interstellar medium due to the presence of intervening dust and gas particles; in near-optical wavelengths (e.g., $ugriz$, objects appear redder than they should.} Since the effects of galactic extinction are systematic and not variable, we have not performed the extinction corrections on our dataset; training with artificial neural networks and radial basis function networks nullifies these systematics \cite{aC04} if the test data are also uncorrected. Leaving the data as-is also allows us to make use of stated one-sigma Gaussian\footnote{The assumption that we can treat stated error bars as following a Gaussian was confirmed by Dr Daniel Vanden Berk, Dept of Astronomy and Astrophysics, Penn State University, in private communication. This assumption is crucial for determination of output confidence in \S\ref{SS:jacobian}.} measurement errors (i.e., photometric noise estimates) for $ugrizJHK$ magnitudes without introducing additional variance for extinction corrections. The utility of these photometric measurement errors leads us to omit the radio and X-ray measurements that lack stated error bars.

        We have arranged the data into five different input sets (with set sizes parenthetical) for training and testing: $ugriz$ (46,420); $C_{ug},C_{gr},C_{ri},C_{iz}$ (46,420); $ugriz,C_{ug},C_{gr},C_{ri},C_{iz}$ (46,420); $ugriz_{6192}$ (6,192 matched to 2MASS); $ugrizJHK$ (6,192); and $C_{ug},C_{gr},C_{ri},C_{iz},C_{zJ},C_{JH},C_{HK}$ (6,192). For each dataset, we have the above 4--9 input magnitudes, corresponding $z_{spec}$, and, for some tests, photometric error values;\footnote{For errors of colours $C_{xy}$, we sum the errors $\sigma_x$ and $\sigma_y$ in quadrature: $\sigma_{C_{xy}} = \sqrt{\sigma_x^2 + \sigma_y^2}$, following Richards et al. (2001b) \cite{gR01b}.} our aim is to investigate how effectively different machine learning techniques can make use of these photometric input data to estimate quasar redshifts.
    \chapter{Artificial Neural Networks for Photo-$z$ Estimation}\label{C:ANNs}
        \section{Basics of Artificial Neural Networks}
            \subsection{Motivation and Network Structure}
            The \emph{artificial neural network} (ANN) has as its model the functioning of the human brain \cite[p. 166]{mN02}: the basic unit of an ANN is the \emph{neuron}, which takes a real-valued input and fires---outputting 0 or 1 or perhaps a value in $(0,1)$---according to an activation function (e.g., a Heaviside step function or a sigmoid function) over the input. These neurons are arranged into a network architecture, typically layered such that all neurons in adjacent layers are connected to each other, output-to-input.

\begin{figure}
            \centering\includegraphics[width=126mm]{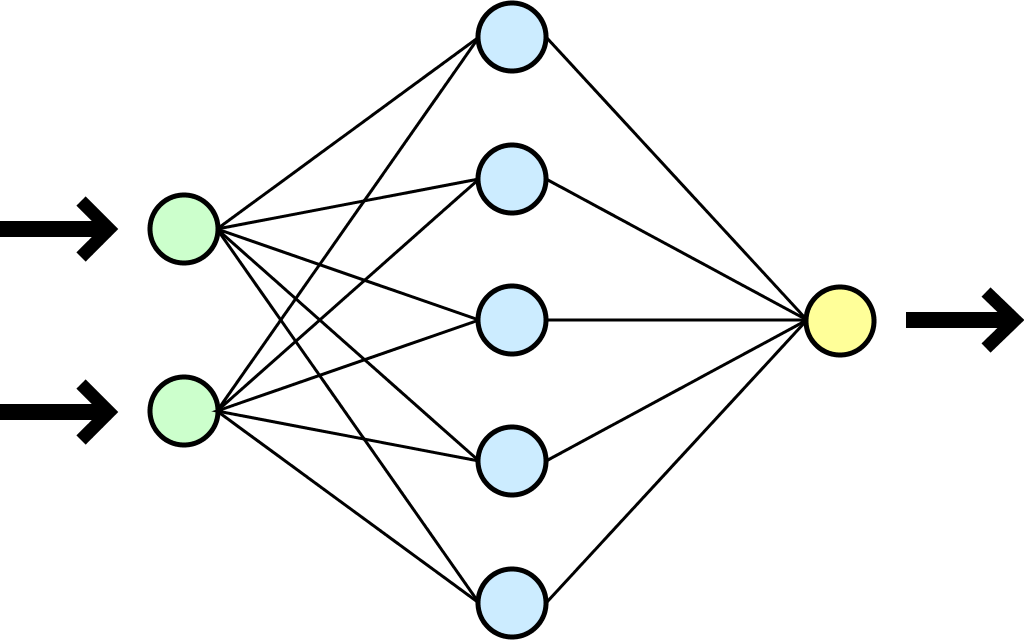}
            \caption[The structure of a two-layer 2:5:1 artificial neural network.]{The structure of a 2:5:1 artificial neural network. From \texttt{http://en.wikibooks.org/wiki/Artificial\_Neural\_Networks}.}\label{F:ANN}
            \end{figure}

            With this ordered structure, as seen in Figure \ref{F:ANN}, neurons in the middle (`hidden') and output layers have multiple inputs, which are weighted in the training stage described below. Additionally, a bias term is added to each neuron's input (in the hidden and output layers) before its output is calculated. Thus, an ANN with one hidden and one output layer (we call this a two-layer ANN, as in \cite[p. 119]{cB95}) is completely specified by its network architecture, two matrices (`vectors') each of weights and biases, and the activation functions used inside the neurons.

            \subsection{Network Training and Error Backpropagation}
            An ANN learns by adapting its weight and bias vectors based on a set of training data of input vectors and corresponding output values. The goal of the training stage is to adapt the weights and biases in the network so as to minimise errors over the training examples; however, one may allow for some small errors over the training set in order to avoid overfitting (memorising) the training data at the expense of the ANN's predictive ability on unseen test sets.

            To properly adjust the weight and bias vectors to fit the training set, ANNs are trained with an algorithm known as \textsc{Backpropagation} \cite[p. 97]{tM97}, which first runs a training input through the ANN, then  `propagates' errors backwards through the network by attributing corrections to individual neurons according to their responsibility in contributing to the overall error \cite[p. 140]{cB95}. A full description of the \textsc{Backpropagation} algorithm can be found in Mitchell (1997) \cite[p. 98]{tM97}.
            \subsection{Function Representation in an ANN}\label{SS:representation}
            It is useful to know how many layers and how many neurons we must have in an ANN in order to properly represent the non-linear functional relationship between photometric magnitudes and redshifts. Bishop (1995) demonstrates that ``any given decision boundary can be approximated arbitrarily closely by a two-layer network having sigmoidal activation functions'' \cite[p. 126]{cB95} and a sufficiently large number of neurons in the hidden layer (`hidden units'). While this does not tell us how many neurons or digits of numerical precision we will require for a particular problem, it does remind us that, in principle, two-layer ANNs are equivalent in representational capability to networks with three or more layers. Accordingly, for simplicity's sake, we will restrict our investigations to two-layer networks.
        \section{Concerns in Redshift Estimation}
            \subsection{The Jacobian Matrix and Output Confidence}\label{SS:jacobian}
            In astrophysical applications, it is important to know not just the best-estimate value, but also to quantify the accuracy of that estimate.\footnote{Cf. \cite{nB07b} and \cite{gR01b}.} Since we have one-sigma (normally distributed) measurement errors ($\sigma_u$, etc.) for each of the $ugrizJHK$ magnitudes in the SDSS quasar dataset, we demonstrate how they can be used to derive error bars for individual $z_{phot}$ outputs in an ANN.

            Given the use of differentiable sigmoid functions in our hidden layer (as opposed to some non-differentiable step functions), we can apply a series of partial derivatives to propagate errors for inputs $x_i$ ($\sigma_{x_i}$, understood in the sense of $\delta x_i$) forwards through the network (similar in form to the \textsc{Backpropagation} algorithm) to arrive at a variance $\sigma_y^2$ for output $y = z_{phot}$. This is done by means of the \emph{Jacobian matrix} $J$, where
            \begin{equation}
            J_i \equiv \frac{\partial y}{\partial x_i}
            \end{equation}
            for a single-output network. The Jacobian matrix thus ``provides a measure of the local sensitivity of the outputs to changes in each of the input variables'' \cite[p. 247]{cB06} according to
            \begin{equation}
            \sigma_y^2 \simeq \sum_i J_i \sigma^2_{x_i}.
            \end{equation}

            We therefore present one-sigma confidence intervals $\sigma_y$ for $z_{phot}$ values estimated with ANN\emph{z}, although we note that similar estimations are possible for radial basis function networks with the differentiable basis functions we present in Chapter \ref{C:RBFNs}. It should also be borne in mind that these stated uncertainties are only those attributable to the presence of photometric noise in the SDSS measurements; they do not take into account degeneracies in the colour-redshift relation or difficulties inherent in estimating quasar $z_{phot}$ values at certain redshifts (see \S\ref{S:CZR}).

            \subsection{Gaussian Weight Distribution and Committees of Networks}\label{SS:committees}
            Before an ANN is trained, its weight and bias vectors are initialised with values close to and normally distributed about 0. Since the particular (local) minima found by weight optimisation algorithms depends on the randomised initial values \cite[p. 255]{cB95}, ANNs with slightly different starting parameters will end up with slightly different representations of the estimated function. In fact, Way \& Srivastava (2006) point out that ``distribution of errors follows a Gaussian'' despite the nonlinearity of the computed function \cite{mW06}. So, rather than training several ANNs and choosing the network with the best performance on the training or test set, multiple trained networks can be combined to form a \emph{committee} of networks; Bishop (1995) demonstrates that these committees will have expected error values of at most the expected error of an individual network \cite[p. 366]{cB95}. This result is due primarily to the averaging out of the normal variance in $z_{phot}$ output created by the Gaussian distribution of initial parameters.

            Because of the expected nonlinearity of the functions represented by ANNs, we do not average the weight or bias vectors of these conjoined networks, but we take the mean\footnote{Median can also be used, cf. \cite{eV04} and \cite{aF02}.} of their $z_{phot}$ output values. In our implementation, each network in a committee differs only in its initial weights and biases before training, and not in architecture, in choice of activation functions, or in the order/selection of training examples from the dataset.\footnote{Cf. again Vanzella et al. (2004) \cite{eV04} who use simple gradient descent. The order of training examples makes no difference in our implementation, as we use batch optimisation methods \cite[p. 240]{cB06} that act on the entire training set simultaneously.}

        \section{Implementation of Artificial Neural Networks}
            \subsection{ANN\emph{z}}\label{SS:ANNz}
            We use two implementations of ANNs in our investigations. The first and primary implementation is a software package by Collister \& Lahav (2004) called ANN\emph{z} \cite{aC04}, nominally presented as a tool specially designed to apply ANNs to the problem of photometric redshift estimation. In fact, ANN\emph{z} is merely a general-purpose ANN tool that implements such advanced techniques as the Jacobian matrix, quasi-Newton optimisation,\footnote{The limited-memory quasi-Newton technique \cite[p. 289]{cB95} used by ANN\emph{z} is a method of optimising hidden-layer weights that is more robust than the scaled conjugate gradient method (see \S\ref{SS:MLPSinNetlab}) but has significantly higher computational cost.} and committees of networks; nothing in its code is specifically designed for photometric redshift estimation.

            ANN\emph{z} will construct a feed-forward ANN (the specific variety of ANN used is known as a multi-layer perceptron, or MLP) of any size, with any number of layers and any number of neurons in each layer. The notation we use, following Firth et al. (2002) \cite{aF02}, specifies the number of neurons in each successive layer; e.g., 4:6:1 specifies a two-layer network with 4 inputs, a single output, and one hidden layer of 6 neurons. This network is then trained on a training set (we have used training sets of a random sample of 60\% of our data) whilst being validated on a validation set (20\%). This validation set allows ANN\emph{z} to avoid overfitting the training data;\footnote{This follows from Mitchell (1997) \cite[p. 111]{tM97} and Bishop (1995) \cite[p. 372]{cB95}.} it fits to the training data for any specified number of iterations but selects the network weights and biases that produce the lowest \emph{root-mean-square} (RMS) deviation---between its estimated $z_{phot}$ and the accurate $z_{spec}$ value---over the \emph{validation} set \cite{aC04}:
            \begin{equation}\label{E:ANNzRMS}
            \sigma_{\text{RMS}} = \sqrt{\langle (z_{phot} - z_{spec})^2 \rangle}
            \end{equation}
            The validation set thus mimics the test set and allows the ANN to test its parameters against unseen data before being applied to the test set (the remaining 20\% of the data), fitting to the training data as closely as possible while still preserving the generalisation capabilities of the network.

            Additionally, ANN\emph{z} estimates variances in its $z_{phot}$ outputs as described in \S\ref{SS:jacobian},\footnote{ANN\emph{z} also includes error due to deviation in committee members' estimations; the program sums this standard deviation in quadrature with the photometric measurement errors, but we have removed this committee error in order to focus on error due to measurement noise.} and readily applies committees of networks as described in \S\ref{SS:committees}. We present results using both of these techniques in \S\ref{S:ANNresults}. Hidden units in ANN\emph{z} use the logistic sigmoid activation function\footnote{See source code at \texttt{http://zuserver2.star.ucl.ac.uk/$\sim$lahav/annz.src.tar.gz}.} \cite[p. 228]{cB06}:
            \begin{equation}
            \sigma(a) \equiv \frac{1}{1 + e^{-a}}
            \end{equation}
            whose output values are restricted to the interval $(0,1)$. On the other hand, units in the output layer use a linear activation function (a simple sum of its weighted inputs and biases; linear transformations are unnecessary given the training process) so as not to restrict the range of possible network outputs. Note that there is no loss of generality with the linear output \cite[p. 127]{cB95}; it does not limit our network's representational capability as put forward in \S\ref{SS:representation}.
            \subsection{MLPs in Netlab}\label{SS:MLPSinNetlab}
            Our second implementation of ANNs (MLPs) is in Netlab,\footnote{\texttt{http://www.ncrg.aston.ac.uk/netlab/}} a toolbox of MATLAB\textsuperscript{\textregistered} methods written by Ian~T. Nabney and Christopher~M. Bishop. Netlab provides a slightly more customisable implementation, allowing us to make use of hyperbolic tangent activation functions in the hidden layer:
            \begin{equation}
            tanh(a) \equiv \frac{e^a - e^{-a}}{e^a + e^{-a}}
            \end{equation}
            which are equivalent in representational power to logistic sigmoid functions \cite[p. 127]{cB95} yet tend to converge faster in training.\footnote{Cf. also \cite[p. 185]{mN02}.} Additionally, Netlab provides the option of using a scaled conjugate gradient\footnote{See \cite[p. 282]{cB95} for a description of this technique.} algorithm to train the weights at $O(NW)$ computational cost, instead of the $O(NW^2)$ cost for a quasi-Newton method \cite[p. 288]{cB95}, where $N$ is the number of training examples and $W$ is the number of adaptive weights. However, MLPs in Netlab cannot have more than one hidden layer and do not include the readily calculated output variance figures according to the Jacobian matrix, as in ANN\emph{z}. Still, the training of MLPs in Netlab gives us a more direct point of comparison to radial basis function networks (Chapter \ref{C:RBFNs}, also implemented in Netlab) with respect to accuracy, efficiency, and convergence.
        \section{Results and Analysis}\label{S:ANNresults}
        For ANN\emph{z}, we use network architectures of x:10:1, x:20:1, x:40:1, and x:100:1 for each dataset, and we form committees of five networks in each case, trained up to 1,000 iterations or until network convergence (whichever is first). We present committee $z_{phot}$ RMS errors and the average (also RMS) error attributable to photometric noise in the measurements, as well as the percentage of redshifts successfully predicted within $\Delta z = [0.1, 0.2, 0.3]$, where $\Delta z \equiv |z_{phot}-z_{spec}|$, for consistency with the literature. For MLPs in Netlab, we use the same network structures as with ANN\emph{z}, but instead present some individual network RMS errors: the minimum, maximum, and mean RMS for networks in each committee, as well as the $\sigma_{\text{RMS}}$ obtained by the committee as a whole, after 500 and 1,000 training iterations. The maximum number of iterations (1,000) was selected after observing the convergence behaviour of x:40:1 and x:100:1 networks training over 3,000 iterations while monitoring the error over the corresponding test set. Errors were found to be minimal between 800 and 1,500 generations; training to a maximum of 1,000 iterations should allow us to avoid overfitting most of the training sets. Finally, we look at plots of $z_{phot}$ vs. $z_{spec}$ and discuss the source of errors in $z_{phot}$ estimates.
            \subsection{Results with ANN\emph{z}}\label{SS:ANNzResults}
            Tables \ref{T:ANNz1}, \ref{T:ANNz2}, \ref{T:ANNz3}, and \ref{T:ANNz4} present the error values achieved with ANN\emph{z} for architectures with 10, 20, 40, and 100 hidden units, respectively.

            \begin{table}
                \begin{tabular}{ | l | c | c | c | c | c | }
                \hline
                Dataset    & $\sigma_{\text{RMS}}$  & RMS\textsubscript{noise} & $\Delta z \leq 0.1$ & $\leq 0.2$ & $\leq 0.3$ \\ \hline
                $ugriz$                             & 0.4241 & 0.1752 & 0.3193 & 0.5444 & 0.6587  \\ \hline
                $C_{ug},C_{gr},C_{ri},C_{iz}$       & 0.4256 & 0.1731 & 0.3313 & 0.5476 & 0.6603  \\ \hline
                $ugriz,C_{ug},C_{gr},C_{ri},C_{iz}$ & 0.4148 & 0.1511 & 0.3372 & 0.5607 & 0.6729  \\ \hline
                $ugriz_{6192}$                      & 0.4940 & 0.2204 & 0.2410 & 0.4140 & 0.5460  \\ \hline
                $ugrizJHK$                          & 0.3718 & 0.1695 & 0.3740 & 0.6120 & 0.7520  \\ \hline
                $C_{ug},C_{gr},C_{ri},C_{iz},C_{zJ},C_{JH},C_{HK}$    & 0.3857 & 0.2793 & 0.3550 & 0.5960 & 0.7190  \\ \hline
                \end{tabular}
                \\[5pt]\caption[ANN\emph{z} results with 5-committees of x:10:1 networks.]{ANN\emph{z} results with 5-committees of x:10:1 networks. RMS\textsubscript{noise} is the estimated error attributable to photometric measurement noise, as discussed in \S\ref{SS:jacobian}.}\label{T:ANNz1}
            \end{table}

            \begin{table}
                \begin{tabular}{ | l | c | c | c | c | c | }
                \hline
                Dataset    & $\sigma_{\text{RMS}}$  & RMS\textsubscript{noise} & $\Delta z \leq 0.1$ & $\leq 0.2$ & $\leq 0.3$ \\ \hline
                $ugriz$                             & 0.4064 & 0.1819 & 0.3766 & 0.5906 & 0.6957  \\ \hline
                $C_{ug},C_{gr},C_{ri},C_{iz}$       & 0.4109 & 0.1763 & 0.3709 & 0.5885 & 0.6943  \\ \hline
                $ugriz,C_{ug},C_{gr},C_{ri},C_{iz}$ & 0.3958 & 0.1470 & 0.3753 & 0.5940 & 0.7075  \\ \hline
                $ugriz_{6192}$                      & 0.4793 & 0.2049 & 0.3140 & 0.4870 & 0.6140  \\ \hline
                $ugrizJHK$                          & 0.3949 & 0.1506 & 0.2690 & 0.5420 & 0.6920  \\ \hline
                $C_{ug},C_{gr},C_{ri},C_{iz},C_{zJ},C_{JH},C_{HK}$    & 0.3621 & 0.2731 & 0.4390 & 0.6530 & 0.7630  \\ \hline
                \end{tabular}
                \\[5pt]\caption{ANN\emph{z} results with 5-committees of x:20:1 networks.}\label{T:ANNz2}
            \end{table}

            \begin{table}
                \begin{tabular}{ | l | c | c | c | c | c | }
                \hline
                Dataset    & $\sigma_{\text{RMS}}$  & RMS\textsubscript{noise} & $\Delta z \leq 0.1$ & $\leq 0.2$ & $\leq 0.3$ \\ \hline
                $ugriz$                             & 0.3970 & 0.1773 & 0.3963 & 0.6100 & 0.7123  \\ \hline
                $C_{ug},C_{gr},C_{ri},C_{iz}$       & 0.4053 & 0.2135 & 0.4019 & 0.6148 & 0.7089  \\ \hline
                $ugriz,C_{ug},C_{gr},C_{ri},C_{iz}$ & 0.3877 & 0.1611 & 0.4091 & 0.6265 & 0.7244  \\ \hline
                $ugriz_{6192}$                      & 0.5048 & 0.2281 & 0.3160 & 0.5080 & 0.6330  \\ \hline
                $ugrizJHK$                          & 0.3730 & 0.1694 & 0.4380 & 0.6480 & 0.7500  \\ \hline
                $C_{ug},C_{gr},C_{ri},C_{iz},C_{zJ},C_{JH},C_{HK}$    & 0.3596 & 0.2791 & 0.4350 & 0.6630 & 0.7570  \\ \hline
                \end{tabular}
                \\[5pt]\caption[ANN\emph{z} results with 5-committees of x:40:1 networks.]{ANN\emph{z} results with 5-committees of x:40:1 networks. Note the marked increase in $\sigma_{\text{RMS}}$ for the $ugriz_{6192}$ set.}\label{T:ANNz3}
            \end{table}

            \begin{table}
                \begin{tabular}{ | l | c | c | c | c | c | }
                \hline
                Dataset    & $\sigma_{\text{RMS}}$  & RMS\textsubscript{noise} & $\Delta z \leq 0.1$ & $\leq 0.2$ & $\leq 0.3$ \\ \hline
                $ugriz$                             & 0.3960 & 0.1853 & 0.3983 & 0.6145 & 0.7129  \\ \hline
                $C_{ug},C_{gr},C_{ri},C_{iz}$       & 0.4067 & 0.2022 & 0.4174 & 0.6228 & 0.7152  \\ \hline
                $ugriz,C_{ug},C_{gr},C_{ri},C_{iz}$ & 0.3854 & 0.1599 & 0.4199 & 0.6380 & 0.7301  \\ \hline
                $ugriz_{6192}$                      & 0.5255 & 0.1162 & 0.1880 & 0.3680 & 0.4980  \\ \hline
                $ugrizJHK$                          & 0.4081 & 0.1458 & 0.3610 & 0.5690 & 0.6990  \\ \hline
                $C_{ug},C_{gr},C_{ri},C_{iz},C_{zJ},C_{JH},C_{HK}$    & 0.3557 & 0.2905 & 0.4500 & 0.6480 & 0.7620  \\ \hline
                \end{tabular}
                \\[5pt]\caption[ANN\emph{z} results with 5-committees of x:100:1 networks.]{ANN\emph{z} results with 5-committees of x:100:1 networks. Note the increasing $\sigma_{\text{RMS}}$ for $ugriz_{6192}$ and $ugrizJHK$.}\label{T:ANNz4}
            \end{table}

            Although RMS errors generally shrink with larger network sizes, the improvement is clearly bounded (using a larger hidden layer does not always decrease error), and in some cases (e.g., $ugriz_{6192}$ and $ugrizJHK$ for 100 hidden units) error actually increases with the larger network. Since ANN\emph{z} does not merely take the final set of weights and biases in training, but instead uses the weights and biases that minimise error over the validation set, we do not see memorisation effects for small datasets in larger networks. Instead, the increase in error for $ugriz_{6192}$ and $ugrizJHK$ is due to the larger networks' inability to train their weights and biases sufficiently over the smaller (order 6,192) datasets. Additionally, in previous (unlisted) runs we obtained consistent RMS errors of over 0.8 for the $ugriz_{6192}$ and $ugrizJHK$ datasets because of an unlucky random distribution of data into of training and validation sets. This is not to suggest that a dataset with poor distributions is necessarily intrinsically unlearnable (Netlab MLPs and RBFNs could learn on the $ugrizJHK$ dataset that ANN\emph{z} could not learn), but dataset size, distribution into training and test sets, and choice of weight optimisation algorithm all contribute to the convergence behaviour of a learning network.

            Also apparent from Tables \ref{T:ANNz1}--\ref{T:ANNz4} is that the breaking down of photometric magnitudes into colours ($ugriz$ into $[C_{ug},$ $C_{gr},$ $C_{ri},$ $C_{iz}]$ and $ugrizJHK$ into $[C_{ug},$ $C_{gr},$ $C_{ri},$ $C_{iz},$ $C_{zJ},$ $C_{JH},$ $C_{HK}]$) tends to improve $\sigma_{\text{RMS}}$ for $ugrizJHK$ and but worsens $\sigma_{\text{RMS}}$ for $ugriz$. It is unclear why this is, but it may be a balancing between the information contained in plain magnitudes and the information made more explicit in their colours: the amount of information made explicit in $[C_{ug},$ $C_{gr},$ $C_{ri},$ $C_{iz},$ $C_{zJ},$ $C_{JH},$ $C_{HK}]$ is more than made explicit in $[C_{ug},$ $C_{gr},$ $C_{ri},$ $C_{iz}]$, possibly compensating for the loss of some information in one of the $ugriz$ magnitudes. That there is information to be drawn from both the magnitudes and their colours is suggested by the fact that $[ugriz,$ $C_{ug},$ $C_{gr},$ $C_{ri},$ $C_{iz}]$ test sets are significantly better predicted than either $ugriz$ or $[C_{ug},$ $C_{gr},$ $C_{ri},$ $C_{iz}]$ sets.

            \subsection{Results with MLPs in Netlab}\label{SS:MLPsResults}
            Since MLPs in Netlab are functionally similar to those trained in ANN\emph{z}, we consider different results in this section: we use committees of five networks of the same architectures as in \S\ref{SS:ANNzResults}, but we present minimum/maximum/mean/committee $\sigma_{\text{RMS}}$ results. Networks were trained up to 500 and 1,000 iterations. This presentation will better illustrate the improvement expected when using committees of networks.

            \begin{table}
                \begin{tabular}{ | l | c | c | c | c | c | }
                \hline
                Dataset    & RMS\textsubscript{min}  & RMS\textsubscript{max} & $\mu_\text{RMS}$ & RMS\textsubscript{com} & iter \\ \hline
                $ugriz$ & 0.5883 & 0.6083 & 0.5972 & \emph{0.5931} & 500 \\ \hline
                $ugriz$ & 0.5751 & 0.5994 & 0.5862 & \emph{0.5812} & 1000 \\ \hline
                $C_{ug},C_{gr},C_{ri},C_{iz}$ & 0.4537 & 0.4607 & 0.4561 & \textbf{0.4459} & 500 \\ \hline
                $C_{ug},C_{gr},C_{ri},C_{iz}$ & 0.4506 & 0.4526 & 0.4519 & \textbf{0.4420} & 1000 \\ \hline
                $ugriz,C_{ug},C_{gr},C_{ri},C_{iz}$ & 0.5717 & 0.5972 & 0.5851 & \emph{0.5793} & 500 \\ \hline
                $ugriz,C_{ug},C_{gr},C_{ri},C_{iz}$ & 0.5637 & 0.5759 & 0.5715 & \textbf{0.5634} & 1000 \\ \hline
                $ugriz_{6192}$ & 0.8169 & 0.8178 & 0.8175 & \emph{0.8174} & 500 \\ \hline
                $ugriz_{6192}$ & 0.8170 & 0.8176 & 0.8173 & \emph{0.8171} & 1000 \\ \hline
                $ugrizJHK$ & 0.4870 & 0.4975 & 0.4928 & \emph{0.4913} & 500 \\ \hline
                $ugrizJHK$ & 0.4722 & 0.4952 & 0.4834 & \emph{0.4783} & 1000 \\ \hline
                $C_{ug},C_{gr},C_{ri},C_{iz},C_{zJ},C_{JH},C_{HK}$ & 0.4116 & 0.4374 & 0.4197 & \textbf{0.4048} & 500 \\ \hline
                $C_{ug},C_{gr},C_{ri},C_{iz},C_{zJ},C_{JH},C_{HK}$ & 0.4078 & 0.4431 & 0.4184 & \textbf{0.4004} & 1000 \\ \hline
                \end{tabular}
                \\[5pt]\caption[MLP/Netlab results with 5-committees of x:10:1 networks.]{MLP/Netlab results with 5-committees of x:10:1 networks. Committee results \emph{better than the mean} are italicised, those \textbf{better than the best individual result} (RMS\textsubscript{min}) are bolded.}\label{T:MLP1}
            \end{table}

            \begin{table}
                \begin{tabular}{ | l | c | c | c | c | c | }
                \hline
                Dataset    & RMS\textsubscript{min}  & RMS\textsubscript{max} & $\mu_\text{RMS}$ & RMS\textsubscript{com} & iter \\ \hline
                $ugriz$ & 0.5775 & 0.5871 & 0.5839 & \textbf{0.5744} & 500 \\ \hline
                $ugriz$ & 0.5488 & 0.5808 & 0.5628 & \textbf{0.5469} & 1000 \\ \hline
                $C_{ug},C_{gr},C_{ri},C_{iz}$ & 0.4456 & 0.4522 & 0.4482 & \textbf{0.4435} & 500 \\ \hline
                $C_{ug},C_{gr},C_{ri},C_{iz}$ & 0.4374 & 0.4419 & 0.4403 & \textbf{0.4335} & 1000 \\ \hline
                $ugriz,C_{ug},C_{gr},C_{ri},C_{iz}$ & 0.5445 & 0.5864 & 0.5733 & \emph{0.5630} & 500 \\ \hline
                $ugriz,C_{ug},C_{gr},C_{ri},C_{iz}$ & 0.5277 & 0.5741 & 0.5572 & \emph{0.5429} & 1000 \\ \hline
                $ugriz_{6192}$ & 0.8165 & 0.8168 & 0.8167 & \emph{0.8166} & 500 \\ \hline
                $ugriz_{6192}$ & 0.8162 & 0.8167 & 0.8165 & \emph{0.8164} & 1000 \\ \hline
                $ugrizJHK$ & 0.4870 & 0.4936 & 0.4901 & \textbf{0.4869} & 500 \\ \hline
                $ugrizJHK$ & 0.4688 & 0.4933 & 0.4799 & \emph{0.4742} & 1000 \\ \hline
                $C_{ug},C_{gr},C_{ri},C_{iz},C_{zJ},C_{JH},C_{HK}$ & 0.4057 & 0.4209 & 0.4130 & \textbf{0.3963} & 500 \\ \hline
                $C_{ug},C_{gr},C_{ri},C_{iz},C_{zJ},C_{JH},C_{HK}$ & 0.3854 & 0.4055 & 0.3978 & \textbf{0.3789} & 1000 \\ \hline
                \end{tabular}
                \\[5pt]\caption{MLP/Netlab results with 5-committees of x:20:1 networks.}\label{T:MLP2}
            \end{table}

            \begin{table}
                \begin{tabular}{ | l | c | c | c | c | c | }
                \hline
                Dataset    & RMS\textsubscript{min}  & RMS\textsubscript{max} & $\mu_\text{RMS}$ & RMS\textsubscript{com} & iter \\ \hline
                $ugriz$ & 0.5700 & 0.6023 & 0.5855 & \emph{0.5783} & 500 \\ \hline
                $ugriz$ & 0.5399 & 0.5798 & 0.5635 & \emph{0.5538} & 1000 \\ \hline
                $C_{ug},C_{gr},C_{ri},C_{iz}$ & 0.4409 & 0.4491 & 0.4448 & \textbf{0.4402} & 500 \\ \hline
                $C_{ug},C_{gr},C_{ri},C_{iz}$ & 0.4366 & 0.4393 & 0.4381 & \textbf{0.4323} & 1000 \\ \hline
                $ugriz,C_{ug},C_{gr},C_{ri},C_{iz}$ & 0.5412 & 0.5693 & 0.5598 & \emph{0.5527} & 500 \\ \hline
                $ugriz,C_{ug},C_{gr},C_{ri},C_{iz}$ & 0.5131 & 0.5486 & 0.5316 & \emph{0.5194} & 1000 \\ \hline
                $ugriz_{6192}$ & 0.8165 & 0.8172 & 0.8169 & \emph{0.8167} & 500 \\ \hline
                $ugriz_{6192}$ & 0.8165 & 0.8174 & 0.8170 & \emph{0.8168} & 1000 \\ \hline
                $ugrizJHK$ & 0.4819 & 0.4990 & 0.4919 & \emph{0.4871} & 500 \\ \hline
                $ugrizJHK$ & 0.4725 & 0.4943 & 0.4830 & \emph{0.4760} & 1000 \\ \hline
                $C_{ug},C_{gr},C_{ri},C_{iz},C_{zJ},C_{JH},C_{HK}$ & 0.3997 & 0.4335 & 0.4105 & \textbf{0.3916} & 500 \\ \hline
                $C_{ug},C_{gr},C_{ri},C_{iz},C_{zJ},C_{JH},C_{HK}$ & 0.3945 & 0.4181 & 0.4054 & \textbf{0.3808} & 1000 \\ \hline
                \end{tabular}
                \\[5pt]\caption{MLP/Netlab results with 5-committees of x:40:1 networks.}\label{T:MLP3}
            \end{table}

            \begin{table}
                \begin{tabular}{ | l | c | c | c | c | c | }
                \hline
                Dataset    & RMS\textsubscript{min}  & RMS\textsubscript{max} & $\mu_\text{RMS}$ & RMS\textsubscript{com} & iter \\ \hline
                $ugriz$ & 0.5608 & 0.5873 & 0.5796 & \emph{0.5750} & 500 \\ \hline
                $ugriz$ & 0.5482 & 0.5589 & 0.5525 & \textbf{0.5477} & 1000 \\ \hline
                $C_{ug},C_{gr},C_{ri},C_{iz}$ & 0.4431 & 0.4514 & 0.4467 & \textbf{0.4415} & 500 \\ \hline
                $C_{ug},C_{gr},C_{ri},C_{iz}$ & 0.4345 & 0.4421 & 0.4381 & \textbf{0.4327} & 1000 \\ \hline
                $ugriz,C_{ug},C_{gr},C_{ri},C_{iz}$ & 0.5481 & 0.5909 & 0.5678 & \emph{0.5634} & 500 \\ \hline
                $ugriz,C_{ug},C_{gr},C_{ri},C_{iz}$ & 0.5215 & 0.5438 & 0.5326 & \emph{0.5277} & 1000 \\ \hline
                $ugriz_{6192}$ & 0.8166 & 0.8169 & 0.8168 & \emph{0.8167} & 500 \\ \hline
                $ugriz_{6192}$ & 0.8167 & 0.8170 & 0.8168 & 0.8168 & 1000 \\ \hline
                $ugrizJHK$ & 0.4779 & 0.4942 & 0.4864 & \emph{0.4822} & 500 \\ \hline
                $ugrizJHK$ & 0.4682 & 0.4751 & 0.4717 & \textbf{0.4680} & 1000 \\ \hline
                $C_{ug},C_{gr},C_{ri},C_{iz},C_{zJ},C_{JH},C_{HK}$ & 0.4094 & 0.4438 & 0.4201 & \textbf{0.4055} & 500 \\ \hline
                $C_{ug},C_{gr},C_{ri},C_{iz},C_{zJ},C_{JH},C_{HK}$ & 0.3905 & 0.4521 & 0.4122 & \textbf{0.3878} & 1000 \\ \hline
                \end{tabular}
                \\[5pt]\caption[MLP/Netlab results with 5-committees of x:100:1 networks.]{MLP/Netlab results with 5-committees of x:100:1 networks. Note the slight increase in RMS\textsubscript{com} over all the datasets including colours $C$.}\label{T:MLP4}
            \end{table}

            It is clear from Tables \ref{T:MLP1}--\ref{T:MLP4} that, as suggested in \S\ref{SS:committees}, MLP committees can be relied upon to produce RMS errors on average no worse than the mean RMS errors of committee members (i.e., the expected $\sigma_{\text{RMS}}$ for an individual network). In a majority of cases, the committees produced results better even than the best performing individual network.

            The RMS errors over some datasets are seen to increase between the x:40:1 and the x:100:1 networks for both 500 and 1,000 iterations. However, if this slight increase in error were attributable to overfitting (memorisation of the training data with the larger hidden layer), errors over these datasets would probably also increase between their 500\textsuperscript{th} and 1,000\textsuperscript{th} iterations. Since none of this error increase is observed, we can conclude that no major overfitting has likely taken place up to 1,000 training iterations.

            We have intentionally left in a $ugriz_{6192}$ dataset that has failed to train properly (owing to the same unlucky distribution into training and test sets described in \S\ref{SS:ANNzResults}). This particular distribution frustrated learning over all MLP network architectures, and will do likewise with RBFN architectures in \S\ref{S:RBFNresults}. The training and test sets could be reassigned, however, as was done to achieve better results in \S\ref{SS:ANNzResults}. The significant improvement yielded by adding the $JHK$ magnitudes to the $ugriz_{6192}$ dataset is best illustrated in Tables \ref{T:ANNz1}--\ref{T:ANNz4}.

            We note also the superiority of nearly all the ANN\emph{z} results to those of MLPs in Netlab. As the training algorithms for ANN\emph{z} and MLP/Netlab are nearly identical (except for their quasi-Newton vs. scaled conjugate gradient methods for weight optimisation and ANN\emph{z}'s minimisation of error over the validation set), and as we do not observe any signs of significant overfitting, we might conclude that ANN\emph{z}'s use of a validation set has allowed it to generalise much more efficaciously than MLP/Netlab. This is to say that, by training over a known training set yet measuring error over an unseen validation set, ANN\emph{z} has more closely approximated the `true' functional relationship between photometric measurements/colours and (spectroscopic) redshift.

            \subsection{Analysis}
            \begin{figure}
            \centering\includegraphics[width=126mm]{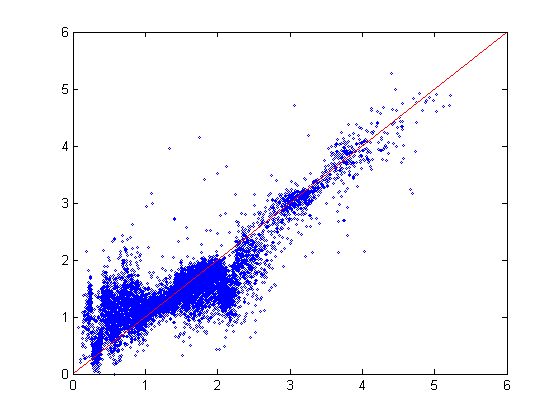}
            \caption[$z_{phot}$ vs. $z_{spec}$ for $ugriz$, 5:10:1, ANN\emph{z}.]{$z_{phot}$ vs. $z_{spec}$ for $ugriz$, 5:10:1, ANN\emph{z}. $\sigma_{\text{RMS}}$ is 0.4241.}\label{F:annzugriz1}
            \end{figure}

            \begin{figure}
            \centering\includegraphics[width=126mm]{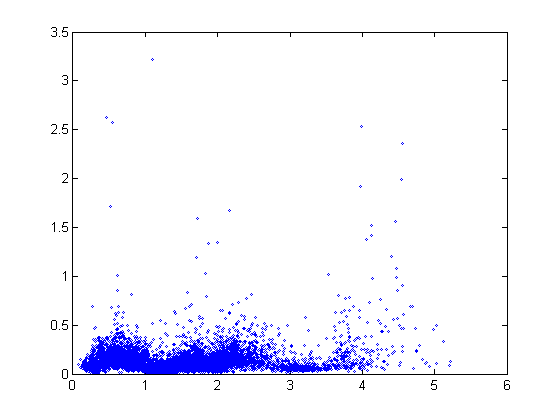}
            \caption[Photometric deviation $\sigma_y$ vs. $z_{spec}$ for $ugriz$, 5:10:1, ANN\emph{z}.]{Photometric deviation $\sigma_y$ vs. $z_{spec}$ for $ugriz$, 5:10:1, ANN\emph{z}. RMS\textsubscript{noise} is 0.1752.}\label{F:annzugriz2}
            \end{figure}

            \begin{figure}
            \centering\includegraphics[width=126mm]{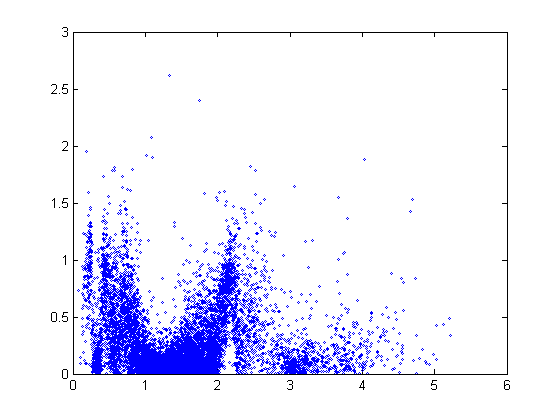}
            \caption{$\Delta z$ vs. $z_{spec}$ for $ugriz$, 5:10:1, ANN\emph{z}.}\label{F:annzugriz3}
            \end{figure}

            \begin{figure}
            \centering\includegraphics[width=126mm]{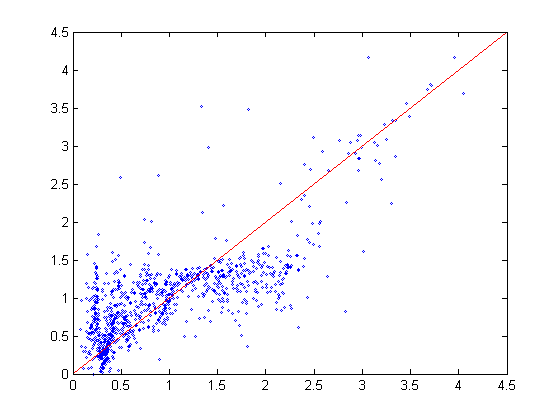}\label{F:annzugriz6192}
            \caption[$z_{phot}$ vs. $z_{spec}$ for $ugriz_{6192}$, 5:10:1, ANN\emph{z}.]{$z_{phot}$ vs. $z_{spec}$ for $ugriz_{6192}$, 5:10:1, ANN\emph{z}. $\sigma_{\text{RMS}}$ is 0.4940.}
            \end{figure}

            \begin{figure}
            \centering\includegraphics[width=126mm]{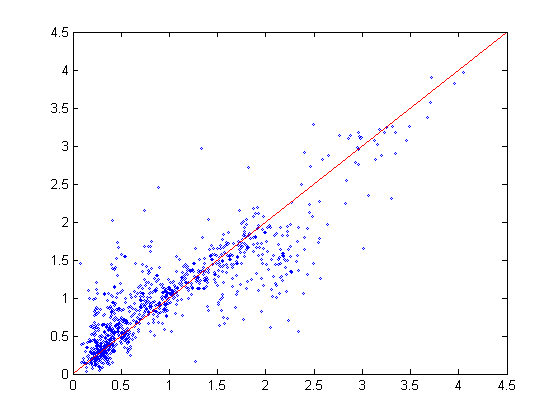}
            \caption[$z_{phot}$ vs. $z_{spec}$ for $ugrizJHK$, 8:10:1, ANN\emph{z}.]{$z_{phot}$ vs. $z_{spec}$ for $ugrizJHK$, 8:10:1, ANN\emph{z}. $\sigma_{\text{RMS}}$ is 0.3718.}\label{F:annzugrizjhk}
            \end{figure}

            Figure \ref{F:annzugriz1} shows a typical $z_{phot}$-$z_{spec}$ plot, with systematic deviations that are also areas of `catastrophic failure' (having very high $\Delta z$). To better understand the source of these deviations, we look at Figure \ref{F:annzugriz2}, which demonstrates that $\sigma_y$ (deviation due to photometric noise; see \S\ref{SS:jacobian}) is not highly correlated with $z_{spec}$; accordingly, the majority of the error relative to $z_{spec}$ is due to the inability of our ANNs to learn the colour-redshift relation. This $\Delta z$ error relative to $z_{spec}$ is plotted in Figure \ref{F:annzugriz3}, which indicates that error is much higher at specific redshifts; the astrophysical explanation for this finding is put forward in \S\ref{S:CZR}.

            Figures 6.5 and \ref{F:annzugrizjhk} present $z_{phot}$-$z_{spec}$ plots for $ugriz_{6192}$ and $ugrizJHK$, for comparison to Figure \ref{F:annzugriz1}. The reduction of dataset size from 46,420 to 6,192 over the $ugriz$ filters creates a noticeable dispersion in the $z_{phot}$ predictions in Figure 6.5. (Notice that the familiar systematic deviations are still manifest in the smaller $ugriz_{6192}$ dataset.) However, once $JHK$ filters are added, $\sigma_{\text{RMS}}$ drops noticeably (Figure \ref{F:annzugrizjhk}), and even some of the catastrophic failure is tempered.

            \begin{figure}
            \centering\includegraphics[width=126mm]{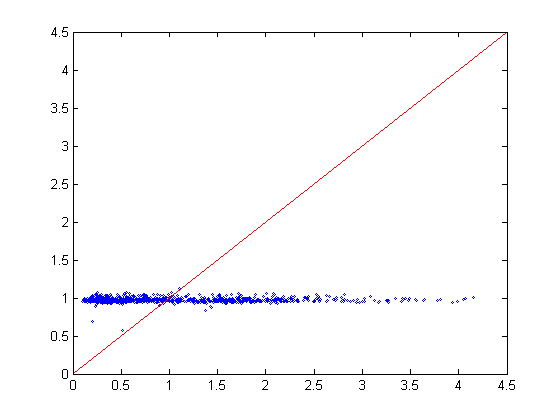}
            \caption[$z_{phot}$ vs. $z_{spec}$ for $ugriz_{6192}$, 5:100:1, MLP/Netlab, after 1,000 iterations.]{$z_{phot}$ vs. $z_{spec}$ for $ugriz_{6192}$, 5:100:1, MLP/Netlab, after 1,000 iterations. The network failed to learn the given dataset over training, and so suggests $z_{phot}$ values closely distributed about unity. $\sigma_{\text{RMS}}$ is 0.8168.}\label{F:mlpugriz6192}
            \end{figure}

            Figure \ref{F:mlpugriz6192} demonstrates the failed learning of the $ugriz_{6192}$ dataset that was poorly distributed into training and test sets. The majority of $ugriz$ magnitudes that were matched to available $JHK$ magnitudes had corresponding $z_{spec}$ values in the interval $(0,2)$. As this was a highly problematic interval in the colour-redshift relation, it is unsurprising that our MLP failed to learn the data properly; what is surprising is the degree of its failure. The $\sigma_{\text{RMS}}$ between $z_{phot}$ and $z_{spec}$ over $ugriz_{6192}$ for a network that output only values of $z_{phot} = 1$ would be 0.8162; the network in Figure \ref{F:mlpugriz6192} achieved $\sigma_{\text{RMS}} = 0.8168$. An investigation of the degeneracy in the colour-redshift relation in the interval $(0,2)$ is carried out in \S\ref{S:CZR}.

    \chapter{Radial Basis Function Networks for Photo-$z$ Estimation}\label{C:RBFNs}
    Similar in structure and representational ability \cite[p. 168]{cB95} to ANNs, \emph{radial basis function networks} (RBFNs) have some important advantages over ANNs, yet are not as widely considered in the astrophysical literature.\footnote{However, cf. \cite{dW08} and \cite{mW06}, where they are referred to as `kernel methods'.}

    The most obvious feature that RBFNs share with ANNs is their network structure: an RBFN is basically identical in form to a two-layer MLP except for its differing activation functions in the hidden layer (see \S\ref{S:RBFs}). An RBFN with $N$ hidden units is able to produce a smooth function that can precisely fit $N$ data points in a training set and can interpolate between the points. However, in our application, we want to approximate the underlying function generating our training data and account for noisy measurements \cite[p. 167]{cB95}; we will sacrifice some accuracy in fitting training data in order to maintain the generalisation ability of our network. Accordingly, we will use far fewer hidden units than training data points, as in an ANN.
        \section{Radial Basis Functions}\label{S:RBFs}
        The heart of an RBFN is, predictably, the \emph{radial basis function} (RBF), used as the activation function in the hidden layer of an RBFN. For an RBFN with $M$ inputs and $N$ hidden units, an RBF is a function $\phi_n(\cdot)$ centred at an $M$-dimensional basis vector $\boldsymbol{\mu}_n$ that varies only with the Euclidean distance (in $M$-space) between the input vector $\mathbf{x}$ and the basis vector $\boldsymbol{\mu}_n$ \cite[p. 299]{cB06}. Thus, the output of an individual hidden-layer neuron in an RBFN is in the form\footnote{The notation is our own, varying slightly from those of \cite{cB95}, \cite{cB06}, \cite{bR96}, \cite{tH01}, \cite{mW06}, and \cite{dW08}: we mean to convey more clearly that the RBF $\phi_n(\mathbf{x})$ is really a parameterised activation function of the two variables $\mathbf{x}$ and $\boldsymbol{\mu}_n$ (the fixed parameter), while $\phi(x)$ is a general kernel function.}  $\phi_n(\mathbf{x}) = \phi(\| \mathbf{x} - \boldsymbol{\mu}_n \|)$ where $\phi(\cdot)$ is one of the kernel functions defined below.

        The kernel functions (weighted about a fixed centre) $\phi$ that we use with our basis functions (forming bases of which network outputs are linear combinations) $\phi_n$ are the Gaussian\footnote{For the Gaussian, the width parameter $\sigma_n$ is fixed in advance along with the centre $\boldsymbol{\mu}_n$.}
        \begin{equation}
        \phi(x) = e^{-({x^2}/{2\sigma^2})}
        \end{equation}
        and the thin-plate spline function\footnote{The spline function $\phi(x) = x^4 \ln(x)$ was also tested, but it produced results less accurate than those of the thin-plate spline function in all tests.}
        \begin{equation}
        \phi(x) = x^2 \ln(x) \text{.}
        \end{equation}
        Note that hyperspherical (radially symmetric) Gaussians are used for simplicity; however, these Gaussians can be generalised to better fit the training data with hyperelliptical basis functions \cite[p. 35]{cB95}. Hyperelliptical basis functions are particularly useful in the presence of irrelevant input variables \cite[p. 184]{cB95} to overcome the curse of dimensionality.
        \section{RBFN Training}\label{S:RBFNtrain}
        Bishop (1995) notes, ``there are many potential applications for neural networks where unlabelled input data [without targets] is plentiful, but where labelled data is in short supply'' \cite[p. 183]{cB95}. This is precisely the case with the redshift estimation problem: photometric $ugriz$ data are readily available and easily acquired, but proper $z_{spec}$ measurements are few and will continue to lag behind by orders of magnitude \cite{nB07b}. While there are many ways to train an RBFN, the following two-stage training procedure is well suited to dealing with the stated problem. The unlabelled $ugriz$ data can be used for unsupervised learning as described below, while the smaller amount of $z_{spec}$-labelled data can be used in the second stage of training. Contrast this to the training procedure of an ANN, then, which will only be able to utilise the smaller, labelled subset of data.
            \subsection{Unsupervised Parameter Optimisation}
            In an RBFN, basis function parameters and second-layer weights could be optimised like those of an ANN with an iterative, supervised training algorithm; however, implementations (like ours) that sacrifice optimal accuracy in favour of computational efficiency will make use of unsupervised learning techniques for part of the RBFN training process.

            The two layers in our RBFNs are trained separately; as our networks have fewer basis functions than training data points, the centres $\boldsymbol{\mu}_n$ and widths $\sigma_n$ of the basis functions $\phi_n$ in the first layer are determined in advance with an algorithm that is blind to the target data of the training set. This means that such an algorithm needs only the training data inputs to fix the first-layer parameters; centres and widths are calculated to represent the estimated mixture density \cite[p. 59]{cB95} of the input vectors.
            \subsection{Linear Determination of Weights and Biases}
            The second-layer weights in our RBFNs are then calculated by a supervised method, but they are not iteratively optimised as in an ANN. Instead, the output values of the training data are used, along with the newly fixed centres and widths of the basis functions, to directly compute the second-layer weights and biases with a set of linear equations \cite[p. 171]{cB95}. As this is a linear problem, it is considerably less computationally intensive than the non-linear optimisation problems in ANN training, for example.
        \section{Implementation of RBFNs in Netlab}
        Our implementation of RBFNs is in Netlab, the MATLAB\textsuperscript{\textregistered} toolbox. As mentioned in \S\ref{S:RBFs}, we use a hyperspherical Gaussian and the thin-plate spline function as our hidden-layer activation functions. Our output is linear, and training is done in two stages as described in \S\ref{S:RBFNtrain}.

        The first training stage uses the `expectation-maximisation' (EM) algorithm to centre the basis functions. Generally speaking, the EM algorithm estimates the $N$ basis function centres as though the input data were composed of $N$ hyperspherical Gaussian distributions mixed amongst each other in the input space \cite[pp. 435ff.]{cB06}. The widths of the Gaussians are then set to some appropriate value; we use the square of the maximum inter-centre Euclidean distance. In the second training stage, the second-layer weights and biases are quickly calculated from the first-layer parameters by minimising sum-of-squared error over the training examples, following a simple matrix algebraic result \cite[p. 92]{cB95}.

        It happens that the EM algorithm is initialised with randomised, normally distributed parameters, and so its outputs, the basis function parameters, are thereby subject to the same normal variance as ANNs in Chapter \ref{C:ANNs}, \S\ref{SS:committees}. Additionally, the linear calculation of second-layer weights and biases is deterministic, so the weights---and therefore the $z_{phot}$ outputs---of the RBFN are normally distributed along with the first-layer parameters. Accordingly, we can use committees of RBFNs in the same way that we use committees of ANNs, taking the mean of the committee members' outputs as our final $z_{phot}$ value.

        \section{Results and Analysis}\label{S:RBFNresults}
\begin{figure}
        \centering\includegraphics[width=126mm]{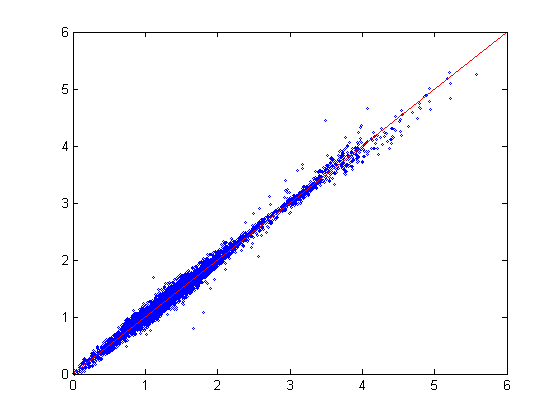}
        \caption{$z_{phot}$ vs. $z_{phot}$ for two individual 5:100:1 TPS RBFNs over $ugriz$.}\label{F:tps100UGRI}
        \end{figure}

        As in \S\ref{SS:MLPsResults}, we present minimum/maximum/mean/committee RMS results. All RBFNs were trained with 100 iterations of the EM algorithm for centring the basis functions. Again, as in \S\ref{SS:MLPsResults}, our committees regularly achieve lower RMS errors than the expected individual network RMS, seen in Tables \ref{T:RBFNg1}--\ref{T:RBFNt4}. Figure \ref{F:tps100UGRI} illustrates the expected Gaussian deviation between committee members, as discussed in \S\ref{SS:committees}.

        The difference in prediction accuracy between RBFNs with Gaussian activations and those with thin-plate-spline (TPS) is often considerable. Although neither Gaussian nor TPS activations are demonstrably superior for any given network architecture, some trends are observed when considering individual datasets.

        \begin{table}
            \begin{tabular}{ | l | c | c | c | c | }
            \hline
            Dataset    & RMS\textsubscript{min}  & RMS\textsubscript{max} & $\mu_\text{RMS}$ & RMS\textsubscript{com} \\ \hline
            $ugriz$ & 0.5717 & 0.5790 & 0.5750 & \emph{0.5720} \\ \hline
            $C_{ug},C_{gr},C_{ri},C_{iz}$ & 0.5526 & 0.5562 & 0.5548 & \emph{0.5533} \\ \hline
            $ugriz,C_{ug},C_{gr},C_{ri},C_{iz}$ & 0.5709 & 0.5776 & 0.5757 & \emph{0.5731} \\ \hline
            $ugriz_{6192}$ & 0.8165 & 0.8169 & 0.8167 & \emph{0.8166} \\ \hline
            $ugrizJHK$ & 0.5081 & 0.5116 & 0.5099 & \emph{0.5085} \\ \hline
            $C_{ug},C_{gr},C_{ri},C_{iz},C_{zJ},C_{JH},C_{HK}$ & 0.5121 & 0.5221 & 0.5169 & \emph{0.5142} \\ \hline
            \end{tabular}
            \\[5pt]\caption[Gaussian RBFN results with 5-committees of x:10:1 networks.]{RBFN results with 5-committees of x:10:1 networks, using the Gaussian activation function with 100 iterations of the EM algorithm for basis centring. Committee results \emph{better than the mean} are italicised, those \textbf{better than the best individual result} (RMS\textsubscript{min}) are bolded.}\label{T:RBFNg1}
        \end{table}

        \begin{table}
            \begin{tabular}{ | l | c | c | c | c | }
            \hline
            Dataset    & RMS\textsubscript{min}  & RMS\textsubscript{max} & $\mu_\text{RMS}$ & RMS\textsubscript{com} \\ \hline
            $ugriz$ & 0.6011 & 0.6213 & 0.6120 & \emph{0.6077} \\ \hline
            $C_{ug},C_{gr},C_{ri},C_{iz}$ & 0.4814 & 0.4973 & 0.4860 & \textbf{0.4809} \\ \hline
            $ugriz,C_{ug},C_{gr},C_{ri},C_{iz}$ & 0.5884 & 0.5942 & 0.5918 & \emph{0.5896} \\ \hline
            $ugriz_{6192}$ & 0.8164 & 0.8175 & 0.8168 & \emph{0.8167} \\ \hline
            $ugrizJHK$ & 0.5217 & 0.5356 & 0.5285 & \emph{0.5256} \\ \hline
            $C_{ug},C_{gr},C_{ri},C_{iz},C_{zJ},C_{JH},C_{HK}$ & 0.5745 & 0.5932 & 0.5857 & \textbf{0.5740} \\ \hline
            \end{tabular}
            \\[5pt]\caption[TPS RBFN results with 5-committees of x:10:1 networks.]{RBFN results with 5-committees of x:10:1 networks, using the thin-plate-spline activation function.}\label{T:RBFNt1}
        \end{table}

        \begin{table}
            \begin{tabular}{ | l | c | c | c | c | }
            \hline
            Dataset    & RMS\textsubscript{min}  & RMS\textsubscript{max} & $\mu_\text{RMS}$ & RMS\textsubscript{com} \\ \hline
            $ugriz$ & 0.5696 & 0.5784 & 0.5737 & \emph{0.5712} \\ \hline
            $C_{ug},C_{gr},C_{ri},C_{iz}$ & 0.5527 & 0.5569 & 0.5551 & \textbf{0.5524} \\ \hline
            $ugriz,C_{ug},C_{gr},C_{ri},C_{iz}$ & 0.5660 & 0.5765 & 0.5703 & \emph{0.5684} \\ \hline
            $ugriz_{6192}$ & 0.8164 & 0.8170 & 0.8167 & \emph{0.8165} \\ \hline
            $ugrizJHK$ & 0.5080 & 0.5092 & 0.5086 & \textbf{0.5079} \\ \hline
            $C_{ug},C_{gr},C_{ri},C_{iz},C_{zJ},C_{JH},C_{HK}$ & 0.5046 & 0.5234 & 0.5134 & \emph{0.5091} \\ \hline
            \end{tabular}
            \\[5pt]\caption{Gaussian RBFN results with 5-committees of x:20:1 networks.}\label{T:RBFNg2}
        \end{table}

        \begin{table}
            \begin{tabular}{ | l | c | c | c | c | }
            \hline
            Dataset    & RMS\textsubscript{min}  & RMS\textsubscript{max} & $\mu_\text{RMS}$ & RMS\textsubscript{com} \\ \hline
            $ugriz$ & 0.6052 & 0.6200 & 0.6134 & \emph{0.6105} \\ \hline
            $C_{ug},C_{gr},C_{ri},C_{iz}$ & 0.4821 & 0.4984 & 0.4859 & \textbf{0.4812} \\ \hline
            $ugriz,C_{ug},C_{gr},C_{ri},C_{iz}$ & 0.5876 & 0.5979 & 0.5903 & \emph{0.5879} \\ \hline
            $ugriz_{6192}$ & 0.8157 & 0.8173 & 0.8164 & \emph{0.8163} \\ \hline
            $ugrizJHK$ & 0.5218 & 0.5312 & 0.5266 & \emph{0.5248} \\ \hline
            $C_{ug},C_{gr},C_{ri},C_{iz},C_{zJ},C_{JH},C_{HK}$ & 0.5244 & 0.6113 & 0.5724 & \emph{0.5591} \\ \hline
            \end{tabular}
            \\[5pt]\caption{TPS RBFN results with 5-committees of x:20:1 networks.}\label{T:RBFNt2}
        \end{table}

        \begin{table}
            \begin{tabular}{ | l | c | c | c | c | }
            \hline
            Dataset    & RMS\textsubscript{min}  & RMS\textsubscript{max} & $\mu_\text{RMS}$ & RMS\textsubscript{com} \\ \hline
            $ugriz$ & 0.5477 & 0.5555 & 0.5503 & \emph{0.5479} \\ \hline
            $C_{ug},C_{gr},C_{ri},C_{iz}$ & 0.5326 & 0.5508 & 0.5378 & \emph{0.5340} \\ \hline
            $ugriz,C_{ug},C_{gr},C_{ri},C_{iz}$ & 0.5367 & 0.5534 & 0.5458 & \emph{0.5427} \\ \hline
            $ugriz_{6192}$ & 0.8176 & 0.8223 & 0.8196 & \emph{0.8193} \\ \hline
            $ugrizJHK$ & 0.4811 & 0.4886 & 0.4854 & \emph{0.4832} \\ \hline
            $C_{ug},C_{gr},C_{ri},C_{iz},C_{zJ},C_{JH},C_{HK}$ & 0.5133 & 0.5473 & 0.5305 & \emph{0.5263} \\ \hline
            \end{tabular}
            \\[5pt]\caption{Gaussian RBFN results with 5-committees of x:40:1 networks.}\label{T:RBFNg3}
        \end{table}

        \begin{table}
            \begin{tabular}{ | l | c | c | c | c | }
            \hline
            Dataset    & RMS\textsubscript{min}  & RMS\textsubscript{max} & $\mu_\text{RMS}$ & RMS\textsubscript{com} \\ \hline
            $ugriz$ & 0.5734 & 0.5834 & 0.5797 & \emph{0.5758} \\ \hline
            $C_{ug},C_{gr},C_{ri},C_{iz}$ & 0.4568 & 0.4649 & 0.4605 & \textbf{0.4561} \\ \hline
            $ugriz,C_{ug},C_{gr},C_{ri},C_{iz}$ & 0.5580 & 0.5652 & 0.5621 & \textbf{0.5569} \\ \hline
            $ugriz_{6192}$ & 0.8177 & 0.8198 & 0.8187 & \emph{0.8183} \\ \hline
            $ugrizJHK$ & 0.5017 & 0.5114 & 0.5057 & \textbf{0.5015} \\ \hline
            $C_{ug},C_{gr},C_{ri},C_{iz},C_{zJ},C_{JH},C_{HK}$ & 0.4778 & 0.4946 & 0.4890 & \emph{0.4825} \\ \hline
            \end{tabular}
            \\[5pt]\caption{TPS RBFN results with 5-committees of x:40:1 networks.}\label{T:RBFNt3}
        \end{table}

        \begin{table}
            \begin{tabular}{ | l | c | c | c | c | }
            \hline
            Dataset    & RMS\textsubscript{min}  & RMS\textsubscript{max} & $\mu_\text{RMS}$ & RMS\textsubscript{com} \\ \hline
            $ugriz$ & 0.5220 & 0.5441 & 0.5295 & \textbf{0.5216} \\ \hline
            $C_{ug},C_{gr},C_{ri},C_{iz}$ & 0.5232 & 0.5650 & 0.5400 & \emph{0.5247} \\ \hline
            $ugriz,C_{ug},C_{gr},C_{ri},C_{iz}$ & 0.5138 & 0.5302 & 0.5208 & \emph{0.5186} \\ \hline
            $ugriz_{6192}$ & 0.8278 & 0.8539 & 0.8372 & \emph{0.8359} \\ \hline
            $ugrizJHK$ & 0.4551 & 0.4674 & 0.4624 & \emph{0.4555} \\ \hline
            $C_{ug},C_{gr},C_{ri},C_{iz},C_{zJ},C_{JH},C_{HK}$ & 0.9373 & 1.0595 & 0.9940 & \textbf{0.8141} \\ \hline
            \end{tabular}
            \\[5pt]\caption{Gaussian RBFN results with 5-committees of x:100:1 networks.}\label{T:RBFNg4}
        \end{table}

        \begin{table}
            \begin{tabular}{ | l | c | c | c | c | }
            \hline
            Dataset    & RMS\textsubscript{min}  & RMS\textsubscript{max} & $\mu_\text{RMS}$ & RMS\textsubscript{com} \\ \hline
            $ugriz$ & 0.5226 & 0.5317 & 0.5290 & \emph{0.5232} \\ \hline
            $C_{ug},C_{gr},C_{ri},C_{iz}$ & 0.4366 & 0.4385 & 0.4379 & \textbf{0.4352} \\ \hline
            $ugriz,C_{ug},C_{gr},C_{ri},C_{iz}$ & 0.4841 & 0.4959 & 0.4878 & \textbf{0.4834} \\ \hline
            $ugriz_{6192}$ & 0.8256 & 0.8276 & 0.8261 & 0.8441 \\ \hline
            $ugrizJHK$ & 0.4640 & 0.4713 & 0.4669 & \textbf{0.4605} \\ \hline
            $C_{ug},C_{gr},C_{ri},C_{iz},C_{zJ},C_{JH},C_{HK}$ & 0.4040 & 0.4181 & 0.4096 & \textbf{0.4014} \\ \hline
            \end{tabular}
            \\[5pt]\caption{TPS RBFN results with 5-committees of x:100:1 networks.}\label{T:RBFNt4}
        \end{table}

\begin{figure}
        \centering\includegraphics[width=126mm]{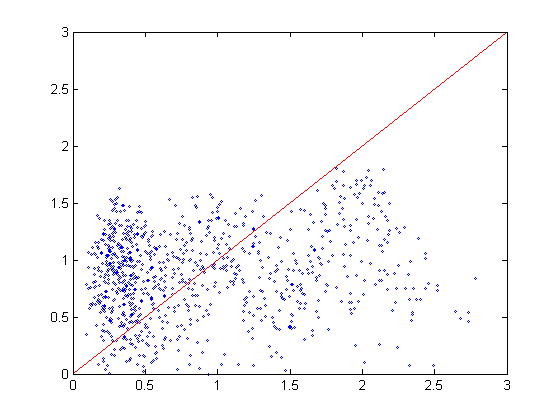}
        \caption[$z_{phot}$ vs. $z_{spec}$ for a 7:100:1 Gaussian RBFN over $\text{[}C_{ug},$ $C_{gr},$ $C_{ri},$ $C_{iz},$ $C_{zJ},$ $C_{JH},$ $C_{HK}\text{]}$.]{$z_{phot}$ vs. $z_{spec}$ for a 7:100:1 Gaussian RBFN over $[C_{ug},$ $C_{gr},$ $C_{ri},$ $C_{iz},$ $C_{zJ},$ $C_{JH},$ $C_{HK}]$. $\sigma_{\text{RMS}}$ is only 1.0595---not dissimilar to other networks in the same committee that converged but had large outliers.}\label{F:gauss100UGRIZJHfail}
        \end{figure}

\begin{figure}
        \centering\includegraphics[width=126mm]{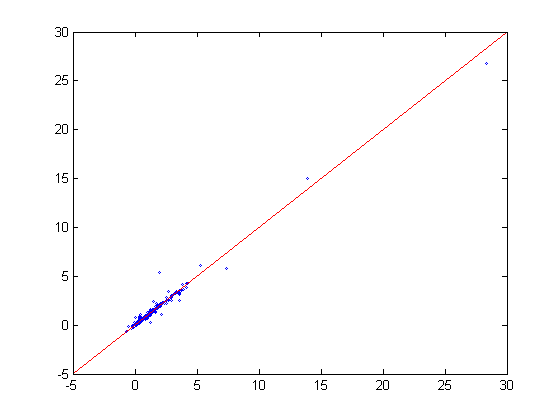}
        \caption[$z_{phot}$ vs. $z_{phot}$ for two individual 7:100:1 Gaussian RBFNs over $\text{[}C_{ug},$ $C_{gr},$ $C_{ri},$ $C_{iz},$ $C_{zJ},$ $C_{JH},$ $C_{HK}\text{]}$.]{$z_{phot}$ vs. $z_{phot}$ for two individual 7:100:1 Gaussian RBFNs over $[C_{ug},$ $C_{gr},$ $C_{ri},$ $C_{iz},$ $C_{zJ},$ $C_{JH},$ $C_{HK}]$. Note the large outliers consistently predicted by both RBFNs.}\label{F:gauss100UGRIZJH}
        \end{figure}

        Looking at committee $\sigma_{\text{RMS}}$ results, the RBFNs with Gaussian activations achieve consistently lower errors than those with TPS activations over the $ugriz$ and $ugrizJHK$ datasets. The opposite is true for the $[C_{ug},$ $C_{gr},$ $C_{ri},$ $C_{iz}]$ dataset, and comparable results are achieved with $[ugriz,$ $C_{ug},$ $C_{gr},$ $C_{ri},$ $C_{iz}]$. Curiously, over the $[C_{ug},$ $C_{gr},$ $C_{ri},$ $C_{iz},$ $C_{zJ},$ $C_{JH},$ $C_{HK}]$ dataset, Gaussians yield significantly higher accuracies in 7:10:1 and 7:20:1 networks, but TPS performs substantially better in 7:40:1 and 7:100:1 networks. In fact, the Gaussian RBFNs in the 7:100:1 networks produce outliers of sufficient magnitude to raise $\mu_\text{RMS}$ to 0.9940, and one network failed to learn the data at all (Figure \ref{F:gauss100UGRIZJHfail}). The 5-committee for these networks, however, is seen to reduce RMS\textsubscript{com} to 0.8141. (A 5-committee calculating the median instead of the mean over these networks' outputs yielded an RMS\textsubscript{com} of 0.8263; some outliers---even $z_{phot} \sim 30$---were consistently produced by committee members, as seen in Figure \ref{F:gauss100UGRIZJH}.)

\begin{figure}
        \centering\includegraphics[width=126mm]{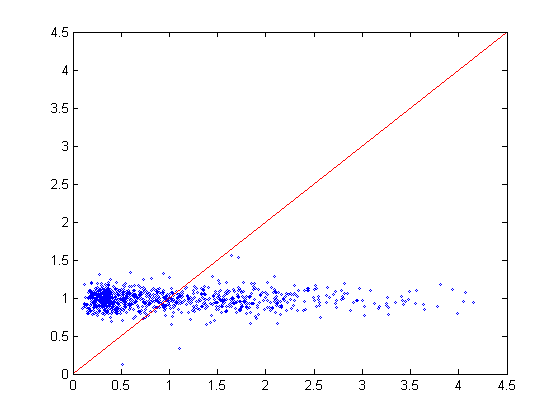}
        \caption[$z_{phot}$ vs. $z_{spec}$ for the 5-committee of 5:100:1 TPS RBFNs over $ugriz_{6192}$.]{$z_{phot}$ vs. $z_{spec}$ for the 5-committee of 5:100:1 TPS RBFNs over $ugriz_{6192}$. $\sigma_{\text{RMS}}$ is 0.8441.}\label{F:tps100ugriz6192}
        \end{figure}

        All 80 RBFNs produced failed to learn the $ugriz_{6192}$ dataset with its original training/test set distribution, which was left as-is. The TPS RBFNs with 100 hidden units (see Figure \ref{F:tps100ugriz6192}) showed the most deviation from outputting only $z_{phot} = 1$, and significantly more deviation than MLPs showed in Figure \ref{F:mlpugriz6192}.

\begin{figure}
        \centering\includegraphics[width=126mm]{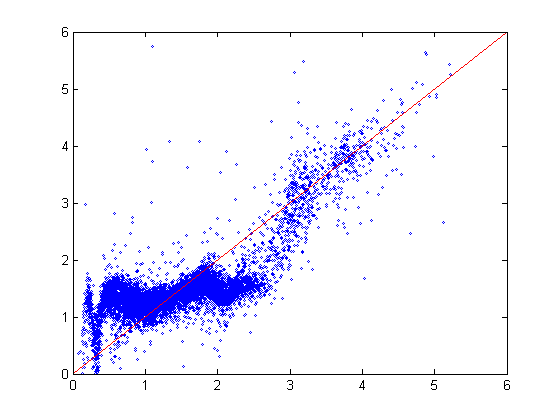}
        \caption[$z_{phot}$ vs. $z_{spec}$ for the 5-committee of 5:100:1 Gaussian RBFNs over $ugriz$.]{$z_{phot}$ vs. $z_{spec}$ for the 5-committee of 5:100:1 Gaussian RBFNs over $ugriz$. $\sigma_{\text{RMS}}$ is 0.5216.}\label{F:gauss100ugriz}
        \end{figure}

\begin{figure}
        \centering\includegraphics[width=126mm]{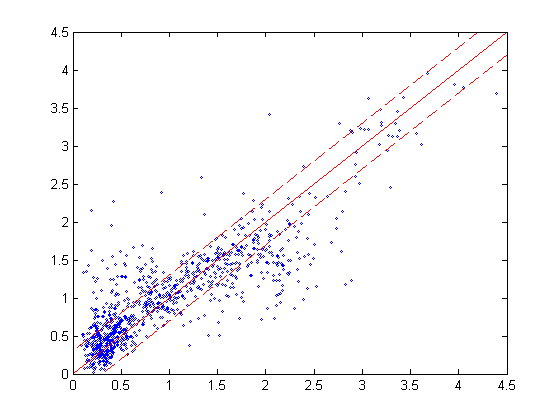}
        \caption[$z_{phot}$ vs. $z_{spec}$ for the 5-committee of 7:100:1 TPS RBFNs over $\text{[}C_{ug},$ $C_{gr},$ $C_{ri},$ $C_{iz},$ $C_{zJ},$ $C_{JH},$ $C_{HK}\text{]}$.]{$z_{phot}$ vs. $z_{spec}$ for the 5-committee of 7:100:1 TPS RBFNs over $[C_{ug},$ $C_{gr},$ $C_{ri},$ $C_{iz},$ $C_{zJ},$ $C_{JH},$ $C_{HK}]$. $\sigma_{\text{RMS}}$ is 0.4014, and 69.5\% of objects are predicted to $\Delta z \leq 0.3$.}\label{F:tps100UGRIZJH}
        \end{figure}

        Finally, Figures \ref{F:gauss100ugriz} and \ref{F:tps100UGRIZJH} illustrate two RBFNs that, while not as accurate as the ANNs of Figures \ref{F:annzugriz1} and \ref{F:annzugrizjhk}, are highly similar in form. This indicates that they suffer from the similar difficulties in learning and generalising on the $ugrizJHK$ data; it also recalls their rough equivalence in representational capability.

        The general symmetry for mild outliers at $\Delta z \sim [1,2]$ about the line $z_{phot} = z_{spec}$ in Figures \ref{F:tps100UGRIZJH} and \ref{F:annzugrizjhk} brings more clearly to light the degeneracy suggested by Richards et al. (2001b) \cite{gR01b} and Weinstein et al. (2004) \cite{mW04}: certain quasars are close matches for two or more redshifts, and the $z_{phot}$-$z_{spec}$ relation illustrates this confusion by plotting them symmetrically about $z_{phot} = z_{spec}$. \S\ref{S:CZR} discusses this degeneracy in detail.

    \chapter{Discussion}\label{C:P2Discussion}
        \section{Systematic Errors}
        In presenting a survey of network-based learning methods for photometric redshift estimation, we have done very little to minimise RMS errors by way of parameter optimisation. Still, we have achieved $\sigma_{\text{RMS}}$ values of 0.3557 and 0.4014 with ANNs and RBFNs, yielding $\Delta z$ values within 0.3 for 76.2\% and 69.5\% of quasars, respectively. These results are comparable to recent standard results in the literature (\cite{sK04}, \cite{mW04}, \cite{tB04}, \cite{xW03}), but they do not compare to the latest results of Ball et al. (to appear), who are able to select a subset of quasars whose redshifts may be estimated with significantly greater accuracy (improving $\sigma_{\text{RMS}}$ from 0.343 to 0.117, and the percentage of quasars within $\Delta z < 0.3$ from 79.8\% to 99.3\%.) \cite{nB07b}.\footnote{Ball et al. (to appear) use statistical techniques that amount to excluding quasars that are near CZR degeneracies in colour-colour space. They exclude 61.1\% of their dataset, but, as this is not even an order-of-magnitude loss and they have managed to remove nearly all catastrophics, their result is exceptional.}

        All published results suffer from the same systematic errors at certain values of $z_{spec}$, however. In fact, with hidden layers of up to 100 units, we were unable even to memorise our training sets after 3,000 training iterations: testing over our training sets yielded the same deviations and regions of catastrophic failure in our $z_{phot}$-$z_{spec}$ relation. This indicates some degeneracy inherent to the training sets; i.e., some sets of photometric magnitudes and colours correspond to multiple spectroscopic redshifts.
        \section{Colour-Redshift Relation and Spectral Lines}\label{S:CZR}
\begin{figure}
        \centering\includegraphics[width=126mm]{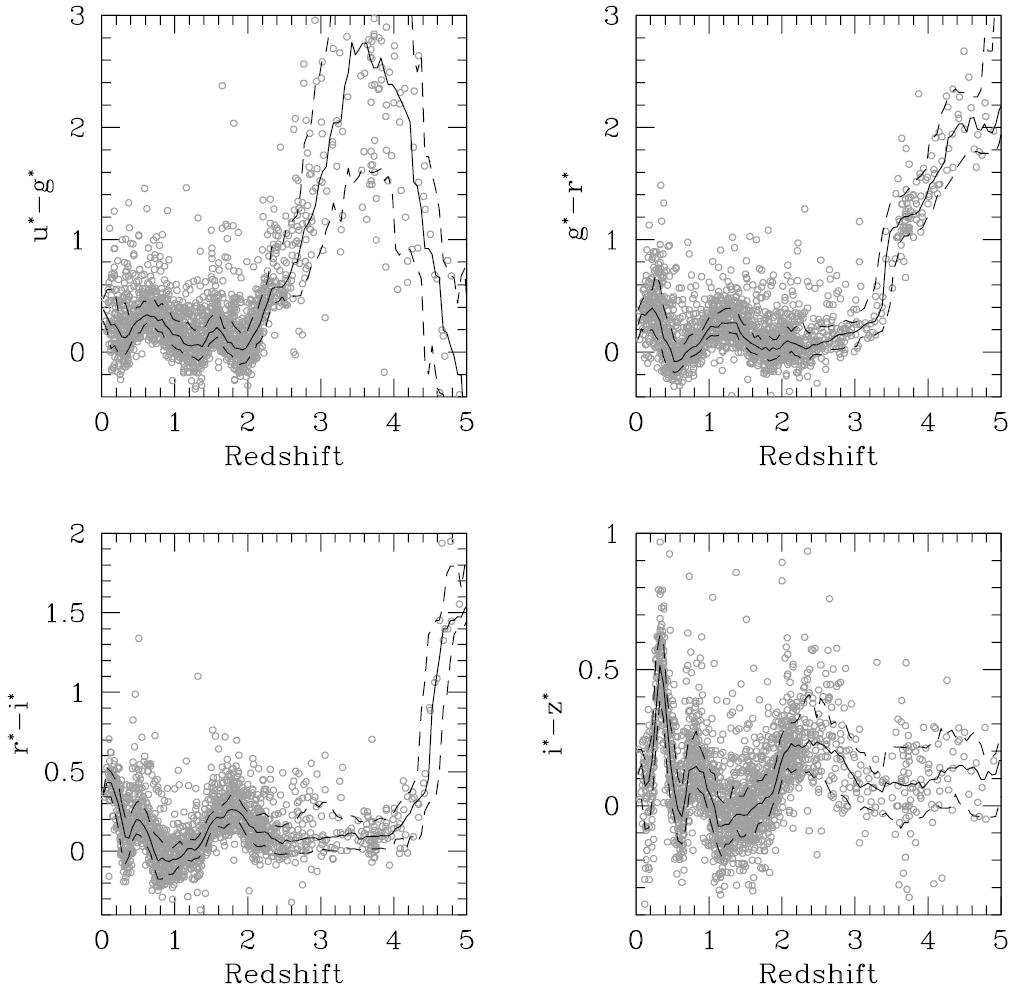}
        \caption[Colour-redshift plots of SDSS quasars indicating median colour as a function of redshift and one-sigma errors.]{Colour-redshift plots of SDSS quasars indicating median colour as a function of redshift (solid curve) and one-sigma errors (dashed curves). The relation is degenerate in colour-redshift space if it is not one-to-one. From Richards et al. (2001b) \cite{gR01b}, adapted from \cite{gR01a}.}\label{F:gR01b}
        \end{figure}

\begin{figure}
        \centering\includegraphics[width=126mm]{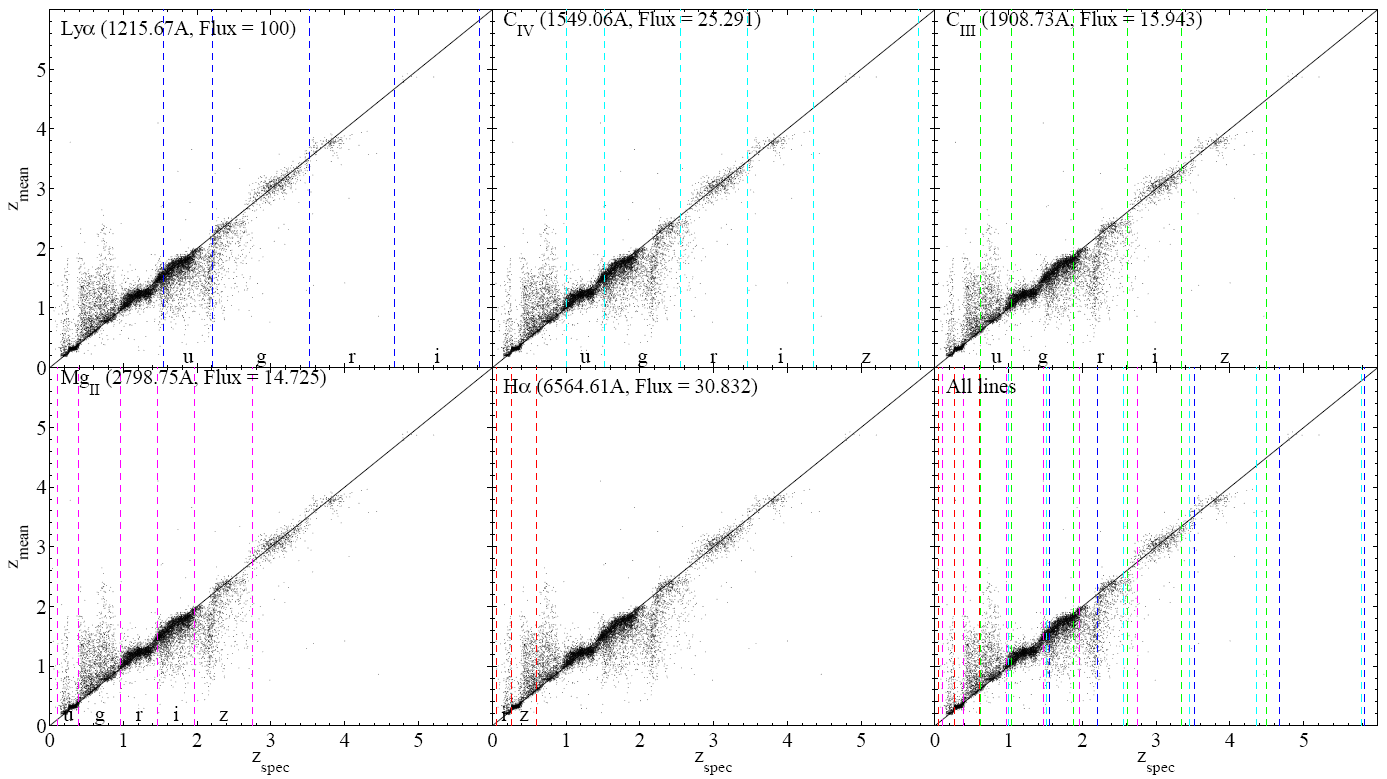}
        \caption{$z_{phot}$-$z_{spec}$ plots indicating where the five brightest emission lines cross $ugriz$ filters. From Ball et al. (to appear) \cite{nB07b}.}\label{F:nB07b}
        \end{figure}

        As discussed in \S\ref{S:PRE}, the colour-redshift relation (CZR) is of primary importance in estimating redshift from photometric measurements. Figure \ref{F:gR01b} \cite{gR01b} plots the CZR for the colours $C_{ug}$, $C_{gr}$, $C_{ri}$, and $C_{iz}$. Richards et al. (2001a) describe in detail the features of the CZR, and find that almost all of the structure can be attributed to emission lines moving in and out of the $ugriz$ filter ranges at different redshifts \cite{gR01a}. This is consistent with the observations of Ball et al. (to appear) who demonstrate that major deviations in the $z_{phot}$-$z_{spec}$ relation fall at redshifts where important emission lines cross SDSS filter boundaries (Figure \ref{F:nB07b}) \cite{nB07b}.\footnote{Cf. also \cite{tB04} for redshift values at which important emission lines cross $ugriz$ transitions.}

\begin{figure}
        \centering\includegraphics[width=126mm]{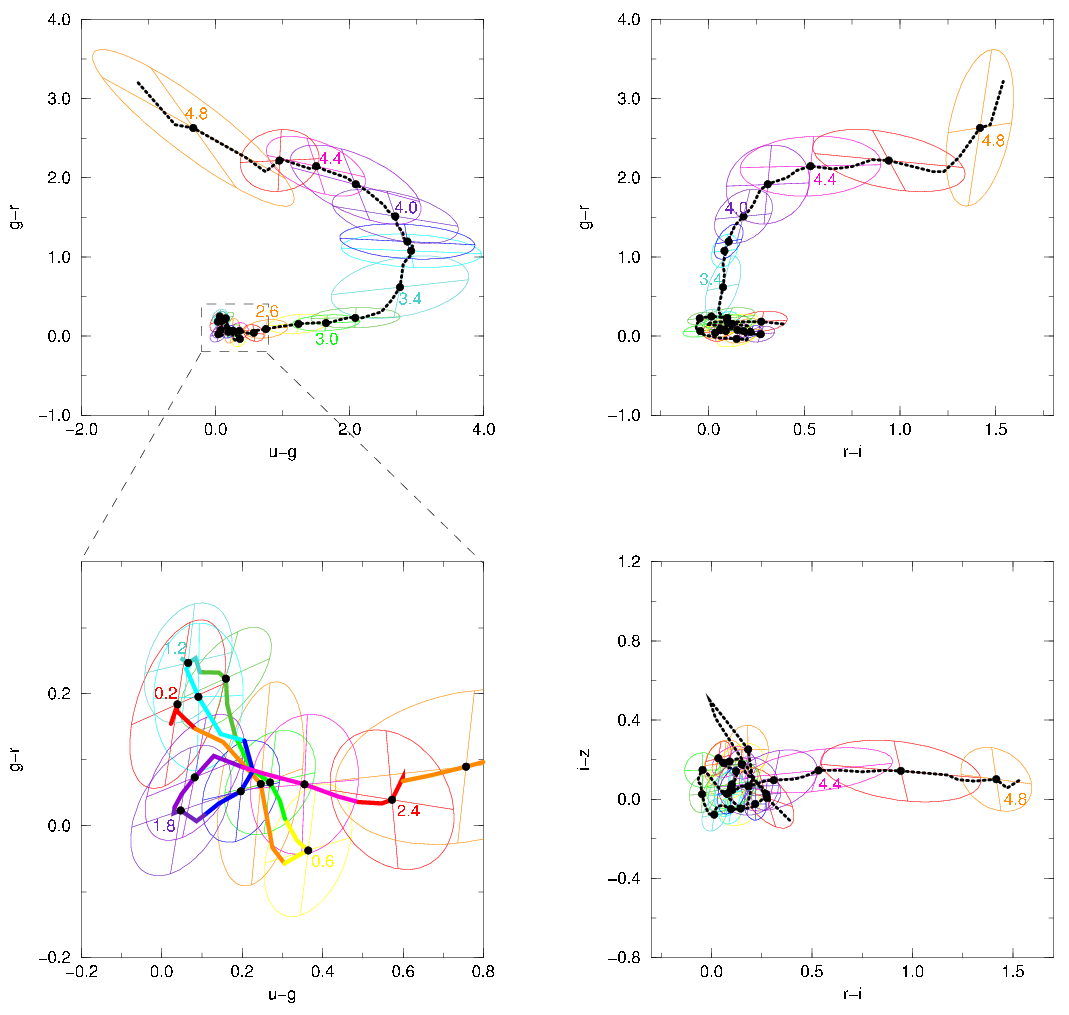}
        \caption[Plots of the CZR in colour-colour space, with one-sigma error ellipsoids for certain (labelled) redshifts. From Weinstein et al. (2004) \cite{mW04}.]{Plots of the CZR in colour-colour space, with one-sigma error ellipsoids for certain (labelled) redshifts. The CZR track is traced by the curves. From Weinstein et al. (2004) \cite{mW04}.}\label{F:mW04}
        \end{figure}

        Further, as illustrated in Figures \ref{F:gR01b} and \ref{F:mW04}, there is significant degeneracy in the CZR, indicating that it is impossible to predict redshift for all quasars as a function of four colours alone. Note that higher-quality photometry will not solve the problem: while some level of photometric certainty is needed to approach the CZR, recall from \S\ref{SS:jacobian} that we were able to deviate our inputs to an ANN by the estimated photometric noise to find the resulting change $\sigma_y$ in $z_{phot}$. As seen in Figure \ref{F:annzugriz2}, photometric noise is not responsible for the majority of the error $\Delta z$. Therefore, as Way \& Srivastava (2006) suggest for empirical reasons \cite{mW06}, photometry improvements alone will not overcome the CZR degeneracy.

\begin{figure}
        \centering\includegraphics[width=118mm]{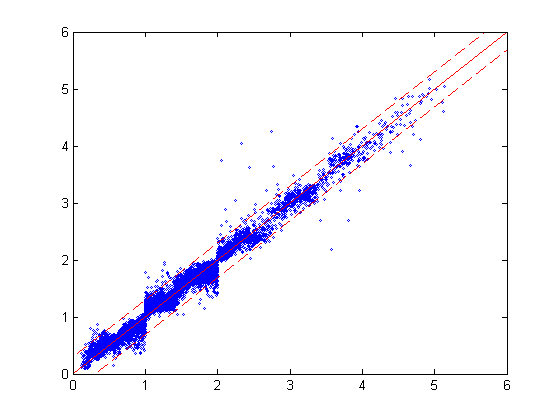}
        \caption[$z_{phot}$ vs. $z_{spec}$ for three separate 5-committees of 4:10:1 MLPs in Netlab, trained on datasets segregated at $z = 1$ and $z = 2$.]{$z_{phot}$ vs. $z_{spec}$ for three separate 5-committees of 4:10:1 MLPs in Netlab, trained on datasets segregated at $z = 1$ and $z = 2$ up to 500 iterations. $\sigma_{\text{RMS}}$ is 0.1283, and 97.4\% of objects fall within $\Delta z \leq 0.3$.}\label{F:zsplit}
        \end{figure}

\begin{figure}
        \centering\includegraphics[width=118mm]{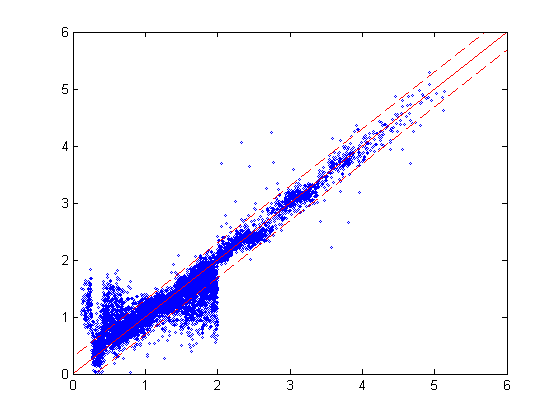}
        \caption[$z_{phot}$ vs. $z_{spec}$ for two 5-committees of 4:10:1 MLPs, segregated at $z = 2$.]{$z_{phot}$ vs. $z_{spec}$ for two 5-committees of 4:10:1 MLPs, segregated at $z = 2$. $\sigma_{\text{RMS}}$ is 0.2857, and 95.8\% have $\Delta z \leq 0.3$.}\label{F:zsplit2}
        \end{figure}

        In order to break the degeneracy, we need additional input parameters. A demonstration of the improvement in accuracy were the degeneracy to be broken is shown in Figure \ref{F:zsplit}. If we had some way of segregating $z < 1$ quasars, $1 < z < 2$ quasars, and $z > 2$ quasars from each other (say, by using non-photometric data or more specialised filters for detecting specific spectral lines as suggested in \cite{nB07b}\footnote{Cf. also Hatziminaoglou et al. (2000) \cite{eH00}, who point out the deficiencies of wide-band filters for quasars.}), a neural network training on this additional information might yield results around $\sigma_{\text{RMS}} = 0.1283$, with 97.4\% of objects within $\Delta z \leq 0.3$. Segregation at $z = 2$ showed most of the usual systematic deviations (Figure \ref{F:zsplit2}), confirming that the major source of errors is a degeneracy in $0 < z < 2$.

        \section{Other Learning Techniques for Photometric Redshift Estimation}
        Other potential improvements to our methods could include the use of cross-variance to avoid overfitting a training set, or weighting in committees of networks \cite[p. 366]{cB95}. Photometric redshift estimation has also been attempted using support vector machines (SVMs) (\cite{yW04} and \cite{dW08}), GAs \cite{nM06}, and random forests \cite{sC07}. We tested SVMs on our datasets using the SVMTorch II facility \cite{rC01},\footnote{SVMTorch II is available at \texttt{http://www.idiap.ch/machine\_learning.php?} \texttt{content=Torch/en\_SVMTorch.txt}. We used a Gaussian kernel with a width of 1.0 and error pipes of 0.01 and 0.001.} but found higher errors in all tests. Still, SVMs are attractive in that they can add extra input parameters with near-linear computational overhead \cite{yW04} and limited dilution of accuracy in the presence of irrelevant data with some SVM algorithms \cite[p. 385]{tH01}.

        Computational considerations should also be taken into account: Hastie et al. (2001) show that, for $N$ training data, $p$ input variables, $W$ weights, $L$ training iterations, $M$ basis functions, and $m$ support vectors, ANNs require $O(NpML)$ operations \cite[p. 367]{tH01}, RBFNs require $O(NM^2 + M^3)$ \cite[p. 190]{tH01}, and SVMs tend to require $O(m^3 + mN + mpN)$ \cite[p. 405]{tH01}.
\part*{Conclusion}
    \chapter{Implications of Research and Future Directions}\label{C:Implications}
    Part I of this dissertation demonstrated that canonical machine learning techniques could be adapted to solve a non-trivial astrophysical problem with only minor modifications. The same techniques could be used in a reverse manner, to approximate the mechanical formulae governing the behaviour of an observed system with known physical parameters. Alternatively, the problem solved could be generalised by using more precise, relativistic physical calculations to extend to the modelling of binary star systems and the search for other planetary star systems.

    Another interesting extension would be to model simultaneously the orbits of several moons with GAs and PSO: in principle, not all moons would be visible within a certain window or from a certain point of view (they would occasionally be obscured by or pass in front of the major planet), and the moons would not be `labelled' individually in each frame.\footnote{This would be similar to using a set of observations like Galileo's (Figure \ref{F:sidnunc}) to automatically determine the total number of moons and the orbital elements of each.} Still, a GA could begin to model the system by using a variable number of satellites (represented in each organism) equal to or greater than the maximum number of moons observed in any one frame, with the satellites ordered within the organism by their semi-major axis distance $a$. Crossover between two organisms with differing numbers of satellites could interact only those satellites whose $a$ values are most similar. Mutation could also introduce or remove new satellites to an organism, and fitness could be measured at each time step by summing the minimum Euclidean distances between estimated moon positions and observed locations, using the location of the major planet as a moon location if there are more satellites represented in the organism than observed in the frame. This would allow the inclusion of frames in which some moons are obscured, and would also penalise organisms that represented more satellites than were present in the physical system. For PSO, multiple distinct swarms could be run simultaneously on the data, with some use of hyperspheres or hypercubes (an enforced minimum distance) in parameter space to prevent the swarms from converging on the same set of orbital elements. Such an extension would require significantly greater computational resource,\footnote{Indeed, the primary reason we have not tested multiple-satellite modelling is that using a \texttt{vector} (a variable-length array) of \texttt{vectors} caused our GA implementation to run remarkably slowly, especially considering the expected increase in computational time that modelling multiple satellites is likely to require.} but this only recalls the growing need for more parallelisable and distributed algorithms for common problems in astrophysical research.

    Part II demonstrated that trivial applications of machine learning techniques can yield results comparable to the most advanced applications of traditional techniques in photometric redshift estimation. As mentioned in earlier chapters, the improvement of these learning techniques---for quasars in particular---will greatly assist research in cosmology \cite{gR01b} and the large-scale structure of the universe \cite{dY00}. It would be useful to determine the relationship between photometric redshift accuracy and quasar magnitude, as the efficient estimation of faint quasar redshifts would assist in studying quasar evolution \cite{gR01b}. It would also be interesting to investigate in more detail the statistical properties of the $z_{phot}$-$z_{spec}$ relation and the CZR degeneracy. The impressive result of Ball et al. (to appear) \cite{nB07b} in selecting quasars according to properties of their $z_{phot}$ probability density functions indicates that there may well be more statistical learning techniques of use in solving the CZR problem.

    Improvements may also be made by calculating the Jacobian matrix for RBFNs as in \S\ref{SS:jacobian} and comparing its $z_{phot}$ deviations $\sigma_y$ with those obtained by ANN\emph{z}, investigating the relationship between computational expense and $\Delta z$ error as regards partially unsupervised RBFNs and (fully supervised) ANNs, incorporating X-ray and radio flux measurements, and experimenting with different activation functions and/or describing their distinct behaviours. However, aside from reducing $\Delta z$ error and ensuring predictable confidence intervals, as not all science applications require minimal redshift errors \cite{gR01b}, serious consideration should also be given to the unsupervised learning properties of RBFNs insofar as they may be of particular advantage in large sky survey science in the future.
\appendix
    \chapter{Source Code: Modelling with GAs and PSO}\label{A:src}
    \renewcommand{\baselinestretch}{1}
        \section{mlastro.cpp}
\small
\begin{verbatim}
// mlastro.cpp

#include "stdafx.h"
#include <stdlib.h>
#include <iostream>
#include <fstream>
#include <string>
#define _USE_MATH_DEFINES
#include <math.h>
#include <vector>
#include <time.h>

using namespace std;

#include "organism.h"
#include "particle.h"

const char* inputXYZ = "X:\\data\\2hAtlasXYZ.txt";
const double EPOCH = 2453006.0;
//2451545 for J2000.0 (Himalia), 2450464.25 for 1997 Jan 16.00 TT (Io),
//2454600 for test, 2453006 for Jan 1 04 (Atlas)
const bool LOG = true; //to turn on or off logging of stdev for convergence
const bool test = false; //if using test set
const bool taper = false; //use tapered vMAX for PSO

class observ {
public:
	double time;
	vector<double> x;
	vector<double> y;
	vector<double> satpang;
	vector<char> visibility;
};

pair<double,double> keplerian(organism o, int sat, double timeInDays) {
	pair<double,double> result;
	result.first=0;
	result.second=0;


	double timeElapsed = timeInDays - EPOCH; //in days from EPOCH
	timeElapsed *= 86400; //now in seconds

	double M; //mean anomaly
	double n; //mean motion, in radians/sec: this is an avg, for circular orbits
	
	//unable to represent o.a[sat]^3:
	n = (double)o.mu;
	n /= o.a[sat];
	n /= o.a[sat];
	n /= o.a[sat];
	n = sqrt(n);
	M = o.m0[sat] + n*timeElapsed;
	//radians + (radians/s)*s, from http://en.wikipedia.org/wiki/Mean_anomaly

	double E; //eccentric anomaly
	E = M;
	for (int i=0;i<8;i++) {
		E = M + o.e[sat]*sin(E);
		//from http://en.wikipedia.org/wiki/Eccentric_anomaly
	}

	double nu; //true anomaly, between -PI and PI
	nu = atan(sqrt((1.+o.e[sat])/(1.-o.e[sat]))*tan(E/2.))*2;
	if (nu<0) nu += 2*M_PI;
	//putting nu between 0 and 2*PI, for theta/phi calculations in keplerian()

	double rho; //distance from major body, in same units as a (metres)
	rho = o.a[sat]*((1.-pow(o.e[sat],2))/(1.+(o.e[sat]*cos(nu))));

	result.first = rho/1000.; //this puts it in km for simplicity's sake
	result.second = nu;

	return result;
}

pair<pair<double,double>,double>
  polarToXYZ(organism o,int sat,pair<double,double> polar){
	pair<double,double> xy;
	pair<pair<double,double>,double> result;
	double x = 0,y = 0,z = 0;
	
	double rho, phi, theta, nu;
	rho = polar.first;
	nu = polar.second;

	double scaleTheta,scalePhi;
	scaleTheta = cos(o.inclin[sat]);
	scalePhi = sin(o.inclin[sat]);

	theta = scaleTheta*(o.w[sat]+nu)+o.node[sat]; //should be 0 to 2*PI
	phi = (M_PI/2)-scalePhi*(o.w[sat]+nu); //should be 0 to 2*PI

	//convert to xyz: in same units as rho (metres, now km)
	x = rho*sin(phi)*cos(theta);
	y = rho*sin(phi)*sin(theta);
	z = rho*cos(phi);


	xy.first = x;
	xy.second = y;
	result.first = xy;
	result.second = z;

	return result;

}

void outputOrganisms(vector<organism>& o) {
	ofstream orgs;
	orgs.open("X:\\data\\organisms.txt");
	char timeStr[9];
	_strtime_s(timeStr);
	orgs << timeStr << endl << endl;
	for (unsigned int i=0;i<o.size();i++) {
		orgs << i+1 << "." << o[i].satellites << endl;
		orgs << o[i].fitness << endl;
		orgs << o[i].a[0] << "," << o[i].e[0] << "," << o[i].inclin[0] << endl;
		orgs << o[i].node[0] << "," << o[i].w[0] << "," << o[i].m0[0];
        orgs << endl << endl;
	}
	orgs.close();
}

double rankFitnessXYZ(vector<organism>& o, const int observations) {
	vector<pair<double,double>> X(observations),Y(observations),Z(observations);
	ifstream input;
	ofstream best;
	ofstream worst;
	string temp;

	input.open(inputXYZ);
	while (!input.eof()) {
		getline(input,temp);
		if (temp.compare("$$SOE")==0) {
			break;
		}
	}
	//file should be at EOF or at $$SOE here
	int count = 0;
	while ((!input.eof())&&(input.peek()!='$')) {
		input >> X[count].first;
		if (!test) getline(input,temp);
		input >> X[count].second >> Y[count].second >> Z[count].second;
		getline(input,temp);
		Y[count].first = X[count].first;
		Z[count].first = X[count].first;
		count++;
	}
	input.close();
	double timeInDays;
	double bestFitness = 0;
	double worstFitness = 1;
	int b,w;
	char timeStr[9];
	for (unsigned int i=0;i<o.size();i++) {
		//iterating over all the organisms in the population
		o[i].fitness = 0;
		for (int j=0;j<observations;j++) {	
			timeInDays = X[j].first;
			pair<pair<double,double>,double> xyz=
              polarToXYZ(o[i],0,keplerian(o[i],0,timeInDays));
			o[i].fitness += pow(xyz.first.first-X[j].second,2) +
				pow(xyz.first.second-Y[j].second,2) + pow(xyz.second-Z[j].second,2);
		}
		o[i].fitness /= observations; //to normalise
		o[i].fitness = 1/o[i].fitness;
		if (o[i].fitness > bestFitness) {
			bestFitness = o[i].fitness;
			b = i;
		}
		if (o[i].fitness < worstFitness) {
			worstFitness = o[i].fitness;
			w = i;
		}
	}
	_strtime_s(timeStr);
	ofstream genfitness;
	genfitness.open("X:\\data\\genfitness.txt", ios::app);
	genfitness.setf(ios::scientific);
	cout << "Best fitness is " << bestFitness << ".\t";
	if (LOG) genfitness << bestFitness << "\t";
	else genfitness << "Best fitness is " << bestFitness << ".\t";
	cout << "Worst is " << worstFitness << ".\t" << timeStr << endl;
	if (LOG) genfitness << worstFitness << "\t" << timeStr << "\t";
	else genfitness << "Worst is " << worstFitness << ".\t" << timeStr << endl;
	genfitness.close();

	best.open("X:\\data\\best.txt");
	worst.open("X:\\data\\worst.txt");
	best << timeStr << endl << "Best fitness = " << bestFitness << endl;
	worst << timeStr << endl << "Worst fitness = " << worstFitness << endl;
	//right now this only writes the first elements
	best << "a:\t" << o[b].a[0] << "\ne:\t" << o[b].e[0] << "\ni:\t";
	best << o[b].inclin[0] << "\nnode:\t" << o[b].node[0] << "\nw:\t";
    best << o[b].w[0] << "\nm0:\t" << o[b].m0[0] << endl;
	worst << "a:\t" << o[w].a[0] << "\ne:\t" << o[w].e[0] << "\ni:\t";
	worst << o[w].inclin[0] << "\nnode:\t" << o[w].node[0] << "\nw:\t";
    worst << o[w].w[0] << "\nm0:\t" << o[w].m0[0] << endl;
	best.close();
	worst.close();
	
	return bestFitness;
}


pair<organism,organism> straightCrossover(organism o1, organism o2) {
	pair<organism,organism> o;
	o.first = o1;
	o.second = o2;

	double crossoverLevel = 0.25;

	double aR,eR,iR,nodeR,wR,m0R; //these are randoms between 0 and 1
	aR = (double)rand()/32768;
	eR = (double)rand()/32768;
	iR = (double)rand()/32768;
	nodeR = (double)rand()/32768;
	wR = (double)rand()/32768;
	m0R = (double)rand()/32768;

	long long itemp;
	double dtemp;

	if (aR < crossoverLevel) {
		itemp = o.first.a[0];
		o.first.a[0] = o.second.a[0];
		o.second.a[0] = itemp;
	}
	else if (eR < crossoverLevel) {
		dtemp = o.first.e[0];
		o.first.e[0] = o.second.e[0];
		o.second.e[0] = dtemp;
	}
	else if (iR < crossoverLevel) {
		dtemp = o.first.e[0];
		o.first.e[0] = o.second.e[0];
		o.second.e[0] = dtemp;
	}
	else if (nodeR < crossoverLevel) {
		dtemp = o.first.inclin[0];
		o.first.inclin[0] = o.second.inclin[0];
		o.second.inclin[0] = dtemp;
	}
	else if (wR < crossoverLevel) {
		dtemp = o.first.node[0];
		o.first.node[0] = o.second.node[0];
		o.second.node[0] = dtemp;
	}
	else if (m0R < crossoverLevel) {
		dtemp = o.first.m0[0];
		o.first.m0[0] = o.second.m0[0];
		o.second.m0[0] = dtemp;
	}


	return o;
}

void mutate(organism& o) {
	organism mutated;
	double mutateLevel = 0.15;
	mutated.create(o.satellites);
	double aR,eR,iR,nodeR,wR,m0R; //these are randoms between 0 and 1
	aR = (double)rand()/32768;
	eR = (double)rand()/32768;
	iR = (double)rand()/32768;
	nodeR = (double)rand()/32768;
	wR = (double)rand()/32768;
	m0R = (double)rand()/32768;

	double mutatePercent;
	mutatePercent = (double)rand()/32768/2 - 0.25 + 1;
    //between -.25 + 1 and .25 + 1 = .75 and 1.25

	if (aR < mutateLevel) o.a[0] *= mutatePercent;
    //a doesn't have strict bounds (besides >0)
	if (eR < mutateLevel) {
		o.e[0] *= mutatePercent;
		while (o.e[0]>=1) o.e[0]--;
	}
	if (iR < mutateLevel) {
		o.inclin[0] *= mutatePercent;
		while (o.inclin[0]>=2*M_PI) o.inclin[0] -= 2*M_PI;
	}
	if (nodeR < mutateLevel) {
		o.node[0] *= mutatePercent;
		while (o.node[0]>=2*M_PI) o.node[0] -= 2*M_PI;
	}
	if (wR < mutateLevel) {
		o.w[0] *= mutatePercent;
		while (o.w[0]>=2*M_PI) o.w[0] -= 2*M_PI;
	}
	if (m0R < mutateLevel) {
		o.m0[0] *= mutatePercent;
		while (o.m0[0]>=2*M_PI) o.m0[0] -= 2*M_PI;
	}

	if (aR < mutateLevel/2) o.a[0] = mutated.a[0];
	if (eR < mutateLevel/2) o.e[0] = mutated.e[0];
	if (iR < mutateLevel/2) o.inclin[0] = mutated.inclin[0];
	if (nodeR < mutateLevel/2) o.node[0] = mutated.node[0];
	if (wR < mutateLevel/2) o.w[0] = mutated.w[0];
	if (m0R < mutateLevel/2) o.m0[0] = mutated.m0[0];

}


pair<double,double> keplerian(particle p, double timeInDays) {
	pair<double,double> result;
	result.first=0;
	result.second=0;

	double timeElapsed = timeInDays - EPOCH; //in days from EPOCH
	timeElapsed *= 86400; //now in seconds

	double M; //mean anomaly
	double n; //mean motion, in radians/sec

	//unable to represent p.a^3:
	n = (double)p.mu;
	n /= p.a;
	n /= p.a;
	n /= p.a;
	n = sqrt(n);
	M = p.m0 + n*timeElapsed;
	//radians + (radians/s)*s, from http://en.wikipedia.org/wiki/Mean_anomaly

	double E; //eccentric anomaly
	E = M;

	for (int i=0;i<8;i++) {
		E = M + p.e*sin(E);
		//from http://en.wikipedia.org/wiki/Eccentric_anomaly
	}


	double nu; //true anomaly, between -PI and PI
	nu = atan(sqrt((1.+p.e)/(1.-p.e))*tan(E/2.))*2;
	if (nu<0) nu += 2*M_PI;
	//putting nu between 0 and 2*PI, for theta/phi calculations in keplerian()


	double rho; //distance from major body, in same units as a (metres)
	rho = p.a*((1.-pow(p.e,2))/(1.+(p.e*cos(nu))));


	result.first = rho/1000.; //this puts it in km for simplicity's sake
	result.second = nu;

	return result;
}
pair<pair<double,double>,double>
  polarToXYZ(particle p, pair<double,double> polar) {
	pair<double,double> xy;
	pair<pair<double,double>,double> result;
	double x = 0,y = 0,z = 0;
	
	double rho, phi, theta, nu;
	rho = polar.first;
	nu = polar.second;

	double scaleTheta,scalePhi;
	scaleTheta = cos(p.inclin);
	scalePhi = sin(p.inclin);

	theta = scaleTheta*(p.w+nu)+p.node; //should be 0 to 2*PI
	phi = (M_PI/2)-scalePhi*(p.w+nu); //should be 0 to 2*PI

	//convert to xyz: in same units as rho (km)
	x = rho*sin(phi)*cos(theta);
	y = rho*sin(phi)*sin(theta);
	z = rho*cos(phi);


	xy.first = x;
	xy.second = y;
	result.first = xy;
	result.second = z;

	return result;

}
void outputSwarm(vector<particle>& p) {
	ofstream swarm;
	swarm.open("X:\\data\\swarm.txt");
	char timeStr[9];
	_strtime_s(timeStr);
	swarm << timeStr << endl << endl;
	for (unsigned int i=0;i<p.size();i++) {
		swarm << i+1 << endl;
		swarm << p[i].fitness << endl;
		swarm << p[i].a << "," << p[i].e << "," << p[i].inclin << endl << p[i].node;
		swarm << "," << p[i].w << "," << p[i].m0 << endl << endl;
	}
	swarm.close();
}
particle rankFitnessXYZ(vector<particle>& p,
  const int observations, particle& pBest) {
	vector<pair<double,double>> X(observations),Y(observations),Z(observations);
	ifstream input;
	ofstream best;
	ofstream worst;
	string temp;

	input.open(inputXYZ);
	while (!input.eof()) {
		getline(input,temp);
		if (temp.compare("$$SOE")==0) {
			break;
		}
	}
	//file should be at EOF or at $$SOE here
	int count = 0;
	while ((!input.eof())&&(input.peek()!='$')) {
		input >> X[count].first;
		if (!test) getline(input,temp);
		input >> X[count].second >> Y[count].second >> Z[count].second;
		getline(input,temp);
		Y[count].first = X[count].first;
		Z[count].first = X[count].first;
		count++;
	}
	input.close();
	double timeInDays;
	double bestFitness = 0;
	double worstFitness = 1;
	int b,w;
	char timeStr[9];
	for (unsigned int i=0;i<p.size();i++) {
    //iterating over all the particles in the swarm
		p[i].fitness = 0;
		for (int j=0;j<observations;j++) {
			timeInDays = X[j].first;
			pair<pair<double,double>,double> xyz =
              polarToXYZ(p[i],keplerian(p[i],timeInDays));
			p[i].fitness += pow(xyz.first.first-X[j].second,2)
				+ pow(xyz.first.second-Y[j].second,2) + pow(xyz.second-Z[j].second,2);
		}
		p[i].fitness /= observations;
		p[i].fitness = 1/p[i].fitness;
		if (p[i].fitness > bestFitness) {
			bestFitness = p[i].fitness;
			b = i;
		}
		if (p[i].fitness < worstFitness) {
			worstFitness = p[i].fitness;
			w = i;
		}
	}
	if (bestFitness > pBest.fitness) pBest = p[b];
	_strtime_s(timeStr);
	ofstream genfitness;
	if (LOG) {
		genfitness.open("X:\\data\\genfitness.txt", ios::app);
		genfitness.setf(ios::scientific);
		genfitness << bestFitness << "\t";
		genfitness << worstFitness << "\t" << timeStr << "\t";
		genfitness.close();
	}

	cout << "Best fitness is " << bestFitness << ".\t";
	cout << "Worst is " << worstFitness << ".\t" << timeStr << endl;
	best.open("X:\\data\\best.txt");
	worst.open("X:\\data\\worst.txt");
	best << timeStr << endl << "Best fitness = " << bestFitness << endl;
	worst << timeStr << endl << "Worst fitness = " << worstFitness << endl;
	best << "a:\t" << p[b].a << "\ne:\t" << p[b].e << "\ni:\t";
	best << p[b].inclin << "\nnode:\t" << p[b].node << "\nw:\t";
    best << p[b].w << "\nm0:\t" << p[b].m0 << endl;
	worst << "a:\t" << p[w].a << "\ne:\t" << p[w].e << "\ni:\t";
	worst << p[w].inclin << "\nnode:\t" << p[w].node << "\nw:\t";
    worst << p[w].w << "\nm0:\t" << p[w].m0 << endl;
	
	best << endl << "Best so far = " << pBest.fitness << endl;
	best << "a:\t" << pBest.a << "\ne:\t" << pBest.e << "\ni:\t";
	best << pBest.inclin << "\nnode:\t" << pBest.node << "\nw:\t";
    best << pBest.w << "\nm0:\t" << pBest.m0 << endl;

	best.close();
	worst.close();
	
	return p[b]; //to aid selection process
}
int _tmain(int argc, _TCHAR* argv[]) {

	const int maxObservations = 31*12+1;
    //42 for abbrev data sets, 51 for test, 31*12+1 for 2h sets
	const int popSize = 100;

	const double replacementRate = 0.7;
	const double mutationRate = 0.15;
	const int generations = 3000000;

	const bool usePSO = false; //else use GA

	int maxSatellites = 1;
	//in a multiple-satellite implementation, this should be at least the length
	  //of the longest x/y vector

	double bestFitness;
	particle pBest,lBest;
	int j;

	ofstream genfitness;
	//genfitness.open("X:\\data\\genfitness.txt");
	//genfitness.close(); //clears file contents

    //creates a vector of _observ_'s of size _maxObservations_
	vector<observ> data(maxObservations);
	cout.setf(ios::scientific); //or fixed
	srand((int)time(0)); //randomise

	vector<organism> pop(popSize);
	vector<particle> swarm(popSize);	
	if (usePSO) {
		cout << "Swarm of " << popSize << " created.\n";
		for (int i=0;i<popSize;i++) swarm[i].create();
		cout << "Swarm initialised.\n";
	}
	else {
		cout << "Organisms created.\n";
		for (unsigned int i=0;i<popSize;i++) pop[i].create(maxSatellites);
		cout << "Organisms initialised.\n";
	}


	pair<double,double> result;
	if (usePSO) result = keplerian(swarm[0],2454600);
	else result = keplerian(pop[0],0,2454600);

	cout<<"r: "<<result.first<<endl<<"v: "<<result.second<<endl;

	pair<pair<double,double>,double> xyz;
	if (usePSO) xyz = polarToXYZ(swarm[0],result);
	else xyz = polarToXYZ(pop[0],0,result);
	cout<<"x: "<<xyz.first.first<<endl;
	cout<<"y: "<<xyz.first.second<<endl;
	cout<<"z: "<<xyz.second<<endl;

	if (usePSO) {
		j = 0;
		while (j<generations) { //and error criteria
			cout << j+1 << ". ";
			if (LOG) {
				genfitness.open("X:\\data\\genfitness.txt", ios::app);
				genfitness << j+1 << "\t";
				genfitness.close();
			}
			lBest = rankFitnessXYZ(swarm, data.size(), pBest);

			if (LOG) {
				genfitness.open("X:\\data\\genfitness.txt", ios::app);
				genfitness.setf(ios::scientific);

				double meanFitness = 0.;
				for (unsigned int i=0;i<swarm.size();i++) {
					meanFitness += swarm[i].fitness;
				}
				meanFitness /= swarm.size();
                //this is now the correct value of 1/meanFitness
	
				double stdev = 0.0;
				for (unsigned int i=0;i<swarm.size();i++) {
					stdev += pow(meanFitness - swarm[i].fitness,2);
				}
				stdev /= swarm.size();
				stdev = sqrt(stdev);
	
				genfitness << meanFitness << "\t" << stdev << "\n";
				//file now reads generation[\t]bestFitness[\t]worstFitness[\t]
				  //time[\t]meanFitness[\t]stdev[\n]
				genfitness.close();
			}

			outputSwarm(swarm);
			for (unsigned int i=0;i<swarm.size();i++) {
				swarm[i].update(pBest,lBest);
			}
			//tapered vMAX enforcement
			if (taper) {
				if (j==5000) for (unsigned int i=0;i<swarm.size();i++) {
					swarm[i].newVmax(1.0);
				}
				else if (j==10000) for (unsigned int i=0;i<swarm.size();i++) {
					swarm[i].newVmax(0.75);
				}
				else if (j==15000) for (unsigned int i=0;i<swarm.size();i++) {
					swarm[i].newVmax(0.5);
				}
				else if (j==20000) for (unsigned int i=0;i<swarm.size();i++) {
					swarm[i].newVmax(0.25);
				}
				else if (j==25000) for (unsigned int i=0;i<swarm.size();i++) {
					swarm[i].newVmax(0.1);
				}
				else if (j==30000) for (unsigned int i=0;i<swarm.size();i++) {
					swarm[i].newVmax(0.05);
				}
			}
			j++;
		}
	}
	else for (j=0;j<generations;j++) { //begin GA loop
		genfitness.open("X:\\data\\genfitness.txt", ios::app);
		cout << j+1 << ". ";
		if (LOG) genfitness << j+1 << "\t";
		else genfitness << j+1 << ". "; //inserts generation count, e.g., "1. "
		genfitness.close();

		//get fitness of all organisms by measuring SSE of XYZ coords
		bestFitness = rankFitnessXYZ(pop, data.size());
		double totalFitness = 0.;
		double probability;
		double selectRand;
		double mutateRand;
		int numSelected = 0;
	
		outputOrganisms(pop);

		//evolve: select, crossover, mutate, update, then re-evaluate fitness
		//select:
		for (unsigned int i=0;i<pop.size();i++) {

			totalFitness+=pop[i].fitness;
			pop[i].selected = false;
		}

		do { //new roulette wheel selection method
			selectRand = ((double)rand()/32768)*totalFitness;
			for (unsigned int i=0;i<pop.size();i++) {
				//probability = pop[i].fitness/totalFitness;
				if (!pop[i].selected) {
					selectRand -= pop[i].fitness;
					if (selectRand <= 0) {
						pop[i].selected = true;
						numSelected++;
						totalFitness -= pop[i].fitness;
						//exclude this organism from future roulette selections
						break;
					}
				}
			}
		} while (numSelected < replacementRate*pop.size());
		int numToCross = pop.size() - numSelected; //numToCross must be even
	
		//copy selected
		vector<organism> popNew;
		for (unsigned int i=0;i<pop.size();i++) {
			if (pop[i].selected) popNew.push_back(pop[i]);
		}

//at this point, popNew contains only selected orgs, not crossed-over ones
//this is where we output to genfitness the stdev and mean of fitness
		if (LOG) {
			genfitness.open("X:\\data\\genfitness.txt", ios::app);
			genfitness.setf(ios::scientific);

            double meanFitness = 0.;
            for (unsigned int i=0;i<popNew.size();i++) {
                meanFitness += popNew[i].fitness;
            }
            meanFitness /= popNew.size();
            //this is now the correct value of 1/meanFitness

            double stdev = 0.0;
            for (unsigned int i=0;i<popNew.size();i++) {
                stdev += pow(meanFitness - popNew[i].fitness,2);
            }
            stdev /= popNew.size();
            stdev = sqrt(stdev);

            genfitness << meanFitness << "\t" << stdev << "\t";
			//file now reads generation[\t]bestFitness[\t]worstFitness[\t]
              //time[\t]meanFitness[\t]stdev[\t]
			genfitness.close();
		}

		//crossover in popNew:
		for (int i=0;i<numToCross/2;i++) {
			int t1,t2;
			t1 = rand() % numSelected;
            //we use numSelected so we're only crossing over old organisms
			t2 = rand() % numSelected;
			pair<organism,organism> newOrg = straightCrossover(popNew[t1],popNew[t2]);
			popNew.push_back(newOrg.first);
			popNew.push_back(newOrg.second);
		}

		//mutate in popNew:
		for (unsigned int i=0;i<popNew.size();i++) {
			mutateRand = (double)rand()/32768;
			if (mutateRand<mutationRate) mutate(popNew[i]);
		}

//here output to genfit the std, mean of fitness of the entire crossed-over pop.
		if (LOG) {
			genfitness.open("X:\\data\\genfitness.txt", ios::app);
			genfitness.setf(ios::scientific);

			double meanFitness = 0.;
			for (unsigned int i=0;i<popNew.size();i++) {
				meanFitness += popNew[i].fitness;
			}
			meanFitness /= popNew.size();
            //this is now the correct value of 1/meanFitness

			double stdev = 0.0;
			for (unsigned int i=0;i<popNew.size();i++) {
				stdev += pow(meanFitness - popNew[i].fitness,2);
			}
			stdev /= popNew.size();
			stdev = sqrt(stdev);

			genfitness << meanFitness << "\t" << stdev << endl;
			//file now reads generation[\t]bestFitness[\t]worstFitness[\t]time[\t]
				//meanFitnessSelected[\t]stdevSelected[\t]meanFitnessAll[\t]stdevAll[\n]
			genfitness.close();
		}

		//update population:
		pop = popNew;

	
	} //end loop
	//return organism with highest fitness


	return 0;
}
\end{verbatim}
\normalsize
        \section{organism.h}
\small
\begin{verbatim}
// organism.h

class organism {
public:
	organism();
	double fitness;
	bool selected;
	void create(int);
	int satellites;
	static const unsigned long long mu = 37931187000000000;
	//standard G for jupiter (126686534000000000) in m^3/s^2
	//test = 330531054456064
	//saturn = 37931187000000000
	//note that orbital period P = (2*pi)/sqrt(mu)*a^(3/2) (in seconds)

	long long a[1];
	double e[1],inclin[1],node[1],w[1],m0[1];
};

organism::organism() {
	fitness = -1.;
	satellites = 0;
}

void organism::create(int maxSize) { //create random organism
	satellites = rand() % maxSize + 1;
	double period;
	for (int i=0;i<satellites;i++) {
		a[0] = (((long long)rand()+1)*8000);
        //Io: 25000, Himalia: 700000, test: 6000, Atlas: 8000
		//a should be on the order of 421,800,000 for Io's orbit.
		//a should be on the order of 11,461,000,000 for Himalia's orbit.
		e[0] = ((double)rand()/32768); //e should be between 0 and 1
		inclin[0] = ((double)rand()/32768*2*M_PI);
		node[0] = ((double)rand()/32768*2*M_PI);
		w[0] = ((double)rand()/32768*2*M_PI);
		m0[0] = ((double)rand()/32768*2*M_PI);

	}
}
\end{verbatim}
\normalsize
        \section{particle.h}
\small
\begin{verbatim}
// particle.h

class particle {
public:
	double c1, c2; //learning factors
	double vMAX; //set vMAX below
	double evMAX; //eccentricity maximum velocity
	particle();
	void create();
	double fitness;
	static const unsigned long long mu = 37931187000000000;
	//standard G for jupiter (126686534000000000) in m^3/s^2
	//test = 330531054456064
	//saturn = 37931187000000000
	
	long long a; //a could be unsigned as well
	double e, inclin, node, w, m0;

	long long aV;
	double eV, iV, nV, wV, m0V;

	void update(particle, particle);
	void newVmax(double);
};

particle::particle() {
	vMAX = 1.0;
	evMAX = 0.1;
	
	fitness = -1.;
	c1 = 2.;
	c2 = 2.;

	//particle positions
	a = 0;
	e = 0.;
	inclin = 0.;
	node = 0.;
	w = 0.;
	m0 = 0.;

	//particle velocities
	aV = 0;
	eV = 0.;
	iV = 0.;
	nV = 0.;
	wV = 0.;
	m0V = 0.;
}

void particle::create() {
	a = ((long long)rand()+1)*8000;
	e = (double)rand()/32768;
	inclin = (double)rand()/32768*2*M_PI;
	node = (double)rand()/32768*2*M_PI;
	w = (double)rand()/32768*2*M_PI;
	m0 = (double)rand()/32768*2*M_PI;
}

void particle::update(particle pBest, particle lBest) {

	double delta[8];
	delta[0] = pBest.inclin - inclin;
	delta[1] = lBest.inclin - inclin;
	delta[2] = pBest.node - node;
	delta[3] = lBest.node - node;
	delta[4] = pBest.w - w;
	delta[5] = lBest.w - w;
	delta[6] = pBest.m0 - m0;
	delta[7] = lBest.m0 - m0;

	for (int i=0;i<8;i++) {
		if (delta[i]>M_PI) delta[i] -= 2*M_PI;
		if (delta[i]<(-1.*M_PI)) delta[i] += 2*M_PI;
	}
	
	//this moves the particles randomly -> one parameter but maybe not another
	/*aV += c1*((double)rand()/32768)*((double)(pBest.a - a))
		+ c2*((double)rand()/32768)*((double)(lBest.a - a));
	eV += c1*((double)rand()/32768)*(pBest.e - e)
		+ c2*((double)rand()/32768)*(lBest.e - e);
	iV += c1*((double)rand()/32768)*delta[0]
		+ c2*((double)rand()/32768)*delta[1];
	nV += c1*((double)rand()/32768)*delta[2]
		+ c2*((double)rand()/32768)*delta[3];
	wV += c1*((double)rand()/32768)*delta[4]
		+ c2*((double)rand()/32768)*delta[5];
	m0V += c1*((double)rand()/32768)*delta[6]
		+ c2*((double)rand()/32768)*delta[7];*/

	//this treats the entire vector as one: particles move towards best values
	  //across all parameters simultaneously
	double rand1, rand2;
	rand1 = (double)rand()/32768;
	rand2 = (double)rand()/32768;
	aV += c1*rand1*((double)(pBest.a - a)) + c2*rand2*((double)(lBest.a - a));
	eV += c1*rand1*(pBest.e - e) + c2*rand2*(lBest.e - e);
	iV += c1*rand1*delta[0] + c2*rand2*delta[1];
	nV += c1*rand1*delta[2] + c2*rand2*delta[3];
	wV += c1*rand1*delta[4] + c2*rand2*delta[5];
	m0V += c1*rand1*delta[6] + c2*rand2*delta[7];

	eV = min(eV,(double)evMAX);
	eV = max(eV,(double)(evMAX*(-1)));
	iV = min(iV,(double)vMAX);
	iV = max(iV,(double)(vMAX*(-1)));
	nV = min(nV,(double)vMAX);
	nV = max(nV,(double)(vMAX*(-1)));
	wV = min(wV,(double)vMAX);
	wV = max(wV,(double)(vMAX*(-1)));
	m0V = min(m0V,(double)vMAX);
	m0V = max(m0V,(double)(vMAX*(-1)));


	a += aV;
	a = max(a,(long long)1);
	
	e += eV;
	e = min(e,0.999999);
	e = max(e,0.);

	inclin += iV;
	while (inclin >= 2*M_PI) inclin -= 2*M_PI;
	while (inclin < 0) inclin += 2*M_PI;
	
	node += nV;
	while (node >= 2*M_PI) node -= 2*M_PI;
	while (node < 0) node += 2*M_PI;
	
	w += wV;
	while (w >= 2*M_PI) w -= 2*M_PI;
	while (w < 0) w += 2*M_PI;
	
	m0 += m0V;
	while (m0 >= 2*M_PI) m0 -= 2*M_PI;
	while (m0 < 0) m0 += 2*M_PI;
}

void particle::newVmax(double newV) {
	vMAX = newV;
}
\end{verbatim}
\normalsize

    \chapter{Source Code: MATLAB Scripts}
        \section{ANN\emph{z} Script}
\small
\begin{verbatim}
% part2annz.m

function [a b c] = part2annz(file)

%for ANNz
fid = fopen(file, 'rt');
data = fscanf(fid, '%f %f %f', [3 inf]);
fclose(fid);

data = data';

count = size(data,1);

for i = 1:count
    data(i,4) = (data(i,1)-data(i,2))^2;
end;

rmse = 0;
del1 = 0;
del2 = 0;
del3 = 0;
for i = 1:count
    rmse = rmse + data(i,4);
    if data(i,4) <= .01, del1 = del1 + 1; end;
    if data(i,4) <= .04, del2 = del2 + 1; end;
    if data(i,4) <= .09, del3 = del3 + 1; end;
end;

rmse = rmse/count;
rmse = sqrt(rmse),

del1 = del1/count,
del2 = del2/count,
del3 = del3/count,

rmsnoise = sqrt(sum(data(:,3).*data(:,3))/count);

errors(row,:) = [rmsnoise rmse del1 del2 del3];
row = row + 1;

close all;

figure('Position', [20 540 560 420]);
a = plot(data(:,1),data(:,2),'o','MarkerSize',2);line([0 6],[0 6],'Color','r');
figure('Position', [620 60 560 420]);
b = plot(data(:,1),data(:,3),'o','MarkerSize',2);
figure('Position', [20 60 560 420]);
c = plot(data(:,1),abs(data(:,2)-data(:,1)),'o','MarkerSize',2);
\end{verbatim}
\normalsize
        \section{MLP/Netlab Script}
\small
\begin{verbatim}
% part2mlp.m

act = 'linear';
nout = 1;
comSize = 5; % committee size
nhidden = 10;
iter = 500;

[x4,t4] = datread('x:\quasartrain-UGRI.dat',4,1,37136);
[x4t,t4t] = datread('x:\quasartest-UGRI.dat',4,1,9284);
[x5,t5] = datread('x:\quasartrain-ugriz.dat',5,1,37136);
[x5t,t5t] = datread('x:\quasartest-ugriz.dat',5,1,9284);
[x6,t6] = datread('x:\quasartrain-ugriz6192.dat',5,1,4954);
[x6t,t6t] = datread('x:\quasartest-ugriz6192.dat',5,1,1238);

[x7,t7] = datread('x:\quasartrain-UGRIZJH.dat',7,1,4954);
[x7t,t7t] = datread('x:\quasartest-UGRIZJH.dat',7,1,1238);
[x8,t8] = datread('x:\quasartrain-ugrizjhk.dat',8,1,4954);
[x8t,t8t] = datread('x:\quasartest-ugrizjhk.dat',8,1,1238);

[x9,t9] = datread('x:\quasartrain-ugrizUGRI.dat',9,1,37136);
[x9t,t9t] = datread('x:\quasartest-ugrizUGRI.dat',9,1,9284);

[x01,t01] = datread('x:\trainz01.dat',4,1,10159);
[x01t,t01t] = datread('x:\testz01.dat',4,1,2540);
[x12,t12] = datread('x:\trainz12.dat',4,1,19166);
[x12t,t12t] = datread('x:\testz12.dat',4,1,4792);
[x2,t2] = datread('x:\trainz2.dat',4,1,7810);
[x2t,t2t] = datread('x:\testz2.dat',4,1,1953);
[x02,t02] = datread('x:\trainz02.dat',4,1,29325);
[x02t,t02t] = datread('x:\testz02.dat',4,1,7332);

trainx = {x4, x5, x6, x7, x8, x9, x01, x12, x2, x02};
traint = {t4, t5, t6, t7, t8, t9, t01, t12, t2, t02};
testx = {x4t, x5t, x6t, x7t, x8t, x9t, x01t, x12t, x2t, x02t};
testt = {t4t, t5t, t6t, t7t, t8t, t9t, t01t, t12t, t2t, t02t};

clear skip;
skip(10, comSize) = 0;
skipActive = 0;
% if skipActive = 1 but skip()() is all 0s, this retrains the MLPs by iter

% skipActive = 1;
% skip(1,:) = 1;
% skip(2,:) = 1;
% skip(3,:) = 1;
% skip(4,:) = 1;
% skip(5,:) = 1;
% skip(6,:) = 1;
% skip(7,:) = 1;
% skip(8,:) = 1;
% skip(9,:) = 1;
% skip(10,:) = 1;

if skipActive==1, savemlps=mlps; end;
if skipActive==0, mlps = cell(6,comSize); end;
y = cell(6,comSize);
rms(6,comSize) = 0;
comOutput = cell(6);
comRMS(6) = 0;
meanRMS(6) = 0;

if skipActive==0
    for j = 1:size(mlps,2)
        mlps{1,j} = mlp(4, nhidden, nout, act);
        mlps{2,j} = mlp(5, nhidden, nout, act);
        mlps{3,j} = mlp(5, nhidden, nout, act);
        mlps{4,j} = mlp(7, nhidden, nout, act);
        mlps{5,j} = mlp(8, nhidden, nout, act);
        mlps{6,j} = mlp(9, nhidden, nout, act);
        mlps{7,j} = mlp(4, nhidden, nout, act);
        mlps{8,j} = mlp(4, nhidden, nout, act);
        mlps{9,j} = mlp(4, nhidden, nout, act);
        mlps{10,j} = mlp(4, nhidden, nout, act);
    end;
end;

for i = 1:size(mlps,1)
    for j = 1:size(mlps,2)
        if skip(i,j)==0
            mlps{i,j} = mlptrain(mlps{i,j}, trainx{i}, traint{i}, iter);
        end;

        y{i,j} = mlpfwd(mlps{i,j}, testx{i});
        rms(i,j) = sqrt(mean((y{i,j}-testt{i}).*(y{i,j}-testt{i})));
        fprintf(1, 'MLP %d, member %d finished.\n', i, j);
    end;
    comOutput{i} = 0;
    for j = 1:comSize, comOutput{i} = comOutput{i} + y{i,j}; end;
    comOutput{i} = comOutput{i} / comSize;
    comRMS(i)=sqrt(mean((comOutput{i}-testt{i}).*(comOutput{i}-testt{i})));
    meanRMS(i) = mean(rms(i,1:size(mlps,2)));
    fprintf(1, 'Finished with MLP committee %d.\n', i);
end;

saverms = [rms comRMS' meanRMS'];
savemlps = mlps;
\end{verbatim}
\normalsize
        \section{RBFN Script}
\small
\begin{verbatim}
% part2rbf.m

act = 'tps';
nout = 1;
comSize = 5; % committee size
nhidden = 20;
options = foptions;
options(1) = 25; % display errors/how often?
options(14) = 100; % number of iterations

[x4,t4] = datread('x:\quasartrain-UGRI.dat',4,1,37136);
[x4t,t4t] = datread('x:\quasartest-UGRI.dat',4,1,9284);
[x5,t5] = datread('x:\quasartrain-ugriz.dat',5,1,37136);
[x5t,t5t] = datread('x:\quasartest-ugriz.dat',5,1,9284);
[x6,t6] = datread('x:\quasartrain-ugriz6192.dat',5,1,4954);
[x6t,t6t] = datread('x:\quasartest-ugriz6192.dat',5,1,1238);

[x7,t7] = datread('x:\quasartrain-UGRIZJH.dat',7,1,4954);
[x7t,t7t] = datread('x:\quasartest-UGRIZJH.dat',7,1,1238);
[x8,t8] = datread('x:\quasartrain-ugrizjhk.dat',8,1,4954);
[x8t,t8t] = datread('x:\quasartest-ugrizjhk.dat',8,1,1238);

[x9,t9] = datread('x:\quasartrain-ugrizUGRI.dat',9,1,37136);
[x9t,t9t] = datread('x:\quasartest-ugrizUGRI.dat',9,1,9284);

trainx = {x4, x5, x6, x7, x8, x9};
traint = {t4, t5, t6, t7, t8, t9};
testx = {x4t, x5t, x6t, x7t, x8t, x9t};
testt = {t4t, t5t, t6t, t7t, t8t, t9t};

clear skip;
skip(6, comSize) = 0;
skipActive = 0;

skipActive = 1;
skip(1,:) = 1;
skip(2,:) = 1;
skip(3,:) = 1;
skip(4,:) = 1;
skip(5,:) = 1;
skip(6,:) = 1;

if skipActive==1, saverbfs=rbfs; end;
if skipActive==0, rbfs = cell(6,comSize); end;
a = cell(6,comSize);
rms(6,comSize) = 0;
comOutput = cell(6);
comRMS(6) = 0;
meanRMS(6) = 0;

if skipActive==0
    for j = 1:size(rbfs,2)
        rbfs{1,j} = rbf(4, nhidden, nout, act);
        rbfs{2,j} = rbf(5, nhidden, nout, act);
        rbfs{3,j} = rbf(5, nhidden, nout, act);
        rbfs{4,j} = rbf(7, nhidden, nout, act);
        rbfs{5,j} = rbf(8, nhidden, nout, act);
        rbfs{6,j} = rbf(9, nhidden, nout, act);
    end;
end;

for i = 1:size(rbfs,1)
    for j = 1:size(rbfs,2)
        if skip(i,j)==0
            rbfs{i,j} = rbftrain(rbfs{i,j}, options, trainx{i}, traint{i});
        end;

        a{i,j} = rbffwd(rbfs{i,j}, testx{i});
        rms(i,j) = sqrt(mean((a{i,j}-testt{i}).*(a{i,j}-testt{i})));
        fprintf(1, 'RBF %d, member %d finished.\n', i, j);
    end;
    comOutput{i} = 0;
    for j = 1:comSize, comOutput{i} = comOutput{i} + a{i,j}; end;
    comOutput{i} = comOutput{i} / comSize;
    comRMS(i)=sqrt(mean((comOutput{i}-testt{i}).*(comOutput{i}-testt{i})));
    meanRMS(i) = mean(rms(i,1:size(rbfs,2)));
    fprintf(1, 'Finished with RBF committee %d.\n', i);
end;

saverms = [rms comRMS' meanRMS'];
saverbfs = rbfs;
\end{verbatim}
\normalsize
    \renewcommand{\baselinestretch}{1.5}
\backmatter

\end{document}